\documentclass[10pt,journal,twoside,final]{IEEEtran}

\usepackage[table]{xcolor}
\usepackage{threeparttable}

\usepackage{comment}
\usepackage{graphicx}
\usepackage{cite}
\usepackage{color,soul}
\usepackage{amsmath}
\usepackage{url}
\usepackage{array}
\usepackage{booktabs}
\usepackage{amssymb}
\usepackage{pifont}
\usepackage{bm}
\usepackage{siunitx}
\usepackage{multicol,multirow}
\usepackage[hidelinks]{hyperref}
\usepackage{enumitem}
\usepackage{svg}
\usepackage{tikz}
\usepackage{hhline}

\usepackage{pgfplots}
\pgfplotsset{compat=newest}
\usetikzlibrary{patterns, patterns.meta}
\usepgfplotslibrary{fillbetween}

\ifCLASSOPTIONcompsoc
\usepackage[caption=false,font=normalsize,labelfont=sf,textfont=sf,subrefformat=parens,labelformat=parens]{subfig}
\else
\usepackage[caption=false,font=footnotesize,subrefformat=parens,labelformat=parens]{subfig}
\fi

\usepackage[acronym,style=mcolindex,stylemods=mcols,nomain,shortcuts,nogroupskip,nopostdot,nonumberlist]{glossaries-extra}

\setabbreviationstyle[acronym]{long-short}
\newacronym{6g}{6G}{sixth generation}

\newacronym{ae}{AE}{autoencoder}
\newacronym{AES}{AES}{advanced encryption standard}
\newacronym{AML}{AML}{adversarial machine learning}
\newacronym{aoa}{AoA}{angle-of-arrival}
\newacronym{aod}{AoD}{angle-of-departure}
\newacronym{awgn}{AWGN}{additive white Gaussian noise}

\newacronym{baae}{BAAE}{Bahdanau attention autoencoder}
\newacronym{BLE}{BLE}{Bluetooth low energy}
\newacronym{bt}{BT}{bagged tree}

\newacronym{cb}{CB}{channel-based}
\newacronym{cfr}{CFR}{channel frequency response}
\newacronym{cfo}{CFO}{carrier frequency offset}
\newacronym{cir}{CIR}{channel impulse response}
\newacronym{cnn}{CNN}{convolutional neural network}
\newacronym{COTS}{COTS}{commercial off-the-shelf}
\newacronym{cr}{CR}{challenge-response}
\newacronym{csi}{CSI}{channel state information}

\newacronym{doa}{DOA}{direction of arrival}
\newacronym{drl}{DRL}{deep reinforcement learning}
\newacronym{dt}{DT}{decision tree}
\newacronym{DUT}{DUT}{devices under test}

\newacronym{pdp}{PDP}{power-delay profile}
\newacronym{el}{EL}{enseamble learning}
\newacronym{elf}{ELF}{extremely low frequency}

\newacronym{fa}{FA}{false alarm}
\newacronym{fft}{FFT}{fast Fourier transform}

\newacronym{gan}{GAN}{generative adversarial network}
\newacronym{glrt}{GLRT}{generalized likelihood-ratio test}
\newacronym{gnn}{GNN}{graph NN}
\newacronym{gnss}{GNSS}{global navigation satellite systems}
\newacronym{gpr}{GPR}{Gaussian process regression}
\newacronym{gps}{GPS}{global positioning system}
\newacronym{gru}{GRU}{gated recurrent unit}

\newacronym{iot}{IoT}{Internet of Things}
\newacronym{iqi}{IQI}{in-phase and quadrature imbalance}

\newacronym{kf}{KF}{Kalman filter}

\newacronym{leo}{LEO}{low Earth orbit}
\newacronym{llm}{LLM}{large language models}
\newacronym{los}{LOS}{line-of-sight}
\newacronym{lora}{LoRa}{long range}
\newacronym{lstm}{LSTM}{long-short term memory}
\newacronym{lt}{LT}{likelihood test}
\newacronym{lte}{LTE}{long-term evolution}
\newacronym{lrt}{LRT}{likelihood ratio test}

\newacronym{knn}{KNN}{K-nearest neighbors}

\newacronym{md}{MD}{misdetection}
\newacronym{mimo}{MIMO}{multiple-input multiple-output}
\newacronym{ml}{ML}{machine learning}
\newacronym{mse}{MSE}{mean square error}

\newacronym{nist}{NIST}{national institute of standards and technology}
\newacronym{nn}{NN}{neural network}
\newacronym{nlos}{NLOS}{non-line-of-sight}

\newacronym{oc-svm}{OC-SVM}{one-class support vector machine}
\newacronym{ofdm}{OFDM}{orthogonal frequency-division multiplexing}

\newacronym{pdf}{pdf}{probability density function}
\newacronym{pla}{PLA}{physical layer authentication}
\newacronym{prn}{PRN}{pseudo-random noise}
\newacronym{puf}{PUF}{physically-unclonable function}

\newacronym{quadriga}{QuaDRiGa}{quasideterministic radio channel generator}

\newacronym{rf}{RF}{radio frequency}
\newacronym{rffi}{RFFI}{radio frequency fingerprint identification}
\newacronym{rff}{RFF}{radio-frequency fingerprint}
\newacronym{ris}{RIS}{reflective intelligent surface}
\newacronym{rl}{RL}{reinforcement learning}
\newacronym{rms}{RMS}{root-mean square}
\newacronym{rnn}{RNN}{recurrent neural network}
\newacronym{rss}{RSS}{received signal strength}
\newacronym{rssi}{RSSI}{received signal strength indicator}

\newacronym{scm}{SCM}{normalized sample covariance matrix}
\newacronym{sdr}{SDR}{software-defined radio}
\newacronym{snr}{SNR}{signal-to-noise ratio}
\newacronym{svm}{SVM}{support vector machine}

\newacronym{tdoa}{TDOA}{time difference of arrival}
\newacronym{tle}{TLE}{two-line element}
\newacronym{toa}{TOA}{time of arrival}

\newacronym{uwac}{UWAC}{underwater acoustic communications}
\newacronym{usrp}{USRP}{universal software radio peripheral}
\newacronym{UWB}{UWB}{Ultra-Wideband}
\newacronym{uwoc}{UWOC}{underwater optical communications}

\newacronym{vae}{VAE}{variational autoencoder}
\newacronym{v2v}{V2V}{vehicle to vehicle}
\newacronym{vlc}{VLC}{visible-light communications}
\newacronym{vanet}{VANET}{vehicular ad-hoc network}

\makeglossaries

\newcolumntype{L}[1]{>{\raggedright\let\newline\\\arraybackslash\hspace{0pt}}m{#1}}
\newcolumntype{C}[1]{>{\centering\let\newline\\\arraybackslash\hspace{0pt}}m{#1}}
\newcolumntype{R}[1]{>{\raggedleft\let\newline\\\arraybackslash\hspace{0pt}}m{#1}}

\newcommand{\etal}{\textit{et al. }}

\newcommand{\cmark}{\ding{51}}%
\newcommand{\xmark}{\ding{55}}%
\newcommand{\omark}{\ding{109}}%

\newlength\figurewidth
\newlength\figureheight
\renewcommand{\arraystretch}{1.15} 

\begin{document}

\title{Physical Layer-Based Device Fingerprinting For Wireless Security: From Theory To Practice}

\author{Junqing~Zhang,~\IEEEmembership{Senior~Member,~IEEE},
Francesco~Ardizzon,~\IEEEmembership{Member,~IEEE},
Mattia~Piana,~\IEEEmembership{Graduate~Student~Member,~IEEE},
Guanxiong~Shen,
and 
Stefano~Tomasin,~\IEEEmembership{Senior~Member,~IEEE}

\thanks{Manuscript received xxx; revised xxx; accepted xxx. Date of publication xxx; date of current version xxx. The work of J. Zhang was supported in part by the UK Engineering and Physical Sciences Research Council (EPSRC) under grant ID EP/Y037197/1 and in part by the UK Royal Society Research Grants RGS$\backslash$R1$\backslash$231435. The work of J. Zhang and S. Tomasin was also supported by the EU COST Action CA22168 - Physical layer security for trustworthy and resilient 6G systems (6G-PHYSEC). The work of M. Piana was funded by the European Commission through the Horizon Europe/JU SNS project ROBUST-6G (Grant Agreement no. 101139068). The work of G. Shen was supported in part by the National Natural Science Foundation of China under Grant 62401138. The work of S. Tomasin was supported by the project ISP5G+ (CUP D33C22001300002), which is part of the SERICS program (PE00000014) under the NRRP MUR program funded by the EU-NGEU.
For the purpose of open access, the authors have applied a Creative Commons Attribution (CC BY) licence to any Accepted Manuscript version arising.
The review of this paper was coordinated by xxx.  
\textit{(Corresponding author: Junqing Zhang.)}}
\thanks{J.~Zhang is with the Department of Electrical Engineering and Electronics, University of Liverpool, Liverpool, L69 3GJ, United Kingdom. (email: junqing.zhang@liverpool.ac.uk)}
\thanks{F. Ardizzon, M. Piana, and S.~Tomasin are with the Department of Information Engineering, University of Padova, Padova, Italy. (emails: francesco.ardizzon@unipd.it; mattia.piana@phd.unipd.it; stefano.tomasin@unipd.it)}
\thanks{G.~Shen is with the School of Cyber Science and Engineering, Southeast University, China. (email: gxshen@seu.edu.cn)}
\thanks{Color versions of one or more of the figures in this paper are available online at http://ieeexplore.ieee.org.}
\thanks{Digital Object Identifier xxx}
}
	
\maketitle

\begin{abstract}
The identification of the devices from which a message is received is part of security mechanisms to ensure authentication in wireless communications. Conventional authentication approaches are cryptography-based, which, however, are usually computationally expensive and not adequate in the Internet of Things (IoT), where devices tend to be low-cost and with limited resources. This paper provides a comprehensive survey of physical layer-based device fingerprinting, which is an emerging device authentication for wireless security. In particular, this article focuses on hardware impairment-based identity authentication and channel features-based authentication. They are passive techniques that are readily applicable to legacy IoT devices. Their intrinsic hardware and channel features, algorithm design methodologies, application scenarios, and key research questions are extensively reviewed here. The remaining research challenges are discussed, and future work is suggested that can further enhance the physical layer-based device fingerprinting.
\end{abstract}



\section{Introduction}


The \ac{iot} is expected to significantly impact our lifestyles. 
According to IoT Analytics, the number of connected devices reached to 18.8 billion in 2024, an increase of 13\% from 2023~\cite{iot_dev_no}. These massively connected IoT devices have transformed our everyday lives with exciting applications such as smart homes, smart cities, connected healthcare, industry 4.0, etc.
Wireless communications are preferred to connect these devices seamlessly. There have been many techniques for IoT, including WiFi (IEEE 802.11), ZigBee (IEEE 802.15.4), \ac{lora}, \ac{BLE}, and narrowband IoT (NB-IoT), to name but a few~\cite{burg2017wireless}.

This revolution requires security at all levels. Security is quite a broad topic, involving confidentiality, integrity, availability, authentication, etc.~\cite{trappe2015low,neshenko2019demystifying}. This article will focus on device authentication, which is the first important step for network security.
The receiver verifies the legitimacy of the received signal by checking specific features in the same signal.
Our current computer and communications networks are protected by cryptography-based approaches, including both symmetric encryption, such as \ac{AES}, and public-key cryptography (PKC) such as Rivest-Shamir-Adleman (RSA).
In particular, authentication is performed using a cryptographic challenge-response protocol based on symmetric encryption or PKC.

However, cryptographic solutions may not be applicable to \ac{iot} devices.
Symmetric encryption requires a key pre-shared, whose refresh turns to be challenging for \ac{iot}~\cite{zhang2020new}.
PKC requires computationally expensive algorithms, which often have severe power and computational limitations~\cite{trappe2015low}, hence they are unsuitable for \ac{iot} devices. In addition, on the eve of quantum computing, PKC may be compromised due to the exponential increase in the computational power of attackers~\cite{cheng2017securing}. 
Due to the above limitations, there is a lack of competent \ac{iot} security solutions, and there have been many notorious security threats to \ac{iot} devices~\cite{neshenko2019demystifying}.


This background is driving the development of lightweight, yet secure technologies for the \ac{iot}. 
{\color{black}
Regarding device authentication, the two most promising non-cryptographic approaches are physical layer-based device fingerprinting~\cite{zeng2010non}, which includes hardware impairments-based \ac{rffi}~\cite{zhang2021radio} and channel-based authentication~\cite{baracca2012physical}. In detail, 
\begin{itemize}
    \item RFFI uses unique hardware impairments as the device identifier. Due to the imperfect manufacturing process, the nominal values of hardware components slightly deviate from their specification. These hardware impairments are unique and stable, which can be exploited as device fingerprints.
    \item Channel-based authentication exploits the channel characteristics through which the signal propagates to identify the source (or, better, its location) at the receiver, taking advantage of the fact that signals transmitted by devices at different locations travel through different channels (i.e., different delays and attenuations for each path). Thus, the propagation environment, rather than the transmitting device characteristics, and the relative position between transmitter and receiver, guarantee the authenticity of the transmitter.
\end{itemize}
}


\subsection{Existing Surveys}\label{sec:comparison}
Here we provide a review of the recent tutorials and survey papers published on similar topics~\cite{xu2015device,sanchez2021survey,chowdhury2022survey,Kumar23device,xie2020survey,bai2020physical,Wang21survey, zhang23survey,illi2024physical,hoang2024physical}. 

\subsubsection{Existing Surveys on Device Fingerprinting}
Paper~\cite{xu2015device} provides a tutorial on fingerprinting at different layers including physical, MAC, and network layers.
We will focus on the physical layer techniques and significantly extend~\cite{xu2015device} by summarizing the recent advances in the area as deep learning has brought several exciting improvements.

Paper~\cite{sanchez2021survey} focuses on device behavior fingerprinting, which is related to not only communication networks-based fingerprints but also in-device fingerprints, e.g., resource usage, software signatures, etc. Moreover, it is not solely concerned with security issues; it also encompasses a significant amount of fault detection content. The authors only briefly introduce the availability of the physical layer device fingerprinting technique but do not provide sufficient details on the latest studies and state-of-the-art schemes.

Survey~\cite{chowdhury2022survey} examined device fingerprinting techniques for resource-constrained \ac{iot} applications. While traffic and impairment-based approaches were considered, the survey did not include wireless channel-based methods. From the perspective of identification algorithms, although the deep learning techniques were mentioned, they were limited to a conceptual overview, with insufficient in-depth profiling of state-of-the-art deep learning-based fingerprinting algorithms.

The work in~\cite{Kumar23device} surveyed numerous available device fingerprints, which span the entire cyber-physical system and encompass various characteristics, including thermal, optical, chemical, magnetic, and electrical aspects. 
However, it only briefly introduced the physical layer device fingerprinting problem and did not discuss the latest technologies. Furthermore, the authors focused on feature selection, while the introduction to the latest authentication algorithms is missing.

\subsubsection{Existing Surveys on Physical Layer Security and Authentication}
Surveys~\cite{xie2020survey,bai2020physical} provide comprehensive coverage of \ac{pla} techniques, with both passive and active approaches. Our survey will focus on the passive approaches as they can be readily applied to our pervasive \ac{iot} devices. Additionally, while the authors already considered the use of \ac{ml} techniques, the coverage of the literature on \ac{ml} solutions for device authentication is only partial, as the use of \ac{ml} has become popular only in recent years.

When looking at physical layer solutions, many techniques require models of specific channels they work on. Existing surveys, such as~\cite{Wang21survey, zhang23survey,illi2024physical} cover physical layer authentication techniques tailored for specific application domains. In particular,~\cite{Wang21survey} considers device fingerprinting for \ac{gnss} antispoofing. Both crypto and physical layer solutions are considered in~\cite{zhang23survey}, but only for satellite Internet. Illi~\etal focus instead on physical layer security solutions and the \ac{iot} ~\cite{illi2024physical}. 

Finally, the survey~\cite{hoang2024physical} reviews both physical layer authentication and secure transmission, and it mainly focuses on channel-based authentication. We will delve into device fingerprinting by covering both hardware impairments-based and channel-based approaches.

\subsubsection{Summary}

{\color{black}
A common shortfall in all existing papers is the absence (or very limited coverage) of experimental results and their derivation, which are crucial for assessing the merits and fostering the implementation of new security approaches.  
Several new techniques have appeared in recent years that are not covered by those surveys, e.g., generative AI for authentication,  reconfigurable wireless environments, e.g., with \acp{ris} and drones for challenge-response authentication at the physical layer, etc.
Lastly, fingerprinting and authentication have been investigated in several domains, including different frequency bands and applications 
 (\ac{iot}, mobile \ac{6g}, WiFi, ...) for radio transmissions, but also in \ac{uwac}. An extensive survey of such domains and their peculiarities for fingerprinting/authentication is missing.
}

\subsection{Survey Aims}
{\color{black}
As summarized in Table~\ref{tab:my-table}, this paper complements and extends the published surveys with a comprehensive review of the physical layer-based fingerprinting for wireless security. 
We will review the design principles of both RFFI and channel-based authentication. We will also compare these two approaches and discuss their integration for more secure authentication mechanisms.
Among the most promising and recent advances in these areas, we mention the availability of new technologies (such as \ac{ris}), the use of new transmission bands that fostered related technologies such as integrated communication and sensing, the experimentation (thus with higher technology readiness level) of physical-layer security mechanisms, and the use of \ac{ml} techniques to secure transmissions by merging information coming from different communication layers.
As unique features of our survey paper, we cover topics from theoretical development to practical implementation and share our experiences and insights on the design considerations of practical implementation.  
Thus, while looking at a specific domain, it will still provide a general framework to discuss solutions across different domains. }
\begin{table}[!t]
\centering
\caption{Comparison with existing surveys. \xmark, \omark, and \cmark mean the topic is not covered, partially covered, and extensively covered.}
\label{tab:my-table}
\begin{tabular}{|c|c|c|c|c|c|c|l|}
\hline
Ref                           & Year & ML       & Exp.   &  Domains          & New Tech.    \\ \hline
 ~\cite{xu2015device}         & 2015 &  \xmark  & \xmark & Wireless Networks & \xmark       \\ 
 ~\cite{sanchez2021survey}    & 2021 &  \cmark  & \xmark & \ac{iot}               & \xmark              \\ 
   ~\cite{chowdhury2022survey} & 2022 &  \cmark  & \xmark & \ac{iot}               & \xmark       \\ 
   ~\cite{Kumar23device}       & 2023 &  \cmark  & \xmark & Cyber-Physical System&
  \xmark       \\
 ~\cite{xie2020survey}        & 2020 &  \omark  & \xmark & Wireless Networks & \xmark       \\ 
 ~\cite{bai2020physical}     & 2020 &  \omark  & \xmark & Wireless Networks & \xmark       \\
 ~\cite{Wang21survey}        & 2021 &  \cmark  & \cmark & GNSS              & \xmark       \\ 
 ~\cite{zhang23survey}       & 2023 &  \cmark  & \xmark & Satellite Internet& \xmark       \\ 
 ~\cite{illi2024physical}     & 2024 &  \cmark  & \xmark & \ac{iot}               & \xmark       \\ 
 ~\cite{hoang2024physical}    & 2024 & \cmark  & \xmark & Wireless Networks  & \xmark       \\ 
\textbf{This}                 & 2025  &  \cmark  & \cmark & \ac{iot}/\ac{6g}/UWANs      & \cmark       \\ \hline
\end{tabular}
\end{table}

\subsection{Survey Structure}
Section \ref{sec:overview} gives an overview of physical layer-based device fingerprint, which is further categorized into two techniques. 
The rest of the survey is comprised of three parts. 
The first part will cover the first technique, which is hardware impairments-based authentication, i.e., RFFI. The second part will describe channel-based authentication.

The first part is on RFFI and spans Sections~\ref{sec:rffi_task} to \ref{sec:rffi_applications}. In particular, Section~\ref{sec:rffi_task} presents the RFFI tasks, while Section~\ref{sec:rf_impairments} models the hardware impairments for both transmitter and receiver. The algorithm design for deep learning-based RFFI is explained in Section~\ref{sec:rffi_design}. For the practical implementation of RFFI, Section~\ref{sec:rffi_applications} describes the key research topics, publicly available datasets, and the investigated scenarios. Section~\ref{sec:exp_methodology} explains the experimental methodologies for RFFI.

The second part is on channel-based authentication and spans Sections~\ref{sec:ChBasedAuth} to ~\ref{sec:cb_results}. In particular, Section~\ref{sec:ChBasedAuth} introduces the definition and the approaches used for \ac{cb}-\ac{pla} and Section~\ref{sec:chFeatures} is devoted to an overview of the channel features exploited for \ac{cb}-\ac{pla}. An in-depth delve into the methodologies used for \ac{cb}-authentication, including both statistical and \ac{ml} approaches, is provided in Section~\ref{authTech}. Lastly, Section~\ref{sec:cb_results} provides an overview of CB-authentication datasets publicly available and existing applications.

The third part provides an overview of challenges and future research activities discussed in Section~\ref{sec:challenge}. The main conclusions are reported in Section~\ref{sec:conclusion}. 

The abbreviations used in this paper can be found in Table~\ref{tab:Abbreviations}.



\begin{table*}[]
\centering
\caption{List of Abbreviations}
\label{tab:Abbreviations}
\begin{tabular}{L{2cm}L{5cm}L{1.5cm}L{2cm}L{5cm}}\hline
Abbreviation & Definition                            &  & Abbreviation & Definition                                   \\\hline
6G           & sixth generation                      &  & LRT          & likelihood ratio test                        \\
AE           & autoencoder                           &  & LSTM         & long-short term   memory                     \\
AML          & adversarial machine   learning        &  & LT           & likelihood test                              \\
AoA          & angle-of-arrival                      &  & LTE          & long-term evolution                          \\
AoD          & angle-of-departure                    &  & MD           & misdetection                                 \\
AWGN         & additive white   Gaussian noise       &  & MIMO         & multiple-input   multiple-output             \\
BAAE         & Bahdanau attention   autoencoder      &  & ML           & machine learning                             \\
BLE          & Bluetooth low energy                  &  & NLOS         & non-line-of-sight                            \\
CB           & channel-based                         &  & NN           & neural network                               \\
CFO          & carrier frequency   offset            &  & OC-SVM       & one-class support   vector machine           \\
CFR          & channel frequency   response          &  & OFDM         & orthogonal   frequency-division multiplexing \\
CIR          & channel impulse   response            &  & pdf          & probability density   function               \\
CNN          & convolutional neural   network        &  & PDP          & power-delay profile                          \\
COTS         & commercial   off-the-shelf            &  & PLA          & physical layer   authentication              \\
CR           & challenge-response                    &  & QuaDRiGa     & quasideterministic   radio channel generator \\
CSI          & channel state   information           &  & RFF          & radio-frequency   fingerprint                \\
DRL          & deep reinforcement   learning         &  & RFFI         & radio frequency   fingerprint identification \\
DT           & decision tree                         &  & RIS          & reflective   intelligent surface             \\
DUT          & devices under test                    &  & RL           & reinforcement   learning                     \\
EL           & enseamble learning                    &  & RMS          & root-mean square                             \\
FA           & false alarm                           &  & RNN          & recurrent neural   network                   \\
FFT          & fast Fourier   transform              &  & RSS          & received signal   strength                   \\
GAN          & generative   adversarial network      &  & SCM          & normalized sample   covariance matrix        \\
GLRT         & generalized   likelihood-ratio test   &  & SDR          & software-defined   radio                     \\
GNN          & graph NN                              &  & SNR          & signal-to-noise ratio                        \\
GNSS         & global navigation   satellite systems &  & SVM          & support vector   machine                     \\
GPR          & Gaussian process   regression         &  & TDOA         & time difference of   arrival                 \\
GPS          & global positioning   system           &  & TLE          & two-line element                             \\
GRU          & gated recurrent unit                  &  & TOA          & time of arrival                              \\
IoT          & Internet of Things                    &  & USRP         & universal software   radio peripheral        \\
KF           & Kalman filter                         &  & UWAC         & underwater acoustic   communications         \\
KNN          & K-nearest neighbors                   &  & UWB          & Ultra-Wideband                               \\
LEO          & low Earth orbit                       &  & V2V          & vehicle to vehicle                           \\
LLM          & large language models                 &  & VAE          & variational   autoencoder                    \\
LoRa         & long range                            &  & VANET        & vehicular ad-hoc   network                   \\
LOS          & line-of-sight                         &  & VLC          & visible-light   communications\\\hline              
\end{tabular}
\end{table*}

\section{Device Fingerprinting at The Physical Layer}\label{sec:overview}

\begin{figure}[!t]
    \centering
    \includegraphics[width = 3.4in]{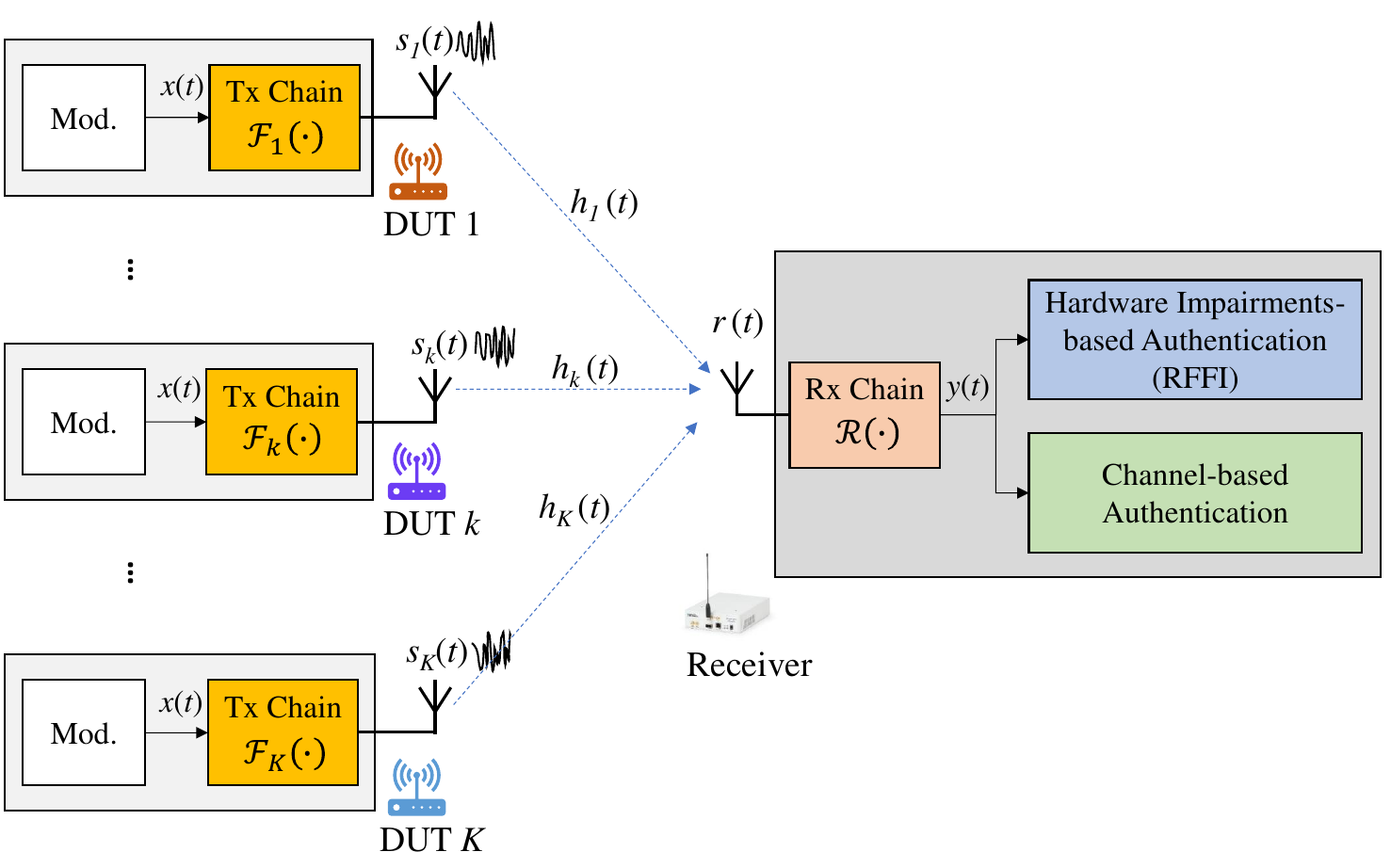}
    \caption{Physical layer-based device fingerprinting system overview. }
    \label{fig:system_overview}
\end{figure}

The authentication on the basis of the signals exchanged at the physical layer comprises security mechanisms that can be classified as hardware fingerprinting or \ac{cb} authentication techniques, which provide lightweight security mechanisms particularly useful in the IoT.
As shown in Fig.~\ref{fig:system_overview}, we will consider a system involving $K$ transmitting IoT devices and a receiver. The IoT transmitter sends packets, which are captured by the receiver. Based on the received signals, the receiver aims to authenticate the transmitter based on its intrinsic hardware impairments and random channel features.

\subsubsection{Transmitter}  For each transmitter, the modulated signal, $x(t)$, passes to the transmitter chain, including the mixer, oscillator, and power amplifier~\cite{wang2016wireless,zhang2021radio}. These hardware components are not perfect due to the variation in the manufacturing process, and their specifications deviate slightly from their nominal values. 
Their effects are collectively represented by $\mathcal{F}(\cdot)$. 
The radio frequency (RF) signal at the transmitter becomes $s(t) = \mathcal{F}(x(t))$.

\subsubsection{Channel}
The RF signal is emitted into the wireless channel, which experiences multipath channel effects, denoted as $h(\tau,t)$, resulting in the received signal as
\begin{align}
    r(t) = h(\tau,t)*\mathcal{F}(x(t)),
    \label{eq:iq}
\end{align}
where $*$ denotes the convolution operation. Note that many IoT devices are mobile; hence the channel impulse response $h(\tau,t)$ is time-varying.


\subsubsection{Receiver} 
The receiver captures the received signal $r(t)$, which is passed to the receiver chain, including the mixer and oscillator too~\cite{zhang2021radio}. The receiver hardware components are not perfect either, and their effects are represented by $\mathcal{R}(\cdot)$. Considering all the above processes, the received signal $y(t)$ can be mathematically written as 
\begin{equation}
    y(t) = \mathcal{R}\Big(h(\tau,t)*\mathcal{F}_k(x(t))\Big) + n(t),
    \label{eq:signal_model}
\end{equation}
where $n(t)$ is the \ac{awgn}.

\subsection{Device Fingerprinting}
As can observed from (\ref{eq:signal_model}), the received signal, $y(t)$, involves both the hardware impairments and channel features, which can be exploited for device authentication.

\subsubsection{Hardware Impairments-Based Authentication}
Due to the manufacturing process, the hardware components are not perfect. Hence, hardware components are subject to impairments, such as mixer imbalance, oscillator imperfection, and power amplifier non-linearities~\cite{wang2016wireless,zhang2021radio}.
These impairments are minute and do not affect the communication functionalities because they can be compensated for by the receiver.
These features are unique and can be used as device identifiers.
RFFI protocols extract the hardware impairments embedded in the signal and infer its corresponding device identity.

\subsubsection{Channel Based Authentication} The channel over which the transmitted signal travels is characterized by reflections, scattering, attenuations, as well as angular / time / Doppler features, and the position of the transmitter and the receiver. \ac{cb} authentication uses this information from the channel to identify the sender of the message (and the channel over which the signal is going). A basic assumption is that devices are slowly moving and the environment is slowly changing; thus, the authentication mechanism checks if different transmissions experience similar propagation channels. Other approaches are also discussed in the following, where the channel can change fast, but its consistent evolution over time provides the authentication feature. 

\begin{figure*}[!t]
    \centering
    \includegraphics[width = 6.8in]{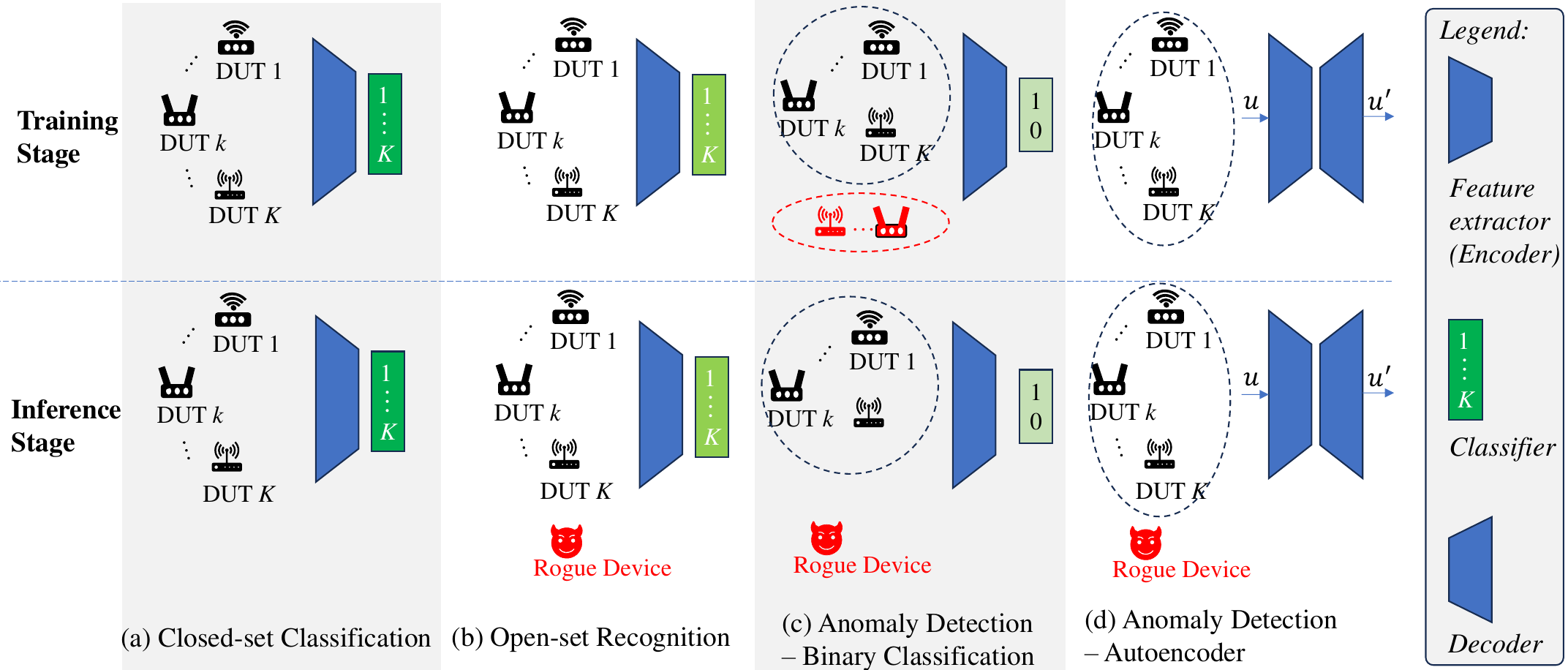}
    \caption{Deep learning-based RFFI tasks.}
    \label{fig:rffi_types}
\end{figure*}

\section{Radio Frequency Fingerprint Identification}\label{sec:rffi_task}
Deep learning has transformed many areas thanks to its powerful automatic feature extraction capability, which has also significantly enhanced RFFI.
To the best knowledge of the authors, the work in~\cite{robyns2017physical} is the first paper applying deep learning to RFFI. Specifically, \ac{cnn} and multilayer perceptron (MLP) are used to classify \ac{lora} devices.
After that, deep learning has attracted massive interest in the RFFI area. Many deep learning approaches, such as \ac{cnn}~\cite{shen2021jsac,shen2021towards,shen2023length}, \ac{rnn} including \ac{lstm}~\cite{shen2021jsac,shen2023length} and \ac{gru}~\cite{shen2023length}, Transformer~\cite{shen2023length}, etc, have demonstrated significant impact, which can alleviate the difficulties of manual feature engineering.


Depending on whether there are rogue devices involved, RFFI can be categorized into closed-set classification, open-set recognition, and anomaly detection~\cite{hanna2020open}, whose implementations are illustrated in Fig.~\ref{fig:rffi_types}.
A deep learning-based RFFI protocol involves two stages, namely training and inference. 
A deep learning model will be trained using a training dataset, $\mathcal{D}_{\rm train}$, and the trained deep learning model will be used for inference in the second stage.

\subsection{Closed-Set RFFI Classification}\label{sec:rffi_dl_closedset}

As shown in Fig.~\ref{fig:rffi_types}(a), there are $K$ legitimate transmitters, a.k.a. \ac{DUT}, to be identified, and no rogue device is considered in the closed-set RFFI classification. The devices in the training and inference stages remain the same, hence the name ``closed-set" comes from. The approach will predict the identity of the \ac{DUT}.  

\begin{figure}[!t]
    \centering
    \includegraphics[width = 2.8in]{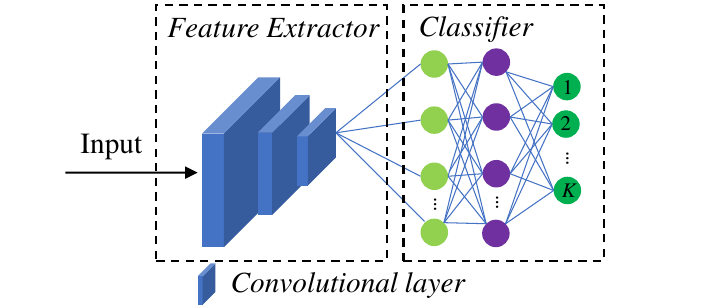}
    \caption{Illustration of a \ac{cnn} architecture.}
    \label{fig:cnn_illustration}
\end{figure}

Close-set RFFI classification is probably the most studied scenario in RFFI, which is a multi-class classification problem. Hence, deep learning is perfect for such tasks. 
A training dataset, $\mathcal{D}_{\rm train} = \{(y_i,\ell_i)\}_{i = 1}^{NK}$, will be constructed, where $\ell_i$ is the device label of the collected $i$-{th} packet and $N$ is the number of packets collected for each \ac{DUT}. The number of packets from each \ac{DUT} should be kept the same, to ensure a balanced dataset. 
A deep learning model can be partitioned into a feature extractor and a classifier.
A \ac{cnn} architecture is given as an example in Fig.~\ref{fig:cnn_illustration}. The feature extraction includes convolutional layers and pooling layers. The classifier is composed of a few fully connected layers, and the last layer has $K$ neurons corresponding to $K$ classes.

In the training stage, the parameters $\Theta$ of the deep learning model $f$ are optimized as
\begin{equation}\label{equ:optimize}
    \Theta = \mathop{\arg\min}_{\Theta} \sum_{(y_i,\ell_i)\in \mathcal{D}_{\rm train}} \mathcal{L}(f(y_i;\Theta),\ell_i),
\end{equation}
where $\mathcal{L}(\cdot)$ is the loss function, e.g., the cross-entropy loss.

In the inference stage, the softmax is used as the activation function, then the last fully connected layer of the classifier will return a list of probabilities $\bm{p} = (p_1, p_2, ..., p_K)$, where $p_k$ represents the probability of the $k$-{th} device, given as
\begin{align}
    p_k = \frac{e^{z_k}}{\sum_{i=1}^{K}e^{z_i}},
    \label{eq:softmax}
\end{align}
where $\bm{z} = (z_1, z_2, ..., z_K)$ is the output of the layer before the softmax activation. 
The final prediction is obtained by locating the maximum probability, given as
\begin{align}
    \hat{\ell} = \arg\max_k(\bm{p}).
    \label{eq:prediction}
\end{align}

\subsection{Open-Set Recognition}\label{sec:rffi_dl_open}
Under closed-set classification, rogue devices will be classified as the legitimate \ac{DUT} with the closest features, which is not acceptable as attackers will be admitted. Therefore, open-set recognition is required.

In open-set recognition, there are $K$ legitimate \ac{DUT}s and rogue devices, as illustrated in Fig.~\ref{fig:rffi_types}.b. Because rogue devices do not appear in the training stage, it is named ``open-set''. We need to first detect whether the \ac{DUT} is legitimate or rogue, then further classify the index for legitimate \ac{DUT}s. 
Open-set recognition can be addressed by the deep learning-based approaches with an adjustment to the softmax activation function. Hence, the training and inference stages will be similar to the closed-set classification. The training dataset can be constructed in the same way as the closed-set classification.

Gritsenko~\etal leveraged the output probabilities of softmax given in (\ref{eq:softmax}) for open-set recognition~\cite{gritsenko2019finding}. Specifically, when the signal is from an unseen rogue device, the confidence level of the neural network prediction is low, hence, the output probability will be smaller than a pre-calculated threshold. In contrast, when it is from a legitimate \ac{DUT}, the neural network can predict as it does in the closed-set classification.

Hanna~\etal adopted a new activation function, the OpenMax~\cite{hanna2020open}. The activation vector $z$ prior to softmax is extended to $K+1$ outputs, given as
\begin{align}
    z_k' = \begin{cases}
z_k\omega_k, & k \in \{1, ..., K\}\\
\sum_{k = 1}^{K}z_k(1-\omega_k), & k = K+1
\end{cases}
\end{align}
where $\omega_k$ is a confidence parameter of the sample belonging to $k$-th class\footnote{Please refer to~\cite{hanna2020open} for the detailed calculation.}
and the additional $K+1$ output refers to the rogue devices. The vector $z_k'$ is then fed into the softmax function, and the prediction can be obtained using (\ref{eq:prediction}). Different from~\cite{gritsenko2019finding} only leveraging the softmax output probabilities, this work exploits the entire activation vector, which is more robust.

Open-set recognition can also be tackled by non-deep learning-based methods.
Shen~\etal designed a \ac{knn}-based method~\cite{shen2021towards}. They created a \ac{rff} database that stores a few RFF features for each legitimate \ac{DUT}. In the inference stage, RFF features will be extracted from the input signal and compared with the features in the database. The attacker is not registered beforehand, hence their features are largely different, which can be detected via a large feature distance. In contrast, the legitimate devices can be identified because there will be a matching feature in the database.

\subsection{Anomaly Detection}\label{sec:rffi_dl_anomaly}
There are $K$ legitimate \ac{DUT}s and rogue devices involved in the inference stage. Different from open-set recognition, anomaly detection only detects whether the \ac{DUT} is legitimate or rogue. Because it is not practical to assume attackers are cooperative, hence, they are not available in the training stage.

Anomaly detection can be achieved by binary classification. As shown in Fig.~\ref{fig:rffi_types}.c, the $K$ legitimate \ac{DUT}s are treated as one class (label 1). A few other \ac{DUT}s will be used to represent rogue devices, which serve as the other class (label 0). The system design will be similar to the closed-set classification, but the number of classes reduced to two. However, in the inference stage, when the rogue device appears, it is supposed to be classified as label 0.

\Ac{ae} is a popular unsupervised deep learning architecture for anomaly detection~\cite{hanna2020open}. An \ac{ae}-based RFFI approach is portrayed in Fig.~\ref{fig:rffi_types}.d. In the training stage,  similar to the binary classification approach, the $K$ \ac{DUT}s are treated as a single class. But differently, there is no other device required.
\Ac{ae} consists of an encoder and a decoder. The encoder first compresses the input, $u$, to a latent feature; the decoder will then try to reconstruct the input signal from the latent feature and output $u'$. The mean square error (MSE) between $u$ and $u'$ is typically used as the reconstruction error. The training process will learn the features of the training data and reduce the MSE.
In the inference stage, if the signal is from the legitimate \ac{DUT}, the trained \ac{ae} can reconstruct the input, and a low MSE will be returned. Otherwise, when the signal is from a rogue device, the MSE will be higher than a threshold, indicating an outlier is detected.

\section{Hardware Impairments for RFFI}\label{sec:rf_impairments}
Due to the variations in the manufacturing processes, the hardware components of the radio devices will not be perfect. Their specifications will deviate from their nominal values slightly, which are referred to as RF hardware impairments.
This section will provide the key parts for the modelling of transmitter and receiver impairments.
The detailed mathematical derivation can be found in~\cite{zhang2021radio}.

\subsection{Transmitter Impairments}
\begin{figure}[!t]
    \centering
    \includegraphics[width = 3.4in]{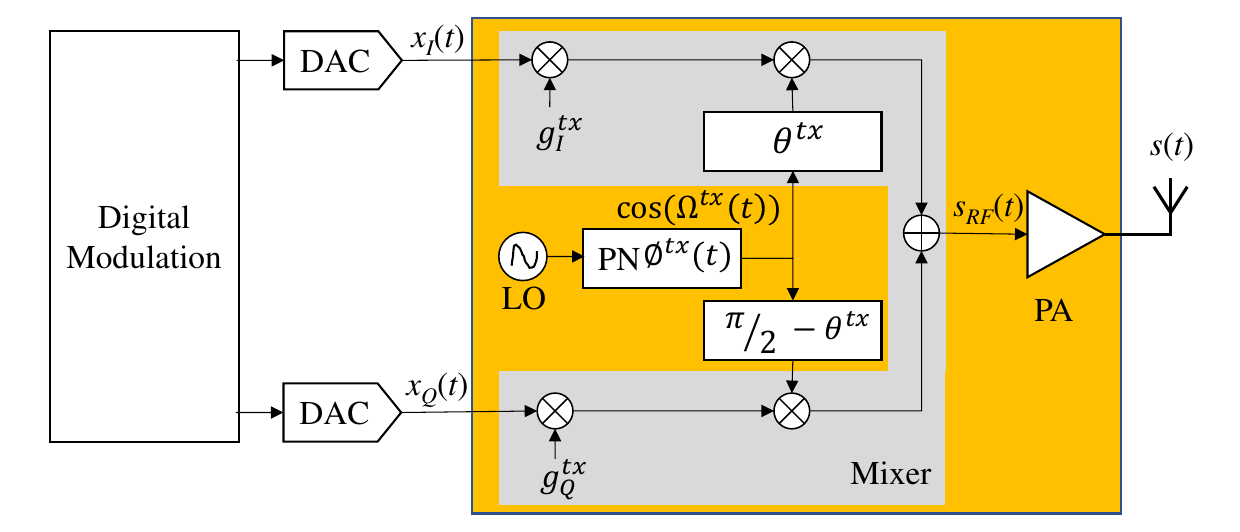}
    \caption{Transmitter impairment model.}
    \label{fig:rffi_tx}
\end{figure}

The architecture of a direct conversion transmitter is portrayed in Fig.~\ref{fig:rffi_tx}. 
Their overall effects are represented as $\mathcal{F}(\cdot)$ in Section~\ref{sec:overview} while their individual effects will be modelled in this section.

The local oscillator (LO) will produce a sinusoidal waveform with a specific carrier frequency. The output frequency is not stable but is subject to temperature and ageing. 
When the system's nominal carrier frequency is $f_c^0$, the real frequency is $f^{tx} = f_c^0 + \Delta f^{tx}$, where $\Delta f^{tx}$ is the offset. Besides the \ac{cfo}, the LO is also subject to phase noise (PN), $\phi^{tx}(t)$.
Considering all the LO imperfections, the carrier phase of the transmitter can be written as
\begin{align}
	\Omega^{tx}(t) & =  2\pi f_c^{tx}t + \phi^{tx}(t)\nonumber \\
    & = 2\pi f_c^0t + 2\pi \Delta f^{tx}t + \phi^{tx}(t).
\end{align}

The mixer will then mix the baseband signal with the carrier wave. However, the mixer is also subject to gain and phase imbalance. Specifically, $g_I^{tx}$ and $g_Q^{tx}$ represent the gain of in-phase (I) and quadrature (Q) branches, respectively; $\theta_{tx}$ denotes the phase imbalance.
Due to the existence of gain and phase imbalance, the RF band signal then becomes~\cite{zhu2013blind}
\begin{align}
	s_{RF}(t) = & g^{tx}_I x_I(t)\cos(\Omega^{tx} + \theta^{tx}) -  g^{tx}_Q x_Q(t)\sin(\Omega^{tx}- \theta^{tx}), \nonumber\\ 
	& = \Re\{s_{BB}(t)e^{j\Omega^{tx}}\},
	\label{eq:srf}
\end{align}
where $x_I(t)$ and $x_Q(t)$ are the baseband data at the I and Q branches, respectively, and
\begin{align}
	s_{BB}(t) = g^{tx}_Ix_I(t) e^{j\theta^{tx}} + jg^{tx}_Qx_Q(t) e^{-j\theta^{tx}}.
	\label{eq:iqmodel}
\end{align}

The RF signal then undergoes the power amplifier, which introduces additional nonlinearities. 
A power amplifier in a narrowband system is usually modelled with memoryless nonlinear effects, including amplitude/amplitude (AM/AM) and amplitude/phase (AM/PM) characteristics~\cite{zhu2013challenges}.
There are several behavioural models, such as the Saleh, Rapp, and Ghorbani models, etc.~\cite{zhu2013challenges}. 

After passing through a power amplifier, the signal becomes 
\begin{equation}	\label{eq:s1}
    \begin{split}
        s(t) &= A(|s_{BB}(t)|)e^{j(\angle s_{BB}(t) + \Omega^{tx}(t) + \Phi(|s_{BB}(t)|))},\\
	& = s'(t)e^{j\Omega^{tx}(t)},
    \end{split}
\end{equation}	
where  $\angle s_{BB}(t)$ is the angle of the baseband signal and
\begin{align}
	s'(t) = A(|s_{BB}(t)|)e^{j(\angle s_{BB}(t) + \Phi(|s_{BB}(t)|))}.
\end{align}

\subsection{Receiver Impairments}
\begin{figure}[!t]
    \centering
    \includegraphics[width = 3.4in]{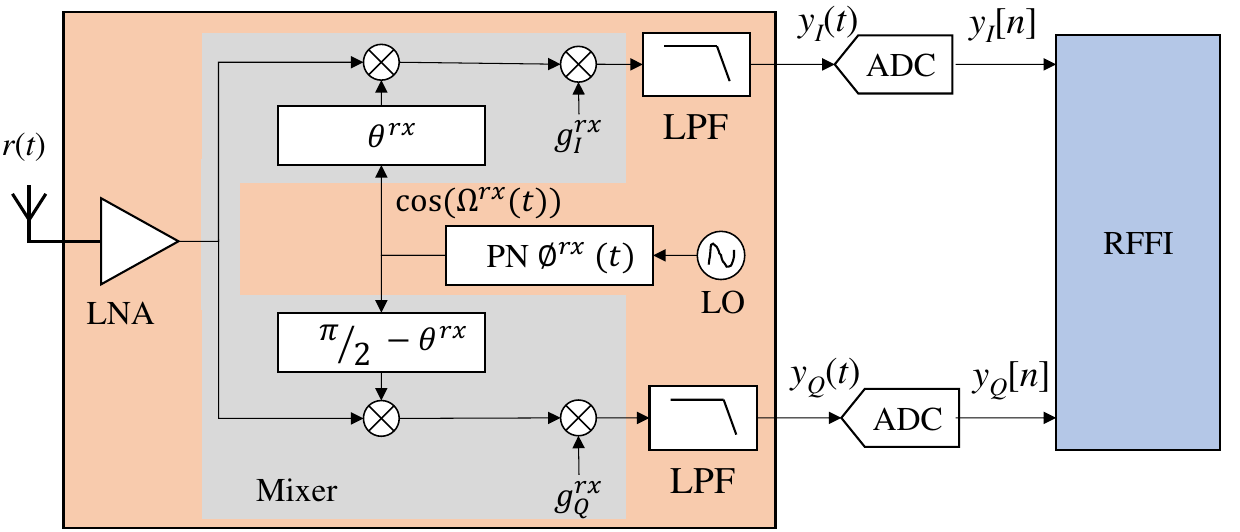}
    \caption{Receiver impairments.}
    \label{fig:rffi_rx}
\end{figure}

Similarly, the receiver will also have RF impairments. Fig.~\ref{fig:rffi_rx} depicts the receiver architecture and its impairments, i.e., receiver LO imperfection and mixer imbalance. Their overall effects are denoted as $\mathcal{R}(\cdot)$ in Section~\ref{sec:overview}. In this section, we will model their individual effects. 

The LO at the receiver is also subject to frequency offset, $\Delta f^{rx}$, and phase noise, $\phi^{rx}(t)$. The receiver carrier can then be written as
\begin{align}
	\Omega^{rx}(t) & =  2\pi f_c^{rx}t + \phi^{rx}(t)\nonumber \\
    & = 2\pi f_c^0t + 2\pi \Delta f^{rx}t + \phi^{rx}(t).
\end{align}

The receiver uses a mixer to mix the received signal, which will downconvert the signal from the RF band to the baseband. Similarly to the transmitter mixer, the receiver mixer also has gain imbalance ($g_I^{rx}$ and $g_Q^{rx}$) and phase imbalance ($\theta^{rx}$).

Considering the effects of the LO imperfection and mixer imbalance, the receiver's carrier wave becomes 
\begin{align}
	C^{rx}(t) = K_{1}^{rx} e^{-j\Omega^{rx}(t)} + K_{2}^{rx} e^{j\Omega^{rx}(t)},
\end{align}
where	$K_{1}^{rx} = (g_I^{rx}e^{-j\theta^{rx}} + g_Q^{rx}e^{j\theta^{rx}})/2$ and $K_{2}^{rx} = (g_I^{rx}e^{j\theta^{rx}} - g_Q^{rx}e^{-j\theta^{rx}})/2$.

The RF signal captured by the receiver can be written as
\begin{align}
	r(t) = (h(\tau,t)*s)(t) = (h(\tau,t)*s')(t)e^{j\Omega^{tx}(t)}.
\end{align}
After the downconversion (by the LO and mixer) and low-pass filter, the received signal at the baseband becomes
\begin{align}
	y(t) &= r(t)C^{rx}(t)\nonumber\\
	 &= K_1^{rx} h(\tau,t)*s'(t)e^{j\Delta \Omega} + K_2^{rx} (h(\tau,t)*s')^*(t)e^{-j\Delta \Omega},
	\label{eq:yt}
\end{align}
where $\Delta \Omega = 2\pi(\Delta f^{tx} - \Delta f^{rx} )t + \phi^{tx}(t) - \phi^{rx}(t)$, $\Delta f = \Delta f^{tx} - \Delta f^{rx}$ is the commonly known \ac{cfo}.

The baseband signal $y(t)$ in (\ref{eq:yt}) possesses all the RF impairments of both the transmitter and receiver. 
The analogue signal is sampled by the analogue-to-digital converter (ADC), which produces a digital sequence, $y[n]$, and is used for RFFI.
The transmitter impairments are the unique hardware features that RFFI explores. Regarding the receiver impairments, when the same receiver is used for collecting training and test datasets, the effects brought by receiver impairments are consistent and can be ignored. However, when different receivers are used, they will indeed affect RFFI performance, which will be reviewed in Section~\ref{sec:key_research_topics_receiver}.

\section{Deep Learning-based RFFI Algorithm Design}\label{sec:rffi_design}
\begin{figure}[!t]
    \centering
    \includegraphics[width = 3.4in]{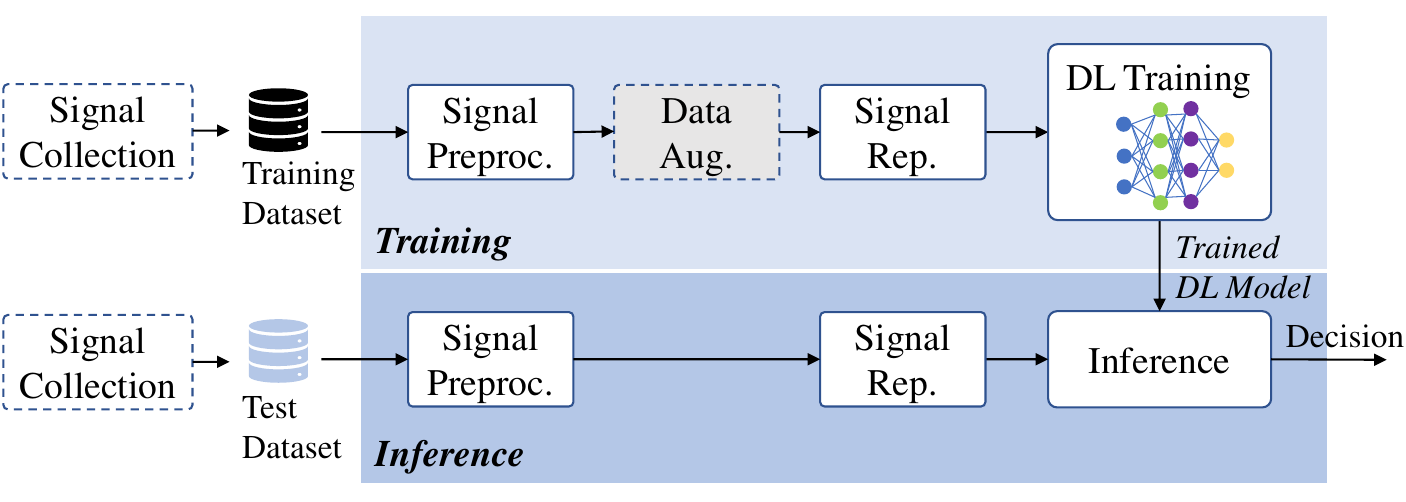}
    \caption{Deep learning-based RFFI algorithm design.}
    \label{fig:rffi_protocol_dl}
\end{figure}

The deep learning-based RFFI algorithm design is shown in Fig.~\ref{fig:rffi_protocol_dl}, including dataset collection, signal preprocessing, data augmentation, signal representation, and deep learning model.
\begin{itemize}
    \item Training Stage: Once a training dataset is created, the sampled signals are processed by signal preprocessing (Section~\ref{sec:preprocessing})
 and then converted to a proper signal representation (Section~\ref{sec:representation}). An additional data augmentation approach is usually adopted to enhance the dataset diversity (Section~\ref{sec:data_augmentation}). The samples are then input into a deep learning model for training, which will produce a trained deep learning model when completed. 
 \item Inference Stage: The signal undergoes the same signal preprocessing and signal representation algorithms, then is input to the trained deep learning model.
\end{itemize}
The deep learning training can usually be done offline, while the inference should be done in real-time in practice, even though many papers do it offline for evaluation purposes.

\subsection{Signal Collection}
The signal data collection is essential to collect IQ samples and build up dedicated datasets. The readers can also opt to use public datasets, which will be summarized in Section~\ref{sec:rffi_applications}.

As introduced in~\cite{shen2023methodology}, most RF fingerprinting studies utilize \ac{sdr} as the wireless receiver to capture IQ samples for fingerprint extraction~\cite{shen2021jsac}. In addition, some WiFi-focused studies have explored the extraction of RF fingerprints from \ac{csi} provided by commercial network interface cards (NICs) or system-on-chips (SoCs), such as the Intel 5300 NIC~\cite{halperin2011tool}, Atheros CSI tool~\cite{Xie:2015:PPD:2789168.2790124,atheroscsi}, Nexmon CSI tool~\cite{gringoli2019free,nexmon:project}, and ESP32 CSI tool~\cite{Hern2006:Lightweight,esp32csi}. More details can be found in Section~\ref{sec:exp_methodology}.


\subsection{Preprocessing}\label{sec:preprocessing}
Signal preprocessing involves power normalization and \ac{cfo} compensation. Power normalization can be done by normalizing the signal power with respect to the \ac{rms} value of the power. 

\ac{cfo} represents the difference between the carrier frequencies of the transmitter and receiver, as embedded in \eqref{eq:cfo}. 
\ac{cfo} has been widely adopted in the literature.
For example, the work in~\cite{hua2018accurate} designed an accurate algorithm to estimate \ac{cfo} from \ac{csi} in WiFi, as \ac{cfo} is twisted with frame detection delay, sample frequency offset, and time of flight.
\ac{cfo} is also used together with other parameters to classify WiFi devices in~\cite{vo2016fingerprinting} and ZigBee devices in~\cite{peng2018design}.

However, \ac{cfo} is subject to temperature drift. In a seven-month study carried out in ~\cite{shen2021jsac}, it is revealed that \ac{cfo} is not suitable as a unique and stable feature. Specifically, the instantaneous \ac{cfo} varies quickly when the device is powered on due to the emitted heat of the device. While the instantaneous \ac{cfo} is varying, the work in~\cite{shen2021jsac} also found that the \ac{cfo} mean value remains relatively stable over the seven-month test period, which is used as an auxiliary approach to calibrate the prediction of the deep learning model. 

In summary, due to the time-varying nature of \ac{cfo}, it is suggested to carry out \ac{cfo} compensation to preprocess the sampled signals, especially for low-cost IoT devices.

\subsection{Data Augmentation}\label{sec:data_augmentation}
Data augmentation is used to augment the training dataset in a simulation manner. It is very time-consuming and labour-intensive to collect a comprehensive training dataset using experiments. In contrast, data augmentation can generate many artificial samples by adding channel and noise effects, which can significantly reduce the data collection overhead~\cite{soltani2020more}.

Specifically, the original training dataset can be constructed by sampling high-quality signals, $\{y_i\}$, which can be achieved by placing the \ac{DUT} and receivers apart with a relatively short distance (e.g., less than 1 meter). We can then augment $\{y_i\}$ by emulating channel and noise as
\begin{align}
    y_i'(t) = (y_i* h'(\tau,t))(t)s + n'(t),
\end{align}
where $ h'(\tau,t)$ is the multipath channel and $n'(t)$ is the AWGN noise, both generated by a simulation model. 

In particular, the multipath channel modelling involves both the \ac{pdp}
 and Doppler shift~\cite{shen2021towards}. The PDP describes the attenuation gains of each channel tap. For example, the exponential PDP can be mathematically given as
 \begin{align}
     P(m) = \frac{1}{\tau_d} e^{-m T_s/\tau_d}, m = 0, 1,..., m_{\max},
 \end{align}
where $\tau_d$ is the \ac{rms} delay spread, $m_{\max}$ is the index of the last tap, and $T_s$ is the sampling interval.
Regarding the Doppler shift, it describes how the channel gain changes over time, with common models such as the Jakes model. 
By incorporating as many PDP and Doppler shift models as possible, data augmentation can significantly enhance the comprehensiveness of the training dataset.

The channel modelling can be achieved by employing the fading channel realization in Matlab~\cite{matlab_fading}. The \texttt{comm.RayleighChannel} and \texttt{comm.RicianChannel} functions provide abundant interfaces to configure PDP and Doppler shift. Furthermore, the Wi-Fi channel models are also available in Matlab~\cite{wlan_channel}, with PDP pre-configured. 
 
Besides the channel effect, AWGN can be added to emulate scenarios with different \ac{snr} levels.

\begin{figure*}[!t]
\centering
\subfloat[]{\includegraphics[width=2.2in]{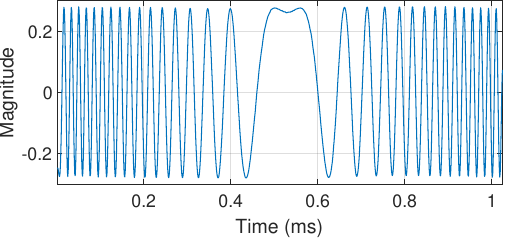}}\hspace{0.05in}
\subfloat[]{\includegraphics[width=2.2in]
{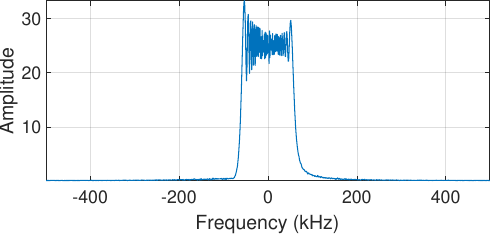}}\hspace{0.05in}
\subfloat[]{\includegraphics[width=2.2in]
{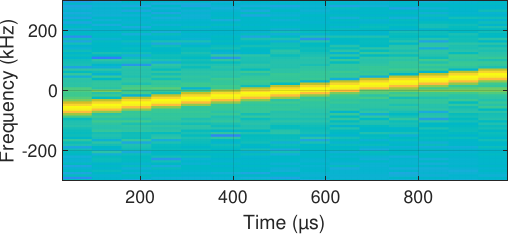}}
\caption{Signal representation: (a) Time domain signals (I branch), (b) \acs{fft} coefficients, (c) spectrogram. }
\label{fig:lora}
\end{figure*}

\subsection{Signal Representation}\label{sec:representation}

The signal captured by the receiver is always in the \textit{time domain} initially, which is named IQ samples in the literature, as shown in (\ref{eq:iq}). {\color{black}Utilizing raw IQ samples is applicable across any wireless protocol. However, as shown in (\ref{eq:iq}), it is a time convolution between the hardware features and the channel, which makes it difficult to separate them in the time domain. Hence, IQ samples tend to be less effective for channel-robust RFFI, as evidenced in~\cite{ma2025wcnc}.}

\textit{Frequency domain} signal is popularly employed, which can be simply obtained by applying \ac{fft} operations to the time domain signal~\cite{shen2021jsac}. The channel effect can be separated from frequency domain signals more easily compared to the time domain counterpart. The \textit{time-frequency domain} spectrogram is a widely employed signal representation in RFFI research~\cite{shen2021jsac, al2021deeplora}. This can be obtained by applying a short-time Fourier transform (STFT) to the time domain signal.
The time-domain IQ samples, frequency-domain \ac{fft} coefficients, and time-frequency domain spectrogram of \ac{lora} preambles are exemplified in Fig.~\ref{fig:lora}.

In addition to these domain transform methods, there are also other specially designed signal representations. For instance, Peng~\etal post-process the constellation figures, generating image-like differential constellation trace figures (DCTFs)~\cite{peng2019deep}. The authors in~\cite{merchant2018deep} subtract the ideal signals from the received ones, creating error signals as neural network inputs. Other available signal representations include bi-spectrum~\cite{ding2018specific}, and Hilbert-Huang spectrum~\cite{zhang2016specific}, etc. 

{\color{black}Aside from the signal representations derived from the steady-state portion of signals, some studies focus on extracting RF fingerprints from the transients that occur when transmitters are powered on or off~\cite{hall2005radio}. However, this approach requires high-end receivers capable of operating at high sampling rates, which can significantly increase the cost of system deployment.}

\subsection{Deep Learning Model}

The deep learning models are capable of extracting unique features from the input signal representations and subsequently predicting the device identity. The design of neural networks should take into account the employed signal representation, as illustrated in the following example. Image-like representations, such as spectrograms~\cite{shen2021infocom,al2021deeplora}, DCTF~\cite{peng2019deep}, and Hilbert-Huang spectrum~\cite{zhang2016specific, pan2019specific}, are suitable for processing with CNNs, while time-domain IQ samples are suitable for processing with 1D CNNs or specially designed complex-valued neural networks~\cite{al2020exposing}. Some studies also utilize MLP to process frequency-domain spectrum~\cite{shen2021jsac,robyns2017physical}. As the captured radio signals exhibit temporal dependencies, sequence models can be employed for RFFI tasks as well. Recent studies have investigated the application of \ac{rnn}, \ac{lstm}, \ac{gru}, and the latest Transformer models~\cite{shen2023length}.


\section{RFFI Key Research Topics, Public Datasets and Applications}\label{sec:rffi_applications}
As shown in (\ref{eq:signal_model}), RFFI performance is affected by channel and noise effects as well as receiver impairments. Therefore, this section reviews the RFFI research activities related to these three areas, namely channel effects elimination, noise mitigation, and receiver distortion mitigation.
In addition, public datasets are critical to the development of deep learning-based RFFI techniques. As summarized in Table~\ref{tab:rffi_dataset}, there are some RFFI datasets shared by the community.
We provide a list of available datasets to evaluate the above three research topics.
Finally, we review the RFFI literature in terms of their application techniques, including WiFi, ZigBee, \ac{lora}, cellular, and Iridium satellites. 

\begin{table*}[!t]
\caption{Summary of Publicly Available RFFI Datasets}
\label{tab:rffi_dataset}
\centering
\begin{tabular}{|L{1.3cm}|l|l|l|L{3cm}|L{4.7cm}|lll|}
\hline
\multicolumn{1}{|c|}{\multirow{2}{1.3cm}{Wireless Technology}} & \multicolumn{1}{c|}{\multirow{2}{*}{Dataset}} & \multicolumn{1}{c|}{\multirow{2}{*}{Paper}} & \multicolumn{1}{c|}{\multirow{2}{*}{\# DUT}} & \multicolumn{1}{c|}{\multirow{2}{*}{Receiver}}               & \multicolumn{1}{c|}{\multirow{2}{*}{Brief Summary}}                                     & \multicolumn{3}{c|}{Evaluation Purpose}                               \\ \cline{7-9} 
\multicolumn{1}{|c|}{}                                     & \multicolumn{1}{c|}{}                         & \multicolumn{1}{c|}{}                       & \multicolumn{1}{c|}{}                            & \multicolumn{1}{c|}{}                                        & \multicolumn{1}{c|}{}                                                                   & \multicolumn{1}{l|}{Channel} & \multicolumn{1}{l|}{Noise}  & Receiver \\ \hline
WiFi                                                       & \cite{wisig_dataset}                          & \cite{hanna2022wisig}                       & up to 150                                        & USRP B210, X310 and   N210 (up to 32)                        & WiFi signals   collected from different days and positions.                             & \multicolumn{1}{l|}{\cmark}  & \multicolumn{1}{l|}{}       & \cmark   \\ \hline
WiFi                                                       &   \cite{kong2024deepcrf_dataset}            & \cite{kong2024deepcrf}                      & 19                                               & Xilinx Zynq ZC706 board + FMCOMMS5   ADI daughter board      & WiFi signals collected from different days and positions (indoor, outdoor, car park). & \multicolumn{1}{l|}{\cmark}  & \multicolumn{1}{l|}{}       &          \\ \hline
WiFi                                                       & \cite{xie2025towards_dataset}                 & \cite{xie2025towards}                       & 10                                               & USRP X310                                                    & WiFi signals collected from different days and positions (indoor, static, mobile).    & \multicolumn{1}{l|}{\cmark}  & \multicolumn{1}{l|}{}       &          \\ \hline
ZigBee                                                     & \cite{shi2023robust_dataset}                  & \cite{shi2023robust}                        & 60                                               & USRP N210                                                    & Outdoor-NLOS, outdoor-LOS,   indoor-NLOS, and indoor-NLOS                               & \multicolumn{1}{l|}{\cmark}  & \multicolumn{1}{l|}{}       &          \\ \hline
LoRa                                                       & \cite{shen2022lora_channel}                   & \cite{shen2021towards}                      & 60                                               & USRP N210                                                    & LoRa preambles. Signals from   different channels available                             & \multicolumn{1}{l|}{\cmark}  & \multicolumn{1}{l|}{\cmark} &          \\ \hline
LoRa                                                       & \cite{shen2023length_dataset}                 & \cite{shen2023length}                       & 10                                               & USRP N210                                                    & LoRa preambles with different   spreading factors                                       & \multicolumn{1}{l|}{}        & \multicolumn{1}{l|}{}       & \cmark   \\ \hline
LoRa                                                       & \cite{shen2024lora_rx}                        & \cite{shen2023towards}                      & 10                                               & USRP N210 (3), B210   (2), B200 mini (2), Pluto (2), RTL (9) & LoRa preambles.   Signals from different receivers and channels available               & \multicolumn{1}{l|}{\cmark}  & \multicolumn{1}{l|}{\cmark} & \cmark   \\ \hline
LoRa                                                       & \cite{shen2024federated_dataset}              & \cite{shen2024federated}                    & 60                                               & USRP N210, B210,   B200, B200 mini, Pluto, RTL               & LoRa preambles.   Signals from different receivers and channels available               & \multicolumn{1}{l|}{\cmark}  & \multicolumn{1}{l|}{}       &          \\ \hline
LoRa                                                       & \cite{deeplora_dataset}                       & \cite{al2021deeplora}                       & 100                                              & USRP N210                                                    & Signals from indoor   and outdoor testbeds, and different days                          & \multicolumn{1}{l|}{\cmark}  & \multicolumn{1}{l|}{}       &          \\ \hline
LoRa                                                       & \cite{oregon_dataset}                         & \cite{elmaghbub2021lora}                    & 25                                               & USRP B210                                                    & Signals from   different days, distances, locations and receivers                       & \multicolumn{1}{l|}{\cmark}  & \multicolumn{1}{l|}{}       &          \\ \hline
Bluetooth                                                  & \cite{jagannath2023embedding_dataset}         & \cite{jagannath2023embedding}               & 10                                               & USRP X300                                                    & Signals collected   from different locations and days                                   & \multicolumn{1}{l|}{\cmark}  & \multicolumn{1}{l|}{}       &          \\ \hline
UWB                                                        & \cite{ardoin2025tracking_dataset}             & \cite{ardoin2025tracking}                   & 13                                               & COST UWB board                                               & Signals collected   from different locations and days                                   & \multicolumn{1}{l|}{\cmark}  & \multicolumn{1}{l|}{}       &          \\ \hline
LTE                                                        & \cite{peng2024hybrid_dataset}                 & \cite{peng2024hybrid}                       & 8                                                & USRP N210                                                    & Signals collected   from different locations and days                                   & \multicolumn{1}{l|}{\cmark}  & \multicolumn{1}{l|}{}       &          \\ \hline
Satellite                                                  & \cite{oligeri2022dataset}                     & \cite{oligeri2022past}                      & 66                                               & USRP X310                                                    & Signals from 66   Iridium satellites                                                    & \multicolumn{1}{l|}{}        & \multicolumn{1}{l|}{\cmark} &          \\ \hline
Satellite                                                  & \cite{2023watch}                              & \cite{smailes2023watch}                     & 66                                               & USRP N210                                                    & Signals from 66   Iridium satellites                                                    & \multicolumn{1}{l|}{}        & \multicolumn{1}{l|}{\cmark} &          \\ \hline
\end{tabular}
\end{table*}

\subsection{Channel Effects Elimination}\label{sec:key_research_topics_channel}

\subsubsection{Research Activities}
The received RF signals are affected not only by the transmitter hardware impairments but also by the wireless channel. In particular, the movement or relocation of wireless transmitters can result in fluctuations in the characteristics of received RF signals, which may subsequently interfere with the RFF extraction process.

The negative impacts of wireless channels have been revealed in numerous recent RFFI studies. The authors in~\cite{al2020exposing} and~\cite{hanna2022wisig} conducted comprehensive experiments to assess the effects of wireless channels on RFFI. Their findings indicate that channel variations can lead to significant performance degradation when fingerprinting WiFi signals. Similar results are also found in fingerprinting wireless signals with narrower bandwidths, such as \ac{lora}~\cite{shen2021towards, elmaghbub2021lora} and ZigBee~\cite{xie2023disentangled}. These studies experimentally demonstrate that the inevitable channel effects can degrade the RFFI performance, presenting a significant challenge that must be addressed.

Recent studies attempt to mitigate the channel effects through two categories of approaches: signal processing and deep learning algorithms. The former category often employs expertise and prior knowledge in wireless communication to design signal processing algorithms to manually separate channel distortions from the received RF signal, constructing channel-robust features for identification purposes. For example, the authors in~\cite{shen2021towards} propose to mitigate the multipath effects in the time-frequency domain by dividing neighbouring columns in a spectrogram, thereby generating a channel-independent signal representation. This method has been demonstrated to be effective in identifying \ac{lora} devices, exhibiting excellent channel-independent properties. 
The work in~\cite{xing2022design} designs a channel-robust WiFi RFF named difference of the logarithm of the spectrum (DoLoS), which is based on the fact that the long training symbols (LTSs) and short training symbols (STSs) in a WiFi packet share similar \ac{cfr}. 
In addition, some studies employ a channel estimation module to approximately measure the \ac{cir} or \ac{cfr}, subsequently utilizing the result to perform channel equalization, thereby largely eliminating the multipath effect~\cite{zheng2019fid, al2020exposing, cekic2020robust, fu2023deep}. However, the channel equalization process can inevitably eliminate some hardware features while eliminating the multipath effect, resulting in the degradation of identification accuracy.

Deep learning algorithms are also employed to mitigate channel effects. More specifically, these approaches aim to enhance the capacity of neural networks, enabling them to extract channel-independent features automatically. The most prevalent method is data augmentation, which has been introduced in Section~\ref{sec:data_augmentation}. In summary, it utilizes a wireless channel simulator to synthesize a greater number of signals exhibiting various channel effects during the neural network training process. This can expand the distribution of training data effectively, aligning it with the test phase to prevent potential performance degradation. The benefits of data augmentation have been illustrated in numerous recent studies~\cite{soltani2020more, shen2021towards, cekic2020robust, al2021deeplora, piva2021tags}, and gradually evolved into a standard procedure in designing RFFI systems. In addition to data augmentation, several studies have attempted to use the latest deep learning methods, such as transfer learning~\cite{tian2022transfer}, domain adaptation~\cite{pan2024equalization}, or disentangled representation learning~\cite{xie2023disentangled} to improve the system's robustness to channel variations.



\subsubsection{Available Datasets}

As discussed, the channel effects, i.e., multipath and Doppler effects, have a considerable impact on the performance of RFFI. It is, therefore, essential to design effective mitigation algorithms.
A number of datasets are available for this purpose, containing signals collected under a variety of positions and channel conditions~\cite{shen2021towards, hanna2022wisig, al2021deeplora, al2020exposing, shi2023robust}. Shen~\etal release a public dataset consisting of 60 \ac{lora} devices. The signal collection is carried out at six locations, including both \ac{los} and \ac{nlos} channel conditions~\cite{shen2021towards}. The authors in~\cite{al2021deeplora} and~\cite{elmaghbub2021lora} also perform experiments in outdoor environments, and the datasets are made available. Shi~\etal collected signals from 60 ZigBee devices, including both indoor/outdoor and \ac{los}/\ac{nlos} scenarios~\cite{shi2023robust}.

In addition to the narrowband \ac{lora} and ZigBee datasets listed above, there are also publicly available datasets within the research community that contain wideband WiFi and \ac{lte} signals. Hanna~\etal collected a large-scale WiFi dataset, consisting of 174 transmitters and 41 \ac{usrp} receivers~\cite{hanna2022wisig}. The experiment is conducted four times, spanning a mont,h within a grid indoor testbed. Additionally, Al-Shawabka~\etal present a large-scale WiFi dataset collected in a grid testbed of 6,000 square ft~\cite{al2020exposing}. The transmitters are 13 \ac{usrp} N210 and seven \ac{usrp} X310, while a \ac{usrp} N210 receiver is used for signal capture. In addition to WiFi, the authors in~\cite{peng2024hybrid} also provide an LTE dataset which consists of signals collected from seven mobile phones. It should be noted that wideband signals are more susceptible to channel variations than narrowband ones.

\subsection{Noise Mitigation}\label{sec:key_research_topics_noise}

\subsubsection{Research Activities}
The propagation of wireless signals over long distances can result in severe attenuation, which in turn leads to a low \ac{snr} condition at the receiver. Given that transmitter hardware impairments are often minute, RFFs are probably submerged in noise. It is therefore difficult to accurately extract them for identification. As wireless systems often operate in RF conditions where the \ac{snr} is less than 20~dB, it is necessary to explore RFFI solutions that are robust to noise contamination. 

Denoising can be leveraged to improve the system's robustness.
Wang~\etal demonstrate that smooth filtering is effective in combating noise contamination~\cite{wang2022radio}, and Xing~\etal conclude that stacking multiple identical symbols is also effective in reducing noise~\cite{xing2018radio}. The authors in~\cite{ohtsuji2019noise} reveal that converting the PHY waveform to a logarithmic power spectral density can improve identification accuracy in low-\ac{snr} environments. Although these manual denoising algorithms are experimentally shown to be effective against noise interference, whether the RFFs are unintentionally eliminated remains unclear. In addition, there are also studies utilizing multiple observations to improve low-\ac{snr} performance, such as merging the identification results of multiple receivers~\cite{shen2023towards, andrews2019crowdsourced, he2020cooperative} or multiple data packets~\cite{shen2023length}.

Apart from these, some studies have attempted to enhance the ability of deep learning models to process low \ac{snr} signals. The authors in~\cite{shen2023length, shen2021asilomar} evaluate different data augmentation strategies, concluding that adding artificial noise to mini-batches during training, i.e., online augmentation, can lead to the most significant improvement. Some studies improve the low-\ac{snr} performance by using specially designed neural networks. For example, the authors in~\cite{wu2021dsln} utilize the dynamic shrinkage learning network, which can integrate denoising capabilities into deep learning models.


\subsubsection{Available Datasets}

RFFs caused by hardware impairments are often faint, and their effective extraction at low SNR conditions is challenging. The authors in~\cite{oligeri2022past} and~\cite{smailes2023watch} present datasets collected from IRIDIUM satellites, which are particularly suitable for low-SNR RFFI research. The IRIDIUM satellites operate in low Earth orbit, at an altitude of approximately 780 kilometers. This results in severe propagation attenuation and makes the signal extremely weak at the ground receiver. In addition to satellites, the \ac{lora} dataset in~\cite{shen2023towards} also contains low-SNR signals. Specifically, the transceiver distances are up to 30 meters, and the signal SNRs are clearly labelled, ranging from 10~dB to 50~dB. Similarly, the authors in~\cite{elmaghbub2021lora} collect \ac{lora} signals at various distances as well, i.e., ranging from 5~m to 25~m. 
Despite the low-quality RF signals provided by the above-introduced datasets, an alternative and efficient method is to add artificial Gaussian noise to high-quality signals, thereby synthesizing low-SNR conditions~\cite{shen2023length, al2021deeplora}. 

\subsection{Receiver Distortion Mitigation}\label{sec:key_research_topics_receiver}
\subsubsection{Research Activities}
While RFFI aims to exploit the transmitter's unique hardware impairments for identification, the receiver impairments will also affect the received signal, as shown in (\ref{eq:signal_model}). Most RFFI studies assume the same receiver is used during the training and inference stages, thus, the receiver effect can be neglected. However, this assumption is not always valid, as the transmitter is frequently served by multiple receivers in practical wireless systems. The authors in~\cite{zhang2021radio} build simulation models to evaluate the effect of receiver impairments. The simulation results demonstrate that the identification accuracy decreases by up to 20\% when the IQ imbalances of the receivers are different between training and test. The performance degradation caused by receiver effects is experimentally validated in~\cite{shen2023towards, elmaghbub2021lora, hanna2022wisig, merchant2019toward}.

{\color{black}To overcome the receiver effect, the authors in~\cite{merchant2019toward, hanna2022wisig} recommend including as many receivers as possible in the training stage, with the aim of improving the model's generalization ability. Moreover, the work in~\cite{shen2023towards} proposes a receiver-agnostic training scheme that employs a gradient reversal layer to direct the deep learning model to learn receiver-independent features. The proposed algorithm was evaluated using 20 SDR receivers, demonstrating excellent generalization ability. The authors in~\cite{zhao2023gan} applied the concept of \ac{gan} to learn receiver-independent features, resulting in an improvement of 20\% in accuracy. Furthermore, recent studies demonstrate that transfer learning and fine-tuning strategies can adapt classifiers trained on one receiver to perform effectively on others~\cite{li2025receiver}.}

\subsubsection{Available Datasets}

As discussed in Section~\ref{sec:key_research_topics_receiver}, the RFFs are not only affected by the transmitter but also by the receiver chain. It is therefore necessary to have datasets containing signals collected by multiple receivers to explore effective receiver-agnostic RFFI solutions. Shen~\etal present a dataset comprising 20 SDR receivers of varying types, spanning from low-end RTL-SDR to high-end \ac{usrp} N210~\cite{shen2023towards}. This dataset was specifically created for the purpose of evaluating receiver effects, and the signals collected at various positions are available. While other \ac{lora} datasets also contain multiple receivers~\cite{shen2024federated, elmaghbub2021lora}, they are incompatible with~\cite{shen2023towards} in terms of number and types of receivers. With regard to WiFi protocols, the WiSig dataset comprises 41 \ac{usrp} receivers, which are suitable for use in research into mitigating WiFi receiver distortion~\cite{hanna2022wisig}.

\subsection{Applications}
There have been many deep learning-based RFFI papers published in the last few years, with applications in WiFi, ZigBee, \ac{lora}, LTE, and satellite communications.

\subsubsection{WiFi}
Recent studies have attempted to apply RFFI to secure WiFi systems, ranging from IEEE 802.11b to IEEE 802.11ax standards~\cite{xing2018radio, del2024fingerprint, al2020exposing, hanna2022wisig, fu2023deep, xing2022design}. Li~\etal design a fractal dimension estimation method to extract features from direct-sequence spread spectrum (DSSS) IEEE 802.11b signals, and use \ac{svm} or \ac{knn} for identification~\cite{xing2018radio}. For wideband \ac{ofdm} signals, the authors in~\cite{al2020exposing} and~\cite{hanna2022wisig} evaluated the channel effects, showing that the multipath effects can significantly degrade the identification performance. To alleviate this problem, Xing~\etal design a DoLoS algorithm, extracting channel-robust features from IEEE 802.11 \ac{ofdm} signals as the neural network input~\cite{xing2022design}. 
The most significant challenge for WiFi RFFI systems is the design of effective algorithms for eliminating channel effects.


\subsubsection{\ac{lora}}

A considerable amount of research has been conducted for designing \ac{lora} RFFI systems~\cite{robyns2017physical, shen2021jsac,shen2021towards,shen2023length,shen2023towards,elmaghbub2021lora,al2021deeplora}. To the best knowledge of the authors,~\cite{robyns2017physical} is the first work attempting to use RFFI to identify \ac{lora} transmitters, which carries out experiments in a transceiver distance of up to 100 meters and achieves 59\% to 99\% accuracy. Furthermore, Shen~\etal conducted a series of research aimed at developing practical and robust \ac{lora} RFFI systems~\cite{shen2021jsac,shen2021towards,shen2023length,shen2023towards, shen2024federated}, and released all the datasets and codes to the public. In~\cite{shen2021jsac}, different signal representations are studied, i.e., IQ samples, \ac{fft} coefficients, and spectrogram, concluding that spectrogram is the most appropriate for \ac{lora} signals because of their frequency-changing property. The work in~\cite{shen2021towards} designed the channel-independent spectrogram to mitigate the channel effects and a three-stage protocol for open-set identification. Afterwards,~\cite{shen2023length} aims at improving identification accuracy in low-\ac{snr} environments, which can be achieved by online augmentation and merging predictions derived from multiple \ac{lora} packets. Finally, the studies in~\cite{shen2023towards} and~\cite{shen2024federated} explore receiver-agnostic and federated RFFI protocols, respectively. In addition to these, the researchers in~\cite{al2021deeplora} and~\cite{elmaghbub2021lora} also carried out in-the-wild experiments to validate RFFI performance. As a low-power, long-range communication technology, the most significant challenge for \ac{lora} RFFI is to design effective algorithms to combat noise contamination.






\subsubsection{ZigBee}

RFFI is also utilized to authenticate ZigBee/IEEE 802.15.4 devices~\cite{merchant2018deep,peng2018design,peng2019deep,xie2023disentangled,yu2019robust,bihl2016feature}. The authors in~\cite{bihl2016feature} construct multiple discriminant analysis (MDA) classifiers to identify ZigBee devices. With the development of deep learning, Merchant~\etal propose inputting error signals, i.e., the difference between received and ideal signals, into CNNs for the identification task~\cite{merchant2018deep}. Peng~\etal design the image-like DCTF feature as the \ac{cnn} input as well~\cite{peng2018design,peng2019deep}. The authors in~\cite{yu2019robust} and~\cite{xie2023disentangled} adopt more advanced deep learning algorithms, using multisampling convolutional neural network (MSCNN) and disentangled representation (DR) learning to construct RFFI systems.


{\color{black}\subsubsection{BLE}
The work in~\cite{givehchian2022evaluating} studied using \ac{BLE} hardware features to track mobile devices. 
The authors designed a non-deep learning-based approach by computing the Mahalanobis distance to track registered devices. 
With comprehensive experiments over 17 mobile devices, involving smartphones, laptops, etc, they revealed it is viable, although sometimes unreliable, to use \ac{BLE} hardware fingerprints to track mobile devices.
The same group later used a CFO obfuscation strategy by modifying CFO~\cite{givehchian2024practical} to prevent such a tracking attack.

Regarding deep learning approaches, Jagannath~\etal designed an embedding-assisted attentional framework and achieved a significant reduction of the memory usage and trainable neural network model complexity~\cite{jagannath2023embedding}.
Yuan~\etal designed a denosing \ac{ae}, with a \ac{cnn} as the backbone, to improve the classification performance under low SNR~\cite{yuan2025robust}. The authors achieved over 75\% accuracy over 10 dB SNR for 18 \ac{BLE} devices.}

{\color{black}\subsubsection{Ultra-Wideband}
\ac{UWB} is usually used for high-resolution localization with a precision at the centimeter level. This is enabled at a cost of high bandwidth, e.g., 500 MHz. 
To the best knowledge of the authors, the work in~\cite{ardoin2025tracking} is the only one studied RFFI for \ac{UWB}. Due to the high bandwidth, it will be challenging using SDR for capturing UWB signals. Instead, the authors employed \ac{COTS} UWB devices to collect CIR measurements and converted them to spectrograms. They designed a Vision Transformer-based deep learning model and achieved over 99\% classification accuracy.}

\subsubsection{Cellular Communications}

Most of the recent RFFI research has focused on identifying devices operating in the unlicensed industrial, scientific, and medical (ISM) bands, and there are only a few studies have investigated its application in cellular systems~\cite{zhuang2018fbsleuth,yang2023led,yin2024multi,peng2024hybrid,reus2020trust}, e.g., GSM, 4G LTE, 5G NR. Zhuang~\etal utilize RFFI technology to identify GSM base stations to detect fake base station (FBS) crimes, which is achieved by extracting modulation errors and statistical features as unique RFFs~\cite{zhuang2018fbsleuth}. The authors in~\cite{yang2023led} propose an RFFI method by applying wavelet decomposition to the 4G LTE demodulation reference signal (DMRS). Yin~\etal capture the transient-on, transient-off, and modulation segment of LTE physical layer random access channel (PRACH) preambles, converting them to DCTF representations separately and designing a multi-channel \ac{cnn} for identification~\cite{yin2024multi}. Peng~\etal combines wavelet transform (WT) coefficient graphs and differential spectrum to extract RFFs from LTE signals and conduct experimental evaluations using commercial phones and SDRs~\cite{peng2024hybrid}. Finally, the authors in~\cite{reus2020trust} create a 5G RFFI system involving four base stations, which demonstrates that RFFI is effective in securing 5G networks as well.





\subsubsection{Satellite Communications}

Recent studies have applied the RFFI technique to satellite identification~\cite{foruhandeh2020spotr,oligeri2022past,smailes2023watch}. For instance, the authors in~\cite{foruhandeh2020spotr} utilize it to detect \ac{gps} spoofing attacks. Specifically, they use multivariate normal distribution (MVN) models to extract features from the captured IQ samples and set a threshold to detect the spoofed GPS signals. Oligeri~\etal targets to identify the \ac{leo} IRIDIUM satellites, which was originally developed by Motorola in the last century~\cite{oligeri2022past}. The authors observe and collect IQ samples from 66 satellites, and then train a \ac{cnn} for identification. The results show that the accuracy is above 80\%. Smailes~\etal also conducts extensive research in fingerprinting IRIDIUM satellites~\cite{smailes2023watch}.

{\color{black}\section{RFFI Experimental Methodologies}\label{sec:exp_methodology}
While it is fundamental to carry out experimental evaluation for RFFI to assess its performance in practical scenarios, there is a lack of explanation of the experimental methodologies in the literature. This section aims to bridge the gap.

Building a testbed is mandatory to carry out an experimental evaluation, therefore this section will cover necessary information for building a testbed, including DUT and receiver. Then, we will discuss the requirements for the dataset collection. 

\subsection{DUT}
There are several DUT options, which include IoT development kits, consumer electronics, and \ac{sdr} platforms.

\textbf{IoT kits} are probably the most commonly used devices. Some examples are given below:
\begin{itemize}
\item WiFi: ESP32~\cite{kong2024csi}
\item BLE: ESP32 and Nordic Semiconductor nRF52840 dongles~\cite{yuan2025robust}
\item LoRa: Pycom LoPy4 \& FiPy (discounted), mbed shield, and Dragino shield~\cite{shen2021towards}.
\end{itemize}
The vendors usually provide example code snippets for transmitting and receiving. Their transmission behavior can be customized, e.g., transmission intervals, transmit power, etc. 

It is desirable to demonstrate that RFFI can work with \textbf{consumer electronics}. We exemplify the following consumer electronics DUTs:
\begin{itemize}
\item WiFi: WiFi dongles~\cite{yin2025noise} and  Nexus 5 smartphones~\cite{gu2024cqp} 
\item BLE: Smartphone, laptop, Apple Watch etc~\cite{givehchian2022evaluating}
\item LTE: Smartphone~\cite{yin2024multi,gu2024cqp}
\end{itemize}
Compared to IoT kits, it is relatively difficult to control transmission parameters accurately. The transmissions can instead be triggered by, e.g., running the ping command for WiFi~\cite{yin2025noise}. For BLE, the device will transmit advertising packets periodically, which can be leveraged~\cite{givehchian2022evaluating}. On the other hand, there is also research reported by using a third-party firmware, nexmon, to inject I/Q imbalance into the baseband signal of smartphones~\cite{gu2024cqp}.

\begin{table*}[!t]
  \centering
    \caption{Summary of SDR Platforms and Their Applications in RFFI}
		\label{tab:sdr}
\begin{tabular}{|l|L{5cm}|l|l|}
\hline
Software                                   & Supported SDR   Platform                                                            & Wireless   Technology                        & Representative   RFFI Papers                                   \\ \hline
\multirow{3}{*}{Matlab   \cite{MatlabSDR}} & \multirow{3}{5cm}{ADALM-PLUTO   SDR, RTL-SDR, USRP SDR and Xilinx® Zynq®-Based Radio} & WiFi (Matlab WLAN   toolbox)                 &   Not reported     \\ \cline{3-4} 
                                           &                                                                                     & LoRa (custom code)                           & USRP N210   \cite{shen2021radio}                               \\ \cline{3-4} 
                                           &                                                                                     & BLE (Matlab Bluetooth toolbox)               & USRP N210   \cite{yuan2025robust}                              \\ \hline
\multirow{2}{*}{GNURadio}                  & \multirow{2}{*}{All SDR platforms support GNURadio}                                 & WiFi IEEE 802.11a/g/p   \cite{gr-ieee802-11} & HackRF One   \cite{li2022radionet}, USRP N210 \cite{gu2024cqp} \\ \cline{3-4} 
                                           &                                                                                     & IEEE 802.15.4 \cite{gr-ieee802-15-4}         & Not reported                                                   \\ \hline
Python \cite{PySDR}                        & ADALM-PLUTO   SDR, RTL-SDR, USRP SDR, HackRF One and BladeRF                        & Custom code                                  & Not reported                                                   \\ \hline
PicoScenes   \cite{PicoScenes}             & USRP SDR and HackRF   One                                                           & WiFi IEEE   802.11a/g/n/ac/ax/be             & USRP N210   \cite{yin2025noise}                                \\ \hline
\end{tabular}
\end{table*}

\begin{table*}[!t]
\centering
\caption{Summary of CSI Tools and Their Applications in RFFI}
\label{tab:csi_tool}
\begin{tabular}{|l|L{7.4cm}|L{4.5cm}|}
\hline
CSI Tool                                                          & Supported   Amendaments and Chipsets/SDR                                                            & Representative   RFFI Papers                          \\ \hline
Intel 5300 CSI   tool \cite{halperin2011tool}                     & IEEE 802.11n for   IWL5300                                                                      & \cite{hua2018accurate,liu2019real,huang2023phyfinatt} \\ \hline
Atheros CSI   tool \cite{Xie:2015:PPD:2789168.2790124,atheroscsi} & IEEE 802.11n                                                                                    & Not reported                                          \\ \hline
Nexmon CSI   tool \cite{gringoli2019free,nexmon:project}          & IEEE 802.11a/g/n/ac   for Broadcom WiFi Chips                                                   & Not reported                                          \\ \hline
ESP32 CSI tool   \cite{Hern2006:Lightweight,esp32csi}             & IEEE 802.11n                                                                                    & Not reported                                          \\ \hline
PicoScenes   \cite{PicoScenes}                                    & IEEE   802.11a/g/n/ac/ax for USRP SDR and HackRF One, AX210/AX200, IEEE 802.11n for QCA9300 and IWL5300                  & AX210   \cite{huang2023phyfinatt}                     \\ \hline
SDR + Matlab   WLAN Toolbox                                       & IEEE   802.11a/g/n/ac/ax for ADALM-PLUTO SDR, RTL-SDR, USRP SDR and Xilinx®   Zynq®-Based Radio & Xilinx® Zynq®-Based   Radio \cite{kong2024csi}        \\ \hline
\end{tabular}
\end{table*}

\textbf{\ac{sdr}} platforms are also employed as DUTs for RFFI. For example, the authors in~\cite{li2022radionet} used 10 HackRF One SDRs and the work in \cite{restuccia2021deepfir} used 20 USRP SDRs. Using SDRs can provide full hardware control. For example, the hardware impairments are reconfigured in~\cite{sankhe2019oracle}. However, their cost is usually higher than IoT kits and consumer electronics.

\subsection{Receiver}\label{sec:rx}
The receiver plays an important role in RFFI, which will perform signal collection and then feed the collected signals for classification. In deep learning-based RFFI, there are two categories of signals for deep learning input, namely I/Q samples and CSI.

\subsubsection{Platform for I/Q Samples Collection}
Most of the RFFI works rely on I/Q samples, which can be captured by \ac{sdr} platforms. There are different ways to access data from \ac{sdr}, including Matlab, GNURadio, Python-based libraries, and PicoScenes, as summarized in Table~\ref{tab:sdr}.

While SDR can capture the I/Q samples, a signal analysis program is required to decode and interpret the data samples. For example, signal synchronization algorithms are required to locate the starting point of the collected packets. MAC address decoding is required for WiFi and Bluetooth to ensure that the captured packets are sent from the target DUT, because there are numerous WiFi and Bluetooth transmissions over the air. Such functions can be either achieved by custom-built codes or available third-party solutions such as Matlab toolboxes, GNURadio implementations (see Table~\ref{tab:sdr}). Regarding PicoScenes, it is a middleware specifically created for WiFi.

\subsubsection{Platform for WiFi CSI Collection}
\Ac{csi} can represent fine-grained channel information. 
While most of the WiFi chipsets do not provide the CSI, there are a few exceptions, as summarized in Table~\ref{tab:csi_tool}. Intel 5300 CSI tool~\cite{halperin2011tool} is probably the most widely used as it is the first CSI tool. However, it can only report channel matrices for 30 subcarrier groups, i.e., every 2/4 subcarriers at 20/40 MHz. Nexmon CSI tool can support up to 80 MHz and return estimated CSI for all the subcarriers, which can significantly increase the extracted information. PicoScenes can support the latest WiFi 6 with up to 160 MHz bandwidth for AX210/AX200. Besides, the MATLAB WLAN toolbox can also provide CSI.

\subsection{Requirement of Dataset Collection}
As indicated in~\cite{al2020exposing}, the training and test datasets are collected on two different days, and channel conditions are similar, but their deep learning-based RFFI cannot work at all~\cite[Fig. 11]{al2020exposing}. This reveals that even slight variations in the channel, noise, or hardware impairments will result in significant consequences. 
In practice applications, the test datasets are highly likely collected on different days from the training datasets. Hence, we need to design a robust RFFI algorithm. In order to demonstrate the robustness of RFFI algorithms, it is always necessary to have training and test datasets collected from different days. 

It is important to avoid overfitting in RFFI. For example, when evaluating RFFI against channel variations, the training and test datasets should not be collected from the same environment or environments with similar channel conditions. In particular, as many different channel scenarios as possible should be covered, e.g., \ac{los} \& \ac{nlos}, static \& mobile, indoor \& outdoor, etc.
}

\section{Channel-Based Authentication}\label{sec:ChBasedAuth}


\Ac{cb} \ac{pla} verifiers leverage the effects of the communication channel for authentication; thus, in this case, the nature of the communication channel itself enables authentication.

\begin{figure}
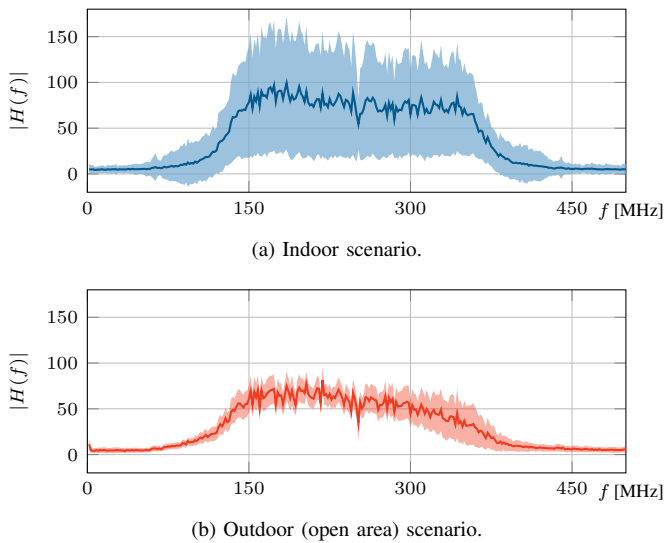

    \centering
    \subfloat[Indoor scenario.]{\input{Pictures/auth/exampleCFR1}}\\
    \subfloat[Outdoor (open area) scenario.]{\input{Pictures/auth/exampleCFR2}}
    \caption{Examples of \ac{cfr}, mean and $1\sigma$ bounds for an indoor and an outdoor scenario.}
    \label{fig:subcarriers_pic}
\end{figure}

In particular, a channel measurement, typically called \emph{channel feature}, is selected. As an example, Fig. \ref{fig:subcarriers_pic} depicts the absolute value of the \ac{cfr} measured using Wi-PoS, an \ac{UWB} hardware platform, with carrier frequency \SI{6.489}{\giga\hertz} and bandwidth of \SI{499.2}{\mega\hertz}, place at the fourth floor of the iGent Tower and in Portus Ganda, both located in Ghent, Belgium \cite{GhazalehDataset}. In particular, we report mean and $1\sigma$ bounds computed over $300$ measurements.  Indeed, while the traces collected in the same environments are related, there are significant changes when the devices are collected in different locations. For instance, the indoor environment is associated with a much higher standard deviation, e.g., due to multipath, than the outdoor environment. This highlights that, indeed, we can exploit traces like these, and thus the channel, to authenticate the devices, as a signal sent by a spoofer placed in a different environment will induce a significantly different \ac{cfr}. The review of the channel features to be used for authentication purposes is reported in  Section~\ref{sec:chFeatures}.

Concerning the authentication mechanism, three main approaches have been studied: the tag-based, channel variation, and \ac{cr} approaches. 
\begin{itemize}
\item The {\em tag-based approach} assumes that the channel features do not change over time and the impersonating attacker is in another location than the legitimate transmitter, enabling the verifier to distinguish between the two transmitting locations by processing the received signal. 
\item The {\em \ac{cr} approach} still considers static channels but it also assumes that the verifier can modify the propagation environment and predict the resulting channel features. Thus, authentication is performed by introducing a random modification (challenge) and comparing the predicted channel features with those obtained from the received signal. 
\item The {\em channel variation approach} is suitable in scenarios where the channel changes, as it includes a prediction of the channel features from previous observations and a comparison between the predicted features and those estimated from the currently received message. 
\end{itemize}
Note that the latter two approaches assume instead that the channel changes, but variations are in part known by the verifier: either change can be predicted (e.g., channel variation) or controlled (e.g., \ac{cr}-authentication) by the verifier.
 
\begin{figure}
     \centering
     \includegraphics[width=3.4in]{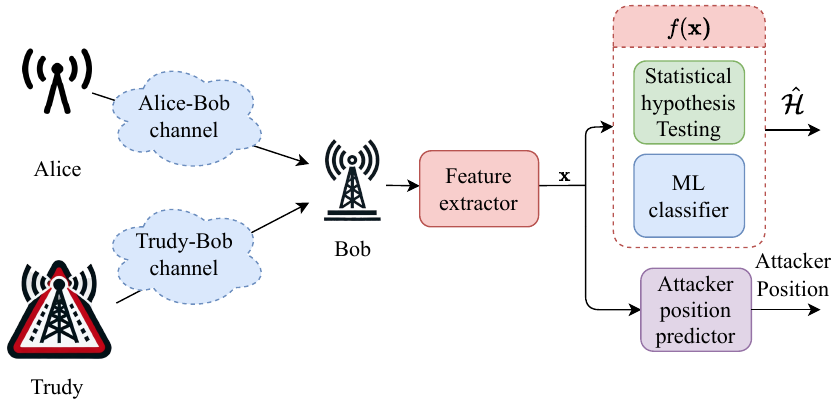}
     \caption{High-level representation of the \acs{cb}-authentication scheme. }
     \label{fig:survey}
 \end{figure}

Fig. \ref{fig:survey} provides a general scheme for \ac{cb} \ac{pla}, in which first the verifier Bob extracts the vector feature $\bm{x}$ from the channel of the received signal and then classifies it as legitimate or not by means of a function $f(\cdot)$. In particular, the classification variable is $\hat{\mathcal{H}} = \hat{\mathcal{H}_0}$ when the received message is considered authentic and $\hat{\mathcal{H}} = \hat{\mathcal{H}_1}$ otherwise. In case the signal is marked as malicious, it may be possible to localize the attacker (see Section~\ref{subsec: ML approach_loc}).  

The following two sections will describe the features and methodologies used for \ac{pla}. Table~\ref{tabCBsol} summarizes the main publications on \ac{cb}-authentication. It provides a categorization based on the application domain (radio communications on air, \ac{uwac}, and \ac{vlc} on air), the decision methodologies (statistical and \ac{ml}), and the channel features used for authentication. 

\subsection{Tag-based PLA}
A typical tag-based \ac{pla} protocol includes two phases: 
\begin{itemize}
    \item {\em Identification Association}: the legitimate transmitting device (Alice) sends a pilot sequence to the verifier device (Bob), which {\em estimates or learns} channel characteristics/behavior, which we call {\em tag}. The pilot transmission is assumed to be authenticated because either we are sure that the attacker is not transmitting or an authentication mechanism using a pre-shared secret between Alice and Bob is implemented. This phase is performed only once and is not repeated for each transmitted packet.   
    \item {\em Identification Verification}: Bob receives a message containing pilot symbols, estimates the channel, and compares such an estimate with the tag obtained in the identification association phase. If the two are {\em compatible}, the message is considered authentic, otherwise, it is discarded as fake. This phase is performed for each message transmission, after the identification association.
\end{itemize}

Tag-based \ac{pla} is subject to two issues: a) the channel may change over time due to the mobility of either the user or surrounding objects, and b) the channel estimate is affected by noise and receiver impairments (e.g., synchronization issues).  Both make the definition of {\em compatibility} between the channels estimated in the two phases problematic. To cope with these issues, two main research paths have been taken. About the first issue, we note that tag-based \ac{pla} is well suited for static channels, while some modifications are needed to make it work under channel variations.

The first research path has looked for channel characteristics that are more robust against both issues to be used for authentication. For example, the number of taps of the \ac{cir} is less time-variant than the full \ac{cir}. Similarly, focusing on the received power and dropping the channel phase information in narrowband transmissions provides a robust approach against synchronization errors.

The second research path focuses on the methodology to assess the message authenticity, taking into account the impairments (interference, noise, channel variations) of the estimated tag. A first set of solutions is obtained by framing tag-based \ac{pla} as a hypothesis testing problem between two hypotheses (the received signal is legitimate or fake) and resorting to statistical tests for its solution: this however, requires the knowledge of the tag statistics. A second set of solutions is obtained by considering tag-based \ac{pla} as a classification problem to be solved using \ac{ml} models: this approach requires a dataset of tag realizations. Solutions mixing the two approaches are also possible, e.g., using a raw statistical approach as the initial test, then a more refined \ac{ml} test.


In a general tag-based \ac{pla} approach, two hypotheses are considered, $\mathcal{H}_0$ and $\mathcal{H}_1$, corresponding to the legitimate and the under-attack case, respectively. To assess the authenticity of the signals, the verifier uses the tag verification function $f(\bm{x})$ and computes the decision via thresholding, i.e., deciding 
\begin{equation}\label{eq:decisionFun}
   D:\quad \hat{\mathcal{H}} = \begin{cases}
        \hat{\mathcal{H}}_0 & f(\bm{x}) \geq \lambda, \\
        \hat{\mathcal{H}}_1 & f(\bm{x}) < \lambda,
    \end{cases}
\end{equation}
where $\lambda$ is a threshold chosen by the verifier, e.g., to minimize the \ac{md} for a target \ac{fa} probability. 

When using a statistical approach, the function $f(\cdot)$ is derived analytically, typically exploiting the statistics of the $\bm{x}$ under $\mathcal{H}_0$ and, eventually, $\mathcal{H}_1$.
More details on this approach are provided in Section~\ref{statapp}.
On the other hand, in the \ac{ml} domain, we aim at classifying the observed tag $\bm{x}$ into the two classes of legitimate and attack messages. We still perform the test with~\eqref{eq:decisionFun}, where now $f(\cdot)$ represents an \ac{ml} model, trained with a dataset of labelled tags, where the label indicates the class to which the tag belongs. 

Comparisons between the statistical and \ac{ml} paradigms have been reported in~\cite{Brighente2019Machine,Senigagliesi2021Comparison,Ardizzon2024Learning}. By properly designing the \ac{ml} model and its training, it is possible to achieve the same performance as the statistical approaches. In general, however, the choice between the two approaches is dictated by the knowledge of statistics or the availability of datasets.

A relevant distinction, common to solutions of both statistical and ML domains, concerns the knowledge about the attacker. In particular, we distinguish
\begin{itemize}
    \item {\em Binary Classification}: the verifier Bob knows the distribution of both Alice and Trudy (statistical domain) or has a labeled dataset with tags belonging to both Alice and Trudy (ML domain). {\color{black} We remark that in the literature, binary classification is also referred to as two-class classification, as it exploits information (i.e., distribution and/or labeled data) of both Alice and Trudy. }
    \item {\em Artificial Dataset}: the verifier has a dataset with tags belonging to Alice but makes some assumptions about Trudy and generates an artificial dataset of Trudy's tags.
    \item {\em One-Class Classification}: the verifier knows only the distribution of Alice (statistical domain) or has a dataset with tags belonging only to Alice. {\color{black}Especially in \ac{ml} contexts, one-class classification is also referred to as an anomaly/outlier detection task.}
\end{itemize}
\textit{Remark}: it is worth pointing out that, differently from the binary class case, in the artificial dataset case, the knowledge about the attacker is only partial, and thus such knowledge does not allow the legitimate party to build a fully reliable Trudy dataset. For instance, we know a region where Trudy may be, but we do not know the exact position; we can then build a dataset with observations from the whole region.



\subsection{Challenge-Response PLA}\label{crplasec}

Modern communication systems enable a partial modification of the electromagnetic propagation environment. For example, a \ac{ris} can be used in a \ac{pla} context and be controlled by the verifier to steer impinging signals in desired directions. Another example is obtained when the verifier is a moving device, e.g., a drone, that can modify the channel from the transmitter by changing its position. \Ac{cr}-\ac{pla} is a mechanism that leverages such control of the electromagnetic environment to strengthen authentication~\cite{9982485}. We define the different conditions of the channel induced by the behaviour of the verifier (e.g., the \ac{ris} configuration or the drone position) as {\em channel configurations}. 

\begin{figure}
    \centering
    \subfloat[Identification association in CR-PLA. \label{fig:subfig1}]{\includegraphics[width=0.9\linewidth]{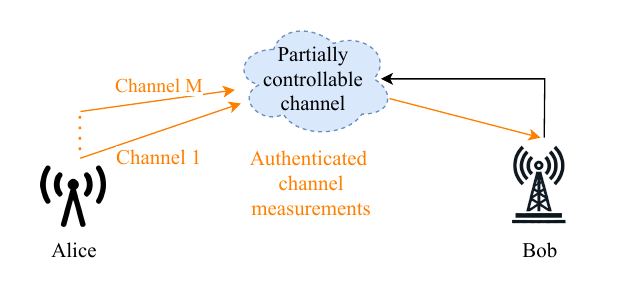}}\\
    \subfloat[Identification verification in CR-PLA. \label{fig:subfig2}]{\includegraphics[width=0.9\linewidth]{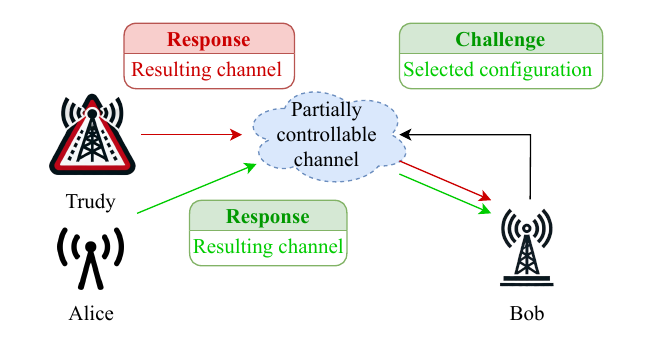}}
    \caption{Challenge-response paradigm scheme.   }
    \label{fig:CR_based_pla}
\end{figure}

As shown in Fig.~\ref{fig:CR_based_pla}, the \ac{cr}-\ac{pla} procedure includes two phases, similar to tag-based PLA, but with different contents:
\begin{itemize}
\item {\em Identification Association:} First, the channel features from the device to be identified are estimated by the verifier {\em for different channel configurations}, namely Channel $1,\ldots,M$. The estimation of the first phase enables the receiver to obtain estimates of the channel also for configurations that have not been explored in this phase, through interpolation algorithms. We must ensure that in this phase the transmission is legitimate (thus no spoofing attack is possible), as it happens in the identification association phase of tag-based \ac{pla}. 
\item {\em Identification Verification:} The identification verification phase is split into two steps, challenge and response.
\begin{itemize}
\item {\em Challenge}: The verifier selects at random a channel configuration before transmission of the message.
\item {\em Response}: The message is transmitted and the verifier estimates the resulting channel from the received signals. Lastly, the verifier compares the estimated channel with the channel predicted for the selected channel configuration, according to the information acquired in the first phase. 
\end{itemize}
\end{itemize}

Note that an attacker to be successful must transmit its signals through the legitimate channel as modified by the receiver (e.g., through the \ac{ris} rather than directly to the receiver) or it must know the instantaneous channel configuration and shape its attack accordingly (see~\cite{10700780} for an in-depth analysis). 

\ac{cr}-\ac{pla} introduces additional randomness to the channel conditions (with respect to tag-based \ac{pla}), thus achieving higher robustness against attacks. The optimization of the defence strategy (choice of the random channel configuration in the Challenge phase) and of the attack strategies has been investigated in~\cite{10615572, 10615677}. 

\subsection{Channel Variation PLA}\label{sec:chVariation}

Channel variation \ac{pla} is an authentication mechanism specifically designed for time-varying channels. 
A typical channel variation protocol works as follows:
\begin{enumerate}
    \item The legitimate transmitter sends $N$ packets to the verifier, which estimates the channel features for each packet ({\em identification association} step).
    \item From the estimated features, the verifier predicts the channel features for future transmissions.
    \item Upon reception of the packet to be verified, the verifier checks the consistency of the predicted and the measured features ({\em identification verification} step).
\end{enumerate}
As in the tag-based \ac{pla}, the main research directions investigate both the choice of robust features and the design of predictors providing a good trade-off between \ac{fa} and \ac{md} probabilities. The predictor can be implemented both via statistical methods or, more recently, via \ac{ml}.

In formulas, given the previous channel observations $[\bm{x}_{-N},\bm{x}_{-N+1},\ldots,\bm{x}_{-1}]$, the prediction function is denoted by $g(\cdot)$ and it provides 
\begin{equation}
{\bm{y}_0} = g\left(\left[\bm{x}_{-N},\bm{x}_{-N+1},\ldots,\bm{x}_{-1}\right]\right) \,,
\end{equation}
and the consistency check for the current observation $\bm{x}_0$ is 
\begin{equation}\label{eq:predictorTest}
    \hat{\mathcal{H}}= \begin{cases}
        \hat{\mathcal{H}_0} & \mbox{if }\| {\bm{y}_0} - \bm{x}_0 \| < \lambda, \\
        \hat{\mathcal{H}_1} & \mbox{if }\| {\bm{y}_0} - \bm{x}_0 \| \geq \lambda.\\
    \end{cases} 
\end{equation}
Variations to \eqref{eq:predictorTest} include tests where no prediction is performed but the threshold is dynamically updated, e.g., as in~\cite{Xiao2016PHY}.

We remark that differently from tag-based \ac{pla}, channel variation \ac{pla} is typically considered as a single-sided hypothesis testing or a one-class classification problem, without any knowledge of the attacker's behavior.



\subsection{Channel-Based Attacker Localization}
When dealing with \ac{pla} attacks, it is important that the receiver first detects the attack and then possibly localizes the attacker. We will also see that some features used for verification are also used for localization. Thus, when implementing \ac{cb}-\ac{pla} protocols, on top of the authentication, it may be possible to add an attacker localization step at a relatively low cost. Thus ,during the rest of the survey, we will also include solutions for attacker localization.  

\begin{table*}
\setlength{\tabcolsep}{10pt} 
\renewcommand{\arraystretch}{1.5} 
\newcolumntype{d}{>{\centering\arraybackslash} p{2.2cm}}
\newcolumntype{t}{>{\centering\columncolor[HTML]{F5F5F5}\arraybackslash} p{2.2cm}}
\newcolumntype{y}{ p{2.5cm}}
\caption{Literature Classification of \ac{cb}-\ac{pla} Mechanisms.}\label{tabCBsol}
\begin{center}
\begin{tabular}{|y| t d | t d | t |}
\hline
\multirow{2}{*}{\textbf{Feature list}} &
\multicolumn{2}{c|}{\textbf{Radio (Air)}} &
\multicolumn{2}{c|}{\textbf{Acoustic Underwater}} &
\textbf{Visible Light (Air)}  \\
\hhline{~*5{-}}
&  Statistical & ML                    &  Statistical & ML                        & Statistical                           \\                     
\hline\hline
\acs{cir} &~\cite{Tan2023Generalized,amin2024potential,Xie2021Physical,Wang2020PhysicalKalman, 10700780,10615572, 10615677, 10436823,piana2024challenge, 6584940, 7501833, Cicek2024impact} &~\cite{Lu2020Reinforcement,Fang2019Learning,Meng2024Multidimensional,Wang2019Deep,Abdrabou2022Adaptive,Meng2023Physical} &~\cite{Li2015Spoofing,Khalid2020Node,Wang2017Physical} &~\cite{Xiao2019Learning,Zhao2023Physical} &~\cite{Brighente2024Physical}     \\
\hline
\acs{cfr} &~\cite{5683463, 4289438,7733021} & ~\cite{song2021enhancing, Liu2018Authenticating,Wang2022Channel,Han2024Model, Chen2021Automated,Xie2022Weighted,pan2019threshold,zhang2023cooperative,chen2023physical,wang2022csi,liao2019security,Qiu2020Learning,Liu2014Practical} &  & &   \\
\hline
\acs{rss} or \acs{snr} &~\cite{Abdrabou2024Game,5934990,piana2024challenge,Kalamandeen2010Ensemble,Wang2020PhysicalKalman, Wang2020NonlinearKalman} &~\cite{Abdrabou2022Physical,Romero2024Spoofing,Xiao2016PHY,Fang2019Learning, 10.1145/3522783.3529522,9321212,Meng2024Multidimensional,Du2023Physical,abdrabou2022authentication,Lu2020Reinforcement, chen2007detecting,yang2012detection,gajbhiye2016rss} &   &  ~\cite{ventura2024authentication}&~\cite{8908721}     \\
\hline
Channel Statistics & ~\cite{He2024enhancing,Joo2024Enhancing} &~\cite{Meng2024Multidimensional} &~\cite{Diamant2019Cooperative,Casari2022Physical}  &~\cite{Bragagnolo2021Authentication, Ardizzon2022Machine,Ardizzon2024RNN} &  \\
\hline
Time Measurements &{\cite{Jedermann2021Orbit,Gao2019Lightweight}} &~\cite{Lu2020Reinforcement}&\cite{Aman2018Impersonation} &  &  \\ 
\hline
\acs{aoa} &~\cite{10.1145/3522783.3529522,Xiong2013SecureArray,ning2020detection,tao2023pilot,xiong2010secure, Bendaimi24how} &~\cite{Wang2018Efficient,10.1145/3522783.3529522,9321212,gao2023esanet} &~\cite{Aman2018Impersonation,Khalid2020Physical} &  &  \\
\hline
Doppler Frequency &{\cite{Yi2021Initial,abdrabou2022authentication,Abdrabou2024Game,Topal2022Physical}}  &~\cite{Abdrabou2022Physical,Fang2019Learning,Du2023Physical}  &  &   &   \\
\hline 

\end{tabular}
\end{center}
\end{table*}

\section{Channel Features For CB-Authentication}\label{sec:chFeatures}

In this Section, we describe the main channel features that are used in \ac{cb}-\ac{pla}. Although most of the works consider specific features, it may be possible to extend their procedures to other features. Moreover, some works do not focus on specific channel features in their schemes but rather aim to design solutions that work for any feature selected by the user. 

For each channel feature, we will also include a discussion on techniques used to localize the attacker, if any.

\subsection{Channel Impulse and Frequency Response}

One of the most popular channel features for \ac{pla} is the entire \ac{cir}. Apart from channel estimation techniques that are common to all approaches, it is possible to distinguish between two main paradigms in their use. In the first approach, raw \ac{cir} is used for authentication, which has been proposed in different contexts, e.g., \ac{6g}~\cite{Meng2024Multidimensional}, \ac{uwac}~\cite{Li2015Spoofing}, and more recently in \ac{vlc}~\cite{Brighente2024Physical}. In the second approach, preprocessing is performed on the estimated \ac{cir}, and a new metric is extracted. Often, such a metric is the result of the comparison between the measured \ac{cir} and a database of previously collected (and trusted) responses. Examples of this approach are for the \ac{uwac} context, the Froebenious norm in~\cite{Xiao2019Learning}, the time reversal-based metric in~\cite{Khalid2020Node}, the maximum and minimum correlation amplitude in~\cite{Zhao2023Physical}, while in the radio context, both the Euclidean distance and the Pearson correlation coefficient in~\cite{Wang2017Physical}. A related approach for WiFi was also proposed in~\cite{Tan2023Generalized} where a channel feature is obtained by comparing the measured \ac{cir} with a locally generated replica. 



The \ac{cir}/\ac{cfr} is also used in \ac{cr}-\ac{pla}. Some of them refer to the \ac{cir} over the \ac{mimo} channels. In this case, by increasing the number of antennas in either or both the transmitter and the receiver, we obtain a richer description of the channel that typically improves the accuracy of the authentication procedure. In particular, \ac{cr}-\ac{pla} has been applied to scenarios with \acp{ris}~\cite{10700780}, where the verifier controls the \ac{ris}, and the channels between the devices and the \ac{ris} are \ac{mimo}. Another context where \ac{cr}-\ac{pla} has been applied, still on the \ac{cir}/\ac{cfr} of the channel is in communications with drones, where the channel variations are due to the movement of the drones, not necessarily equipped with multiple antennas~\cite{10436823,piana2024challenge}. Similar approaches of \ac{cr}-\ac{pla} have been proposed in other frameworks. For example,~\cite{6584940} proposes a scheme in which the \ac{cir} is used to hide both the challenge and the response of the attacker using wiretap coding techniques; an extension of this approach includes the use of artificial noise~\cite{7501833}.  

An alternative to \ac{cir} is \ac{cfr}, which can be easily obtained from the \ac{cir} by \acs{fft}, or is immediately available in communication systems operating in the frequency domain, e.g., the \ac{ofdm}.  Examples of application domains where \ac{cfr} is used for \ac{cb}-\ac{pla} are \ac{iot}/industrial \ac{iot}~\cite{Chen2021Automated,Xie2022Weighted,pan2019threshold,zhang2023cooperative,chen2023physical}, cellular systems~\cite{Qiu2020Learning,liao2019security}, or WiFi~\cite{5683463, 4289438,song2021enhancing, Liu2018Authenticating, 7733021}. 
Other works exploit a dataset of previously collected \acp{cfr} to derive a metric, e.g., in~\cite{pan2019threshold} the authors compare the performance when using the whole \ac{cfr} matrix as input rather than the difference between a reference channel matrix and a measured one. Several works exploit a database of previously collected \acp{cfr} to predict the current one, which is then compared to the measured \ac{cfr} to verify authenticity, e.g.,~\cite{Han2024Model}. The \ac{cfr} in an \ac{ofdm} system can also be interpreted as an image and then used to feed deep learning models~\cite{wang2022csi}.
 
Although \ac{cir} and \ac{cfr} provide a complete description of the channel, resulting in a very detailed channel feature for \ac{pla}, their estimate is subject to several limitations that either reduce the accuracy of the authentication or require additional processing. Indeed, the main problem is related to synchronization, which may differ upon reception of different messages. Synchronization errors lead to time and phase shifts in the baseband equivalent \ac{cir}, and correspondingly to phase changes in \ac{cfr}. Moreover, the estimate of each channel tap is affected by estimation noise that may significantly change the resulting \ac{cir} and \ac{cfr}.

\subsection{Received Power}

To address the issues of \ac{cir} and \ac{cfr} features, it is possible to use {\em channel parameters} either directly estimated from the received signal or extracted from the estimated \ac{cir} and \ac{cfr}, which are robust to impairments. The first case is given by the received power, which is a channel parameter basically unaffected by small synchronization errors. Many receivers already compute the received power, for example, for signal quality assessment, range, or power control; thus, no extra effort is required to obtain the input of \ac{pla} mechanisms. Note that the received power is also denoted as \ac{rss}, while the estimated \ac{snr} is analogous to the received power, apart from a normalization of the noise power.

Excluding the fading component, the received power is related to the path loss and thus the relative distance between the transmitter and Bob. Moreover, in a context where Alice-Bob's distance is known, Bob can compute the reference power to be compared against the measured one for authentication purposes. Examples of this approach include the use of the norm between the measured and the expected \ac{rss}~\cite{Xiao2016PHY}, \ac{snr}~\cite{10.1145/3522783.3529522}, received power~\cite{Abdrabou2024Game}, and the detection of anomalous path gains via \ac{ml}~\cite{Wang2018Efficient}, eventually also including dynamic scenarios~\cite{Romero2024Spoofing}.

Thus, many works propose the use of this already-available feature for \ac{pla} purposes in different contexts such as WiFi, \ac{vanet}, \ac{uwac}, and \ac{vlc}, as now discussed.
For instance, in WiFi, the variations of \ac{rss} due to the movement of devices have been exploited in~\cite{Kalamandeen2010Ensemble} to determine whether two pairing devices are in physical proximity to each other, thus authenticating their proximity. In~\cite{Lu2020Reinforcement}, an \ac{rl} mechanism is used in the \ac{vanet} context to dynamically adjust the authentication test threshold based on the previous results, including false alarm rates and authentication policy parameters.

Concerning the \ac{uwac} context, the \ac{scm}, i.e., the power covariance measured at different receivers, is used in~\cite{ventura2024authentication} to estimate the transmitter position, and later for channel variation \ac{pla}. 

In the \ac{vlc} context, a first study of \ac{cb} authentication is~\cite{8908721}, where the attacker transmits when the legitimate transmitter is idle; then \ac{los} direct current channel gain is used as an authentication feature.
Then,~\cite{Brighente2024Physical} applies the \ac{cr}-\ac{pla} technique to \ac{vlc}. In this case, a \ac{ris} that operates on visible light signals is configured randomly to enable the receiver photodetector to authenticate the transmitter.  

\paragraph*{Attacker Localization} The \ac{rss} can be used further for attacker localization~\cite{chen2007detecting,yang2012detection,gajbhiye2016rss}, typically exploiting ML, as we will detail later in Section~\ref{subsec: ML approach_loc}. However, \ac{rss} is known to be not adequate for localization as \ac{rss}-based localization methods may be vulnerable to beamforming attacks~\cite{wang2013analysis}. 



\subsection{Channel Statistics}

Beyond synchronization and estimation noise issues, in some contexts, the use of \ac{cir} and \ac{cfr} becomes problematic by fast channel variations. In this case, it is better to use as a channel feature the {\em statistics of the channel} rather than its instantaneous realization.  

Concerning the industrial \ac{iot} context,~\cite{Du2023Physical} uses the mean and variance of the subcarrier amplitude, carrier phase, and carrier frequency computed over the whole CIR. 

{\color{black}A wide range of features has been tested in \cite{Joo2024Enhancing} in the \ac{UWB} context, with results suggesting skew and kurtosis of the tap's magnitude, maximum peak-to-earlier peak ratio (MPEP), and the peak-to-average power ratio (PAPR) to be the most promising features.}

This approach is also popular in the \ac{uwac} context. In particular, in~\cite{Diamant2019Cooperative}, while looking for features that are stable over time but not over space, several channel features have been investigated. The study concluded that the best performance was achieved using the number of channel taps, the average tap power, relative \ac{rms} delay, and smoothed received power. Such features also proved their effectiveness in the following works in authentication~\cite{Bragagnolo2021Authentication, Ardizzon2022Machine}. In a dynamic \ac{uwac} context, the power-weighted arrival delay has been proposed to track the evolution of the channel; therefore, anomalous changes were associated with the start of an attack~\cite{Casari2022Physical}. The work was extended in~\cite{Ardizzon2024RNN}, where the new feature was integrated with those derived in~\cite{Diamant2019Cooperative}. 

\subsection{Time Measurements}
In many scenarios, power-related features are not usable, as they are either too predictable by the attacker or too variable to be used for \ac{cb} authentication, e.g., due to fading. An alternative is offered by time-based features, such as \ac{toa}, \ac{tdoa}, or even the estimated transmitter clock bias.  Still, it is worth pointing out that, while effective, the use of the \ac{tdoa} requires the presence of multiple synchronized receivers. {\color{black}The \ac{tdoa} is used in the satellite context in~\cite{Jedermann2021Orbit} and in industrial/\ac{UWB} communication in~\cite{Gao2019Lightweight}}, while the \ac{toa} has been used in the \ac{uwac} in~\cite{Aman2018Impersonation}, along with the \ac{aoa}. 

\subsection{Angle of Arrival}
For a receiver equipped with multiple antennas, the \ac{aoa} is another channel parameter to be used as a channel feature for \ac{pla}. Indeed, the \ac{aoa} is related to the transmitter position; thus, exploiting, for instance, a prior knowledge of the legitimate transmitter position, it is possible to discriminate between the legitimate transmitter and the spoofer just by looking at the \ac{aoa}.  

The \ac{aoa} has been used in \ac{uwac} in both~\cite{Aman2018Impersonation} and~\cite{Khalid2020Physical} and  in~\cite{Xiong2013SecureArray} to profile the client's WiFi network. For IoT authentication,  \ac{aoa} has been exploited in~\cite{Wang2018Efficient}.

In~\cite{gao2023esanet}, the authors exploit the massive-\ac{mimo} geometrical channel to extract an image of angle and delay and then adopt ML techniques to authenticate the transmitters. 

\paragraph*{Attacker Localization} It is also possible to use the \ac{aoa} of the received signal to localize the attackers~\cite{yilmaz2015survey}. SecureAngle~\cite{xiong2010secure} is a framework to estimate the signal's \ac{aoa} and create \ac{aoa}-based signatures to identify the legitimate users. If a user's signal does not belong to the authorized signatures pool, it gets rejected and localized by using the \acp{aoa} of the direct path of its signal, estimated by multiple access points. Pilot spoofing attacks are addressed in~\cite{ning2020detection} and~\cite{tao2023pilot}: in such attacks, a spoofer corrupts the initial channel estimation phase by sending the same pilot sequence as the legitimate users at the same time. In particular, ~\cite{ning2020detection} employs an uplink and downlink training phase to detect and localize an attacker using the spatial spectra on the received signals and exploiting the reciprocity of uplink and downlink channels in time-division-duplex (TDD) systems. Still, due to the duration of the training phases, the method is vulnerable to environment changes (e.g., a moving attacker)~\cite{tao2023pilot}. Thus,~\cite{tao2023pilot} proposes an uplink joint detection and localization of an attacker via sequential Bayesian inference (i.e., by considering the time correlation on the estimated quantities).

\subsection{Doppler Frequency}

The use of the Doppler frequency shift is particularly popular in the satellite communication context. Indeed, analogously to the power-based approaches, a receiver that knows the position and velocity of both itself and the satellites can compute the relative velocity and thus the Doppler shift. Such an estimate can then be compared to the measured Doppler for \ac{pla} for authentication purposes.
This approach has been used, for instance, in~\cite{Yi2021Initial}, where the receiver computes its position and velocity via \ac{gnss}, and in~\cite{Topal2022Physical}, where the authors tackle the problem of inter-satellite link authentication.

Often, the Doppler frequency shift is used in pair with the \ac{rss},e.g.,~\cite{Abdrabou2022Physical, Abdrabou2024Game}.

\paragraph*{Attacker Localization}  Doppler frequency is also used to localize the attacker. The scenario considered in~\cite{sanders2020localizing} sees a \ac{v2v} communication system attacked by a fixed or mobile terminal that is spoofing a \ac{gnss} signal. The vehicles use commercial \ac{gnss} receivers to measure the spoofer signal Doppler frequency. Next, the vehicles share their local measurements with the others, and by combining them, they localize the attacker. Note that, as all the vehicles are locked on the same spoofing signal, no additional synchronization among them is required.

{\color{black}
\subsection{CB-Authentication With RISs}\label{CBRIS}

If a verifier-controlled \ac{ris} is available in the network, specific solutions can be implemented. 

In \cite{He2024enhancing}, a \ac{glrt} technique is used, but the second-order statistics of both the legitimate and the attack channel are known. In this case, the configuration of the \ac{ris} is fixed, and both the direct channel and the channel through the \ac{ris} are estimated for the hypothesis testing procedure.  In \cite{Cicek2024impact}, the impact of residual hardware impairments on authentication mechanisms in the presence of a \ac{ris} is investigated.  In \cite{Bendaimi24how}, it is proposed to use the \acp{aoa} of the direct and cascaded links at Bob and the effective \acp{aod} at the \ac{ris}. The sparsity of the direct channel and also the unique double-structured sparsity of the beamspace cascaded channel are exploited as authentication features. 

In all these works, the configuration of the \ac{ris}, i.e., the setting of the phase of the elements, is fixed and typically optimized to maximize the communication performance. However, as already mentioned in Section~\ref{crplasec}, the possibility to control the propagation characteristics of the channel with a \ac{ris} allows a new mechanism for authentication exploiting the \ac{cr} approach. In fact, the use of \acp{ris} for this purpose was introduced in the first paper of the topic \cite{9982485}, and the security and communication performance were then studied in \cite{10700780}. Specific attacks and defense strategies (both for the control of the \acp{ris} by the verifier and for suitable beamformers to be used by the attacker to increase the chances of success) have been investigated in \cite{10615572} and \cite{10615677}.
}

\section{Identification Verification Methodologies}\label{authTech}
In this Section, we describe in detail the identification verification phase of \ac{cb}-\ac{pla} mechanisms, i.e., the part where the verifiers check that the currently received message is authentic. The description is organized into three parts related to statistical approaches, binary classification approaches, and one-class classification approaches. Lastly, we also provide a survey of techniques for localization based on \ac{ml} models.

\subsection{Statistical Approaches}\label{statapp}
With statistical approaches, we assume to have available \ac{pdf} of the channel feature in either or both the legitimate and under-attack conditions. The authentication process is then seen as a hypothesis testing problem, and the test function is obtained from the \acp{pdf}.

\subsubsection{Tag-based Authentication}
Concerning the tag-based authentication, referring to binary hypothesis testing, the \ac{lrt} is shown (by the Neyman-Pearson theorem) to minimize the missed detection probability for a fixed false alarm~\cite{Neyman1933Problem}. It provides the test function
\begin{equation}\label{eq:LRT}
   V:\quad f(\bm{x}) = \frac{p(\bm{x}|\mathcal{H}_0 )}{p(\bm{x}|\mathcal{H}_1)}\,,
\end{equation}
where $p(\bm{x}|\mathcal{H}_i)$ is the \ac{pdf} of the tag in case $\mathcal{H}_i$ computed in $\bm{x}$. Such a test has been used in several works, such as~\cite{Khalid2020Physical, Khalid2020Node, Topal2022Physical}. In particular, in  \cite{Khalid2020Physical}, the verifier Bob, upon receiving a new message, computes the Mahalanobis distance between the current observation and a database of previously collected \acp{aoa} containing both legitimate and non-legitimate samples.

Still, \eqref{eq:LRT} has a major drawback as it requires the verifier to know, or at least assume to know, both legitimate and under-attack tag statistics, which may be a strong assumption in many practical applications.

In a single-sided testing problem, where only the tag statistics in legitimate conditions are known, the \ac{lt} is typically employed, which provides the test function
\begin{equation}\label{eq:LT}
    f(\bm{x}) ={p(\bm{x}|\mathcal{H}_0 )}\,.
\end{equation}
In the specific case of a Gaussian-distributed vector, e.g., when the measurement is affected by \ac{awgn}, \eqref{eq:LT} becomes 
\begin{equation}\label{eq:GaussLT}
   f(\bm{x}) =  \|\bm{x} - \bm{x}'\| \,,
\end{equation}
where $\bm{x}'$ is the expected observation, which is used as a reference. Such an approach has been used, for instance, in~\cite{Jedermann2021Orbit}, where the \ac{rms} error between the measured and the expected \ac{tdoa} is thresholded. 
Still, it is worth noting that the \ac{lt} is typically sub-optimal with respect to the \ac{lrt}, but do not assume any knowledge of Trudy's attack statistical distribution. Such an approach has been used in~\cite{Aman2018Impersonation,Diamant2019Cooperative,Yi2021Initial, 8908721, Xiao2016PHY,5683463,7733021,5934990,Kalamandeen2010Ensemble, 4289438}.

Alternative tests to \eqref{eq:LT} have been considered, for instance, resorting to the Pearson correlation between the different observation sequences~\cite{Xiong2013SecureArray, Kalamandeen2010Ensemble}. The Pearson correlation factor between the scalar feature sequences $x_i$, $i=1, \ldots, n$, and $x'_i$, $i=1, \ldots, n$, is  \begin{equation}\label{correlation}
r = \frac{\sum_{i=1}^n \left(x_i -\mu(x) \right)\left(x'_i - \mu(x')\right) }{\sqrt{\sum_{i=1}^n \left(x_i - \mu(x) \right)^2} \sqrt{\sum_{i=1}^n\left(x'_i - \mu(x')\right)^2}},
\end{equation}
where $\mu(x) = \frac{1}{n}\sum_{j=1}^n x_j$ and 
$\mu(x') = \frac{1}{n}\sum_{j=1}^n x'_j$. For example, in~\cite{Kalamandeen2010Ensemble}, {\em witness} external devices check if two devices that should be paired are in proximity by controlling if the \ac{rss} are correlated. This approach is based on the assumption that close-by transmitters will yield correlated time series of \ac{rss} to a receiver. 

Such tests are often integrated with more complex protocols. A common scenario considers a verifier that coordinates several independent receivers or channels and has to aggregate the local decisions to perform authentication. The local decision is often performed considering either~\eqref{eq:LT} or~\eqref{eq:GaussLT}, and then the verifier has to design a function to perform the final decision. A major difference between these works is that while some share with the verifier the soft output, others share only a local decision, i.e., the binary output of the combination between \eqref{eq:LT} and~\eqref{eq:decisionFun}. 
In this context, in~\cite{Topal2022Physical}, $6$ channels are considered, and several methods have been investigated to aggregate the local decision, in particular, OR, AND, and majority rule.
On the other hand, a distributed test is considered in~\cite{Diamant2019Cooperative}, where soft information is provided by the devices, and the aggregation is performed by weighting the local observation, considering, for instance, the distance between each receiver.

A different approach is proposed in~\cite{Li2015Spoofing} where the authentication is framed as a game, where the legitimate party utility function is a mixture of \ac{fa} probability, \ac{md} probability, and spoofing cost; the legitimate party and the attacker have to choose the test threshold value and the spoofing probability respectively. A similar approach has been proposed in the satellite context in~\cite{Abdrabou2024Game}. 

\subsubsection{Channel Variation PLA}

The most popular statistical method for channel variation \ac{pla} involves the use of \acp{kf} predictors. 
In detail, considering a characteristic $\bm{z}_i$ to be tracked (e.g., the user distance or velocity), typically called \textit{state}, its time evolution is modeled as 
\begin{equation}\label{eq:stateTransEq}
    \bm{z}_i = \bm{A}_i \bm{z}_{i-1} + \bm{w}_i\;,
\end{equation}
where $\bm{A}_i$ is the state transition matrix at time-step $i$, and $\bm{w}_i \sim \mathcal{N}(\bm{0},\bm{Q}_i)$ represents the process noise (assumed to be Gaussian).
On the other hand, the \textit{measurement}, which for us is the channel feature, and the state are related 
\begin{equation}\label{eq:measurementEq}
    \bm{x}_i = \bm{B}_i \bm{z}_i + \bm{r}_i\;,
\end{equation}
where $\bm{B}_i$ is the observation matrix and $\bm{r}_i \sim \mathcal{N}(\bm{0},\Sigma_i)$ models the observation noise.

The Kalman filter has two operating modes: \textit{prediction} and \textit{model update}. 
During the former, it computes the a priori state estimate and its covariance matrix, respectively, as 
\begin{subequations}
\begin{flalign}
    \tilde{\bm{z}}_{i|i-1} &= \bm{A}_i \tilde{\bm{z}}_{i-1|i-1} \label{eq:predictionKF}\,,\\
    \bm{P}_{i|i-1} &=  \bm{A}_i \bm{P}_{i-1|i-1} \bm{A}_i ^{\rm T}\,.
\end{flalign}
When a new channel feature is provided $\hat{\bm{x}}_i$, Bob can refine its model as 
\begin{flalign}
    \bm{e}_i &= \hat{\bm{x}}_i -  \bm{B}_i \tilde{\bm z}_{i|i-1}\\
    \bm{C}_i &= \bm{B}_i  \bm{P}_{i|i-1} \bm{B}_i ^{\rm T} + \Sigma_i\\
    \bm{G}_i &= \bm{P}_{i|i-1} \bm{B}_i^{\rm T} \bm{C}_i^{-1} \\
    \hat{\bm z}_{i|i} &= \hat{\bm z}_{i|i-1} + \bm{G}_i \bm{e}_i\\
    \bm{P}_{i|i} &= (\bm{I} - \bm{G}_i \bm{B}_i) \bm{P}_{i|i-1} \;,
\end{flalign}
\end{subequations}
where $\hat{\bm{z}}_{i|i}$ and $\bm{P}_{i|i}$ are the a posteriori state estimate and its covariance, respectively, while $\bm{G}_i$ is often called \textit{Kalman gain}.
Finally, the prediction error $\bm{e}_i$ is called \textit{innovation} of the Kalman filter and can be used for security purposes. In particular, Bob computes
\begin{equation}\label{eq:innovation}
    \beta_n = \bm{e}_i^{\rm T} \bm{C}_i^{-1} \bm{e}_i \;,
\end{equation}
which Bob uses as input for authenticity verification. We remark that, differently from the general model of the Kalman filter, we have no control input. The interested reader may look for a more formal introduction of the \ac{kf} in~\cite{Kay:1993}.
Several variations can be introduced to the \ac{kf}. For instance, in the so-called extended \ac{kf} non-linear equation, replace either or both \eqref{eq:stateTransEq} and \eqref{eq:measurementEq}.

The main issue of statistical approaches, such as the \ac{kf}, is that they require an analytic model. For instance, \acp{kf} typically requires an a priori knowledge of the measurement evolution over time and the relation between the measurement and the true state (e.g., the transmitter position), e.g., $\bm{B}_i$ and $\bm{A}_i$ used in the (linear) \ac{kf} in \eqref{eq:stateTransEq} and \eqref{eq:measurementEq}, respectively. However, such models are often hard to obtain in practice, as they depend on both the features themselves and the environment, thus limiting the set of possible input features.

In the \ac{uwac} context, in~\cite{Casari2022Physical}, a set of receivers is tracking the power-weighted arrival delay using a \ac{kf}. The \ac{kf} so-called \emph{innovation}, which measures the discrepancy between the value predicted by the \ac{kf} and the observed measurement, is monitored, thus associating high innovation values with the start of a spoofing attack. Such an approach was generalized in~\cite{Ardizzon2024RNN}, replacing the \ac{kf} with a \ac{lstm} \ac{nn}.
A similar approach has also been considered in~\cite{Wang2020PhysicalKalman, Wang2020NonlinearKalman}, targeting the V2X scenario.



As an alternative to the KF approach, in the \ac{mimo} radio context,~\cite{Xie2021Physical} considered a scenario where the \ac{cir} evolves following a Gauss-Markov process. The tests evaluate (the norm of) the difference between the current and the previous \ac{cir} at different transmitter-receiver pairs, considering both the cases where each pair observes statistically independent channels, and where the observed channels are correlated. Still, we notice that the check is still related to \eqref{eq:innovation}, eventually considering the covariance to be a diagonal matrix.

Differently, the work in~\cite{Yin2021Online} considers an online adaptive method where the threshold is dynamically adjusted by the verifier, according to its previous statistics.
Another adaptive method has been proposed in~\cite{Tan2023Generalized}, where the tag symbol setup is dynamically adapted, following a water-filling approach where the power associated with each tag symbol is adjusted to match a predefined detection probability.



\subsection{ML Approaches - Binary Classification}\label{ML-two}

In \ac{ml}-based solutions, it is assumed to have a dataset of tag samples under one or both the legitimate and attack conditions. The authentication problem is framed as a classification problem, and the test function is replaced by an \ac{ml} trained with the available dataset(s).

We first consider the binary classification solutions, where two datasets (of legitimate and attack conditions) are available to train the \ac{ml} model. Note that the dataset of attack samples can also be artificial, i.e., generated by the verifier under suitable assumptions, when no real data are available, as discussed in Section~\ref{MLartif}. The availability of the Trudy dataset is related to some assumptions, i.e., expected position, type of device, or resulting channel feature.
To the best knowledge of the authors, no work has considered the use of two-class \ac{ml} techniques for channel variation \ac{pla} and \ac{cr}-\ac{pla}. Thus, all results are relative to the tag-based \ac{pla}. 

In~\cite{Brighente2019Machine} it has been proven that a sufficiently complex \ac{nn} trained with a sufficiently large dataset, containing both positive (Alice) and negative (Trudy) samples, when used in \eqref{eq:decisionFun} achieves the same performance as the optimal \ac{lrt}.

The work in~\cite{pan2019threshold} compares the performance of four standard classification algorithms, namely \ac{dt}, \ac{svm}, \ac{knn}, and \ac{el}, in particular \acp{bt}. The last achieved the best performance in both simulations and tests, but at a higher computational cost with respect to the other solutions.

In~\cite{Wang2019Deep}, a \ac{cnn} is used to extract the relevant features from the channel observation, and then a \ac{rnn} is expected to learn the spectral dependencies between the extracted features. In~\cite{gao2023esanet}, the authors propose to use the sparse nature of the channel in a massive \ac{mimo}-\ac{ofdm} communication system to first build an angle-delay image that is efficient for \ac{nn} training. Then, they exploit you-only-look-once (YOLO), an advanced single-stage object detection network, to capture the angle-delay features from the received signal, and finally, a lightweight neural network to perform the classification.  

We remark that \ac{iot} is a relevant use case for \ac{pla}, where devices have limited computing capabilities; therefore, \ac{ml} solutions can be particularly useful.
For example, in~\cite{Wang2018Efficient}, a two-step authentication mechanism for \ac{iot} devices in 5G networks is proposed. The first step aims to detect anomalies in the virtual \ac{aoa} and path gains of all the \ac{iot} devices in the cell; the second step provides an authentication mechanism based on \ac{ml}. In particular, the anomaly detected in the first step compares the number of communications at the MAC layer with those identified at the physical layer. If the anomaly is detected, a \ac{ml} is used to decide if a single communication (at the MAC layer) comprises more than one path at the physical layer to confirm the anomaly. The work in~\cite{Xie2022Weighted} exploits the presence of multiple devices at the edge to build a collaborative authenticator. In particular, the \acp{cfr} associated with a \ac{mimo} system are divided among different devices of an edge network, where a subclassifier makes a first classification, which is interpreted as a vote. Finally, the decision is taken by aggregating the single votes, each associated with a proper weight.  Different techniques are proposed  in~\cite{chen2023physical,liao2019security} to reduce the computation load, thus allowing power-constrained devices to perform \ac{pla}. In~\cite{chen2023physical}, a convolutional denoising \ac{ae} is used to preprocess the raw \ac{csi} measurements to reduce the dimension of the features, eliminate noise, and extract key features. After the pre-processing, a weighted \ac{knn} algorithm classifies the extracted features and authenticates the transmitters, which reduces the computational overhead compared to other ML approaches. \cite{liao2019security} aims at reducing the computation overhead in training the \ac{nn}, thus proposing three gradient descent algorithms to accelerate training.


\subsection{ML Approaches - Artificial Dataset}\label{MLartif}

When the dataset available for training contains only channel feature samples under legitimate conditions, two approaches are possible. One solution provides that first, an {\em artificial dataset} of attack channel features is generated, and then the binary classification approach is used for training on the available legitimate and artificial attack datasets. A second solution uses only the available dataset, which is denoted as a one-class classification approach. In this section, we consider the solution based on the artificial dataset, while in Section~\ref{ML-one}, we detail the one-class classification approach. 

\subsubsection{Tag-based Authentication}
The solution based on the artificial dataset is employed in \cite{Du2023Physical, Chen2021Automated}.

A \ac{dt} is exploited to perform authentication in~\cite{Du2023Physical}, in the absence of an attack dataset. In particular, they train the classifier using positive unlabeled data, where only positive (i.e., legitimate) data is used, but part of the data is labeled as non-legitimate and left out during this first training. Then, the procedure is repeated, changing positive and unlabeled data until a robust enough classifier has been trained.

An automated labeling strategy is proposed in~\cite{Chen2021Automated}, which comprises an offline and online procedure. They specifically look for clone or Sybil attack data samples, which are later used to train the more refined \ac{svm}-based (online) procedure.


\subsubsection{Channel Variation PLA}
The solution based on the artificial dataset is employed in~\cite{Wang2017Physical, Zhao2023Physical, wang2024transferable, Romero2024Spoofing}.

An extreme learning approach is proposed in~\cite{Wang2017Physical}, where the input contains both previous trusted observations and the observation to be verified; then, the network is trained to check the coherence between the two. An artificial dataset is generated to improve the network classification performance. In particular, the artificial dataset has the same distribution as the legitimate one but contains samples that are uncorrelated with the (previous) legitimate ones.

An artificial dataset is considered in~\cite{Zhao2023Physical}, and it contains samples from nodes that are far from the receiver as negative. Then \ac{svm} is used to build a classifier. The features, e.g., the maximal time-reverse resonating strength, are then specifically chosen to minimize the impact of the channel time variability.

Using channel measurements relative to a specific location of devices as an authentication feature makes the authentication very scenario-dependent; thus, channel time-varying patterns as scenario-independent features are used in ~\cite{wang2024transferable} to authenticate devices even in uncalibrated scenarios, including both \ac{cir} and \ac{cfr} as input to the \ac{ml} model.

The use of a \ac{gnn} was considered in~\cite{Romero2024Spoofing}, where the verifier has to decide whether the \acp{rss} measured from several receivers within a frame comes only from Alice or from multiple transmitters, that is, Alice and Trudy. In particular, a two-step approach is proposed. First, a \ac{nn} checks whether two consecutive transmissions come from the same position. The latter step has to split the received frame sequence into regions associated with the same transmitter. In particular, in the first step, the training dataset collects pairs of consecutive transmissions where i) in the legitimate case, the transmission comes from transmitters in the same position, while ii) in the under-attack case, the pair contains samples coming from different randomly sampled positions. 

\subsection{ML Approaches - One-class Classification}\label{ML-one}

We now consider one-class classification solutions, where only a legitimate dataset is used to train the classifier.

\subsubsection{Tag-based Authentication}  The first set of solutions collects anomaly detection models, such as \acp{ae} or \ac{oc-svm}. For example, a \ac{oc-svm} is considered in~\cite{Abdrabou2022Adaptive}, with both magnitude and IQ samples of \ac{cir} measured at different antennas as input. {\color{black}\Ac{oc-svm} has been also considered \ac{UWB} \cite{Joo2024Enhancing}.} In~\cite{wang2022csi}, a \ac{nn} determines the position of a device (from a set of discrete positions) from the observed \ac{csi}. If the confidence metric associated with the predicted position is below a threshold, the device is considered not authentic. Indeed, this authentication mechanism boils down to a region location verification, i.e., it verifies that the device is in a set of pre-approved positions. 
The combination of an \ac{ae} for dimensionality reduction and \ac{vae} for its generative capabilities has been considered in~\cite{Meng2023Physical}.





\paragraph*{Cooperative One-Class Classification}
When multiple detectors are available, cooperative \ac{cb}-\ac{pla} can be applied.  In fact, multiple verifiers can cooperate in the final decision by merging the collected data (or local decisions) to form a distributed authenticator and improve the security performance.

Two-step approaches are considered in~\cite{Liu2014Practical,song2021enhancing,Meng2024Multidimensional}, where first clustering-based methods are used to detect the presence of outliers within the dataset. Then the authentication is completed by using \ac{svm}\cite{Liu2014Practical}, an ensemble of \acp{ae}~\cite{song2021enhancing}, or graph learning~\cite{Meng2024Multidimensional} approaches.

In~\cite{Bragagnolo2021Authentication}, the authors compared binary \ac{nn} and one-class \ac{ae} for \ac{uwac} networks. The same techniques have also been considered in~\cite{Ardizzon2022Machine}, where the authentication process is performed in two steps: a first pre-elaboration on each device, with a \ac{nn}, and then a second central elaboration made on the pre-elaborated data, using a second \ac{nn}.


\subsubsection{Channel Variation PLA}
A kernel-based prediction method is proposed in~\cite{Fang2019Learning}, and \ac{gpr} is considered in~\cite{Wang2022Channel}.
The use of a Bahdanau attention \ac{ae} predictor is considered to predict the current \ac{cfr} in~\cite{Han2024Model}. Compared to a more traditional \ac{ae}-based predictor, this architecture includes an attention mechanism to take into account the effects of fading and Doppler shift. 

A second solution aims at extending the solution developed for the static to the dynamic context, thus assuming the distribution to change slowly over time. For example, in~\cite{Qiu2020Learning}, the concatenation of the legitimate features (tested) before and the measurement under test is fed to \ac{cnn}, whose output represents the probability that the whole input matrix belongs to Alice. \Ac{oc-svm} is used for classification in the satellite context in~\cite{Abdrabou2022Physical} but is progressively updated to take into account the evolution of the statistic over time.

A third alternative is to optimize the test threshold. 
The authentication problem has been formulated as a zero-sum game, where \ac{rl}, in particular, Q-learning and Dyna-Q, have been used to optimally set the threshold for \ac{lt} in~\cite{Xiao2016PHY}.
In~\cite{Lu2020Reinforcement}, a \ac{drl} mechanism is used for the \ac{vanet} context where Bob sets the authentication threshold and Trudy plays with the attack rate.


The tag-based solution in~\cite{Liu2014Practical} has been extended in ~\cite{Liu2018Authenticating} to also take into account mobile users by adding a processing block that monitors the temporal correlation between subsequent \ac{csi} blocks. 

In the context of \ac{uwac}, Casari~\etal investigate the use of \ac{oc-svm} and an \ac{ae} to fuse the local \ac{kf}'s innovations~\cite{Casari2022Physical}, later extended in~\cite{Ardizzon2024RNN}, where a \ac{rnn} allowed to track features that cannot have been analytically modelled and thus tracked by \ac{kf}.
In~\cite{zhang2023cooperative}, a framework considers federated learning among cooperating edge devices: a group of edge devices is selected using a Q-learning-based adaptive search procedure and collaborates to form an authenticator. Unlike previous works, two threats are examined: the presence of external attackers (i.e., regular spoofers) and internal threats, the latter represented by malicious nodes injecting false parameters that might lead to convergence failure or convergence to a wrong model.
A similar approach is also proposed in~\cite{ventura2024authentication}, where first, a \ac{cnn} estimates the transmitter positions from the \ac{scm}, and next, a \ac{rnn} predictor tracks the transmitter position.

\subsection{ML Approaches - Attacker Localization} \label{subsec: ML approach_loc} 

Various approaches for the location of the attacker have been considered using \ac{ml} models.
In~\cite{chen2007detecting}, the authors use the K-means clustering algorithm to detect and locate attackers. In particular, after a training phase, they combine \ac{rss} readings from multiple access points and divide them into clusters in the signal space. If there are multiple transmitters at the same time (i.e., a legitimate party and one or multiple spoofers), then the resulting point in the signal space will be far from the centroids of the legitimate clusters, and thus an anomaly is detected. After the spoofer is detected, the cluster centroids are used to localize it. The authors experimentally tested their approach on WiFi and Zigbee networks, reaching $P_{\rm md}>0.95$ with $P_{\rm fa}<0.05$. 

A drawback of~\cite{chen2007detecting} is that it cannot localize multiple attackers. Thus, in~\cite{yang2012detection}, the authors propose the IDOL (Integrated Detection and Localization) framework, capable of detecting and localizing multiple attackers. In~\cite{yang2012detection}, three types of algorithms were tested to locate the attackers: nearest neighbor matching in signal space, probability-based, and multilateration; while in~\cite{gajbhiye2016rss}, the authors used a discriminant-adaptive neural network to perform the same task.


\section{CB-Authentication Public Datasets and Applications}\label{sec:cb_results}

This Section describes the common methodologies to test the performance of the proposed CB-\ac{pla} mechanisms, considering both simulation tools and experimental datasets.

{\color{black}
\subsection{Simulation and Experimental Methodologies}
The methodology for CB-\ac{pla} mirrors the one used for RFFI, described in Section~\ref{sec:exp_methodology}.
It typically involves the collection of two datasets. The first is used to set up the detector. For instance, in \ac{kf}-based channel variation authentication, the first dataset is used to adapt the filter parameters during the initial transient (e.g., \cite{Casari2022Physical,ventura2024authentication}). In \ac{ml}-based solutions, such a dataset is instead used to train the detector. 
The second dataset is used for testing, to evaluate the performance of the trained detector. 
We remark that, while formally two datasets need to be collected, often only one is collected, which is then split into two. These operations need to be performed carefully. 
For instance, training on samples that are collected close to each other in time helps the detector to learn the channel stationarity, and thus, we neglect the evolution of the channel over time. On the other hand, as it happens when overfitting, this also makes the detector less robust to variations. Thus, to make the detector more robust to temporal variations, it is advisable to split the dataset randomly. 

Finally, it should be noted that when testing CB-authentication solutions, the hardware impairments are typically neglected, implicitly treating them as estimation noise. Indeed, even if costly, a better practice would involve the collection of multiple datasets, each collected with a different transmitter/receiver hardware pair, which would make the detector, trained on the merged dataset, truly device-independent. Alternatively, future works should include an estimation and correction step to correct the hardware impairments or, even better, a joint RFFI \& CB-based authentication, which allows the detector to exploit both techniques at the same time, as detailed in Section~\ref{sec:challenge}.
}

\subsection{Simulation Tools and Setups}

\paragraph*{WiFi}
A broad set of simulation tools is used for the WiFi context. A simple model provides independent Gaussian distributed channel taps~\cite{Xie2021Physical}, while other solutions, such as~\cite{Fang2019Learning}, consider generating more realistic \acp{cfr}, including an exponential \ac{pdp}. Other parameters have been set according to the IEEE 802.11a specification. Another alternative is to consider geometric models, e.g., ray tracing tools, as in~\cite{4289438}. 

\paragraph*{Vehicular}
Specifically targeting the V2X context, simulations have been performed in~\cite{Wang2020PhysicalKalman,Wang2020NonlinearKalman}, with communication parameters adapted from the SAE J2945/1 standard using Matlab. In ~\cite{Wang2020PhysicalKalman}, the authors simulated two traffic scenarios: straight and intersection. In the first, Alice and Bob are driving straight on the same road, while in the second, Eve follows Alice who is driving crosswise with respect to Bob. On the other hand, ~\cite{Wang2020NonlinearKalman} considers instead a more abstract model, where the \ac{rss} is modeled after a log-normal distribution.   





\paragraph*{Cellular Wireless}
While some papers generate the features via statistical models (e.g.,~\cite{Wang2018Efficient}), three simulators are popular in the literature:
\begin{itemize}
\item \Ac{quadriga} channel simulator~\cite{Jaeckel2014QuaDRiGa}: it has been used in~\cite{Wang2022Channel,wang2022csi}. In particular, in~\cite{Wang2022Channel} the simulation includes also the movement between transmitter and receiver, with parameters set to simulate the ground city macrocell in the Berlin survey in Germany (BERLIN UMa).
\item WINNER II channel model~\cite{Kyosti2008WINNER}: it is used for instance in ~\cite{Abdrabou2022Adaptive} to model a \acrlong{nlos} scenario, with users moving at different velocities.
\item MATLAB 5G toolbox channel: this was used to test the \acrlong{baae} proposed in~\cite{Han2024Model} operating in the 5G FR1 n78 band.
\end{itemize}


\paragraph*{Underwater Acoustic}  
Concerning \ac{pla} in the \ac{uwac} context, the most popular solution involves the Bellhop ray-tracing simulator~\cite{bellhop,Morozs2020Channel}, used for instance in  ~\cite{Diamant2019Cooperative, Bragagnolo2021Authentication,Casari2022Physical,Ardizzon2024RNN,ventura2024authentication}. 
Such a tool also includes the description of several environmental parameters, such as sound speed profile and bathymetry. For instance, among others, the San Diego Bay area was considered in~\cite{Diamant2019Cooperative, Casari2022Physical, Ardizzon2024RNN, ventura2024authentication}.

An alternative simulator used in~\cite{Zhao2023Physical} has been described in~\cite{Qarabaqi2013Statistical}.
Finally, in~\cite{Aman2018Impersonation}, the performance is evaluated considering both an \ac{awgn} channel and a coloured noise channel with and without frequency-dependent path loss, respectively.




\paragraph*{Satellite} 
Different channel models are considered in the satellite communication, including both \ac{awgn}~\cite{Abdrabou2022Physical} and Rician fading channel~\cite{Yi2021Initial}.
Specifically concerning satellite orbit datasets, \ac{tle} datasets have been used in both~\cite{Jedermann2021Orbit, Abdrabou2024Game}. In the former, the dataset was derived from~\cite{NORAD19CelesTrak}, while in the latter via the Ansys STK \cite{ansysSTK}.


{\color{black}
\paragraph*{UWB}
Concerning \ac{UWB}, a MATLAB simulation is performed in \cite{Gao2019Lightweight}, modeling the \ac{UWB} signals as the first derivative Gaussian pulses, and the channel is modeled as \ac{awgn}.
To test the performance, \cite{Joo2024Enhancing} considers both simulation and experimental tests. The simulations have been performed in MATLAB with receivers implemented following the IEEE 802.15.4z standard and channels compliant with the IEEE 802.15.4a standard. In particular, both \ac{los} and \ac{nlos} have been included. 
}

\subsection{Experimental Setups and Public Datasets}
Table \ref{tab:ch_dataset} collects a list of datasets available online that may be used to develop/test new channel-based authentication techniques, classified by wireless technology and measured channel features. In the remaining part of the section, we discuss the use of simulation data, experiments, and datasets in wireless technology.
\begin{table*}[!t]
\caption{Summary of Publicly Available Datasets for CB Authentication}
\label{tab:ch_dataset}
\centering
\begin{tabular}{|C{2cm}|l| C{2.5cm}|c|L{6cm}|}
\hline
\textbf{Wireless Technology} & \multicolumn{1}{c|}{\textbf{Dataset}} & \multicolumn{1}{c|}{\textbf{Paper}} & \multicolumn{1}{c|}{\textbf{Features}} & \multicolumn{1}{c|}{\textbf{Brief Summary}} 
\\ \hline \hline
WiFi, IoT, Industrial IoT & \cite{candell2016radio} & \cite{Chen2021Automated, Du2023Physical, Meng2023Physical, pan2019threshold,chen2023physical,wang2024transferable}  & CSI &  Data collected in industrial environments and open-area sites by NIST \\ \hline
 & \cite{wangning2025sls} & \cite{9321212,10.1145/3522783.3529522} & SNR &  SLS SNR traces collected from the communication between AP and clients at mmWave frequencies \\ \cline{2-5}
WiFi & \cite{Lohan2017Wi-Fi} & \cite{Romero2024Spoofing} & \ac{rss} &  Fingerprints collected in an indoor scenario from 21 different Android devices \\\cline{2-5}
 &  \cite{AlQahtani2023RSSI}& -- &  \ac{rss} & Measured from two Raspberry PIs, placed at various distances from one another \\ \hline
\multirow{2}{*}{UWAC} & \cite{Walree2016Watermark} &  --  & CIR & Experiment performed in Kauai (Hawaii) in 2011. \\ \cline{2-5}
 & \cite{Jianchun2019Channel} & \cite{Zhao2023Physical} & CIR & Long-range experiment performed in the Mediterranean Sea in 2019.  \\ \hline
\end{tabular} 

\end{table*}

\paragraph*{WiFi} 

Several works have provided experimental results on WiFi networks. First, a public dataset containing \ac{rss} and channel measured from WiFi access points, placed at increasing distance from a reference transmitter \cite{AlQahtani2023RSSI}.

 In~\cite{li2020sdr}, a \ac{sdr} platform for the WiFi PHY layer has been implemented and \ac{cb} \ac{pla} is performed on the \ac{csi}, \ac{rss}, and frequency offset. In~\cite{5683463}, the \ac{csi} was considered in a typical indoor scenario with fixed locations of the users.


A dedicated prototype has been developed to test the performance of the \ac{aoa}-based \ac{pla} solution described in~\cite{Xiong2013SecureArray}. The developed access point has two FPGA platforms, with four radio front ends and four antennas each. The clients are two Soekris boxes, equipped with Atheros IEEE 802.11g radios.


For the proximity-based \ac{pla} solution of~\cite{Kalamandeen2010Ensemble}, experiments were performed with ten Nokia N800 Internet Tablets, showing that the proposed solution can reliably detect attackers as close as two meters away from legitimate devices.

Experimental results on the clustering-based approach of \ac{cfr} for authentication have been reported in~\cite{Liu2018Authenticating, Liu2014Practical} where an IEEE 802.11n WiFi network was considered, with two laptops (Lenovo T500 and T61) serving as monitors that collect the wireless packets.  
A commercial wireless Linksys E2500 access point is the device to be authenticated, transmitting 10 packets/second. For each packet, the \ac{cfr} relative to 30 subcarriers is extracted with equal spacing among the 56 subcarriers of a 20 MHz channel. 

In~\cite{song2021enhancing} experimental results are reported with a commercial WiFi device, Huawei TAS-AN00 operating as a station, transmitting at a rate of 100 packets/second in \SI{20}{\mega\hertz} WiFi the channel on \SI{2.4}{\giga\hertz}. 

The experimental results reported in~\cite{7733021} have been performed on the Microsoft Sora SDR, reaching a false positive and false negative ratio of $10^{-3}$.


In~\cite{9321212}, a WiFi operating at mmWave (60~GHz band) is considered, with reference to the IEEE 801.11ad standard. The considered feature is the \ac{snr} trace obtained at the receiver in the sector level sweep (SLS) process, and an ML approach is used to authenticate the message. Talon AD7200 routers and MG360 WiGig USB Adapters are used to perform experiments in a meeting room, achieving a sum of \ac{md} and \ac{fa} probability less than 1\%. In~\cite{10.1145/3522783.3529522} experimental results for authentication are presented, based on the dataset of~\cite{9321212}. 

In~\cite{5934990}, experiments are conducted for the authentication based on the verification of \ac{snr} series observed at Alice and Bob, through a statistical method.  Alice, Bob, and Trudy are implemented on Dell E5400 laptops, which usethe  Intel iwl5300 chipset, operating IEEE 802.11g with channel one in the \SI{2.4}{\giga\hertz} frequency, with a transmission rate of 12~Mbps and transmission power of 15~dBm. 



Many works perform dedicated experiments~\cite{Xiao2016PHY,Wang2019Deep,Tan2023Generalized,Qiu2020Learning}, deploying three or more \acp{usrp} in an indoor environment mimicking an office or industrial context, with parameters following the standards, e.g., IEEE 802.11a/g and IEEE 802.11n/ac.
  
The \ac{gnn}-based solution proposed for the channel variation in~\cite{Romero2024Spoofing} in the artificial dataset training framework, exploits the dataset from~\cite{Lohan2017Wi-Fi}, a publicly available WiFi fingerprint dataset which collects fingerprints collected with 21 devices in an indoor scenario.






\paragraph*{IoT \& Industrial IoT} 
As in the WiFi context, many works only perform dedicated experiments, using again \acp{usrp} deployed in an industrial-like environment, e.g.,~\cite{Yin2021Online, Xie2022Weighted}, eventually also in a cooperative setting, such as \cite{zhang2023cooperative}.

In \cite{liao2019security}, the \ac{iot} nodes are emulated using \acp{usrp}, placed in various positions in an indoor environment.

A popular dataset for testing in the industrial \ac{iot} context is~\cite{candell2016radio}, described in~\cite{Candell2017Industrial,quimby2017nist}. For instance, both the \textit{open area test site} and the \textit{automotive assembly plant} scenario datasets were used in~\cite{Meng2024Multidimensional}. The dataset was also used for training and testing in~\cite{Chen2021Automated, Du2023Physical,Meng2023Physical,pan2019threshold,chen2023physical,wang2024transferable}.
Many works use both the NIST dataset and a dataset from dedicated experiments. 
A first example is \cite{chen2023physical}, where the experimental dataset was collected using three Lenovo X220 laptops (Alice, Bob, and Trudy) placed approximately 2–\SI{4}{\meter} apart transmitting at \SI{2.4}{\giga\hertz}, using IEEE 802.11n protocol, with 3 transmitting antennas, 2 receiving antennas, and 30 subcarriers using Linux \ac{csi} Tool. In~\cite{wang2024transferable}, the performance of the channel variation-based \ac{ml} approach was tested using~\cite{quimby2017nist} for the static scenario, then dedicated experiments were performed using two \acp{usrp} to account for the dynamic one. 

\paragraph*{Underwater Acoustic}
Due to the lack of a standard channel model and the impact of the environmental condition on the measurement, experiments, and proof of concept, often called \textit{sea trials} are common in \ac{uwac} studies~\cite{Diamant2019Cooperative, Ardizzon2022Machine, Bragagnolo2021Authentication}, eventually re-using datasets from previous experiments, such as in~\cite{Bragagnolo2021Authentication}, where Bragagnolo~\etal used the Hadera (Israel) dataset from~\cite{Diamant2019Cooperative}. 
However, only a few datasets are actually publicly available. For instance, a popular dataset is the KAM11, which was only published with the Watermark simulator~\cite{Walree2016Watermark}. An example of a public dataset is instead the LR19~\cite{Jianchun2019Channel} used, for instance, in~\cite{Zhao2023Physical}.

Another alternative is to perform tests in (typically indoor) pools: in~\cite{Xiao2019Learning} an experiment is run by collecting measurements from a  $25\times6 \times \SI{1.6}{\meter^3}$ non-anechoic pool where $9$ transmitters and $1$ receiver were deployed.




\paragraph*{Satellite}
Since not many works consider \ac{pla} for satellite, very few experiments have been reported yet. An exception is~\cite{abdrabou2022authentication}, where the Abdrabou~\etal collected real \ac{leo} satellite data using the system toolkit.

{\color{black}
\paragraph*{UWB}
Dedicated experiments have been performed to test the performance of the solution proposed in \cite{Joo2024Enhancing}. In particular, two nRF52840-DK boards have been used to implement the legitimate users, while a NUCLEO-Z429 is used for the attacker, all equipped with Qorvo DWM3000 modules. 1000 \acp{cir} have been collected in both static and dynamic scenarios. 
}

\section{Challenges and Future Research} \label{sec:challenge}

\subsection{Challenges for RFFI}

Despite significant development in deep learning-driven RFFI technology, numerous challenges remain unresolved. This section elaborates on these challenges and presents a summary based on the most recent studies.

\subsubsection{Lack of Capacity Evaluation}

The term `capacity' is used to describe the maximum number of wireless devices that can be accurately distinguished by analysis of their RFFs, which is critical for an authentication technique. Most RFFI studies use commercially available wireless transmitters, but most involve fewer than tens or dozens of devices, and few large-scale experiments have been conducted. To the best of the authors' knowledge, the work in~\cite{al2020exposing} presents the experiment with the largest number of wireless devices, up to 10,000. However, this large dataset has not been published, and researchers in this community cannot use it to explore the maximum capacity of the RFFI technique. {\color{black}A few studies have attempted to provide a theoretical analysis of the user capacity in RFFI systems \cite{wang2016user,wang2017user}. Nevertheless, achieving an accurate prediction of the user capacity remains a significant challenge, particularly for deep learning-driven RFFI systems.} There is still a need for large-scale experimental evaluation and theoretical analysis to assess the capacity of RFFI.

\subsubsection{Lack of Stability Evaluation}

As an identifier used for authentication, the stability of RFFs is critical. However, there are rare studies that systematically investigated the stability of RFFI systems. In particular, the characteristics of RF components can change slightly due to variations in the surrounding environment, such as temperature and humidity, and hardware ageing. The authors in~\cite{robyns2017physical, shen2021jsac} indicate that the oscillator frequency is sensitive to temperature variations and that \ac{cfo} compensation at the receiver side can improve system stability. However, these studies only focus on the stability of the frequency offset function resulting from the oscillator impairments, while the other hardware characteristics are not investigated. {\color{black}Moreover, the authors in~\cite{saeif2023day} experimentally demonstrate that the RF fingerprints of SDR transmitters exhibit significant variations when transitioning between on and off states. However, the evaluation of wireless transmitters beyond SDRs remains an open area for exploration.} The comprehensive evaluation of the RFF stability and the design of robust feature extraction algorithms represent crucial directions for future research.

\subsubsection{Lack of Benchmark Datasets}

The RFFI research community does not have a benchmark dataset that is as widely used as ImageNet in computer vision. This limits comparisons among studies. In addition, researchers without RF hardware and experience in designing wireless signal acquisition systems cannot efficiently engage in RFFI research. As discussed in Section~\ref{sec:rffi_applications}, some studies have released public datasets detailing the collection environments and hardware setup~\cite{elmaghbub2024no, shen2023methodology, hanna2022wisig}. However, most of these datasets still do not meet benchmark requirements in terms of dataset size, device population, and diversity in channel conditions. The collection and publication of large-scale benchmark datasets of various wireless protocols remains an urgent need in the field of RFFI.

{\color{black}
\subsubsection{Limited Studies on Adversarial Machine Learning Attacks and Defense}
Cutting-edge RFFI schemes heavily rely on deep learning. However, recent research in the ML community has revealed that deep learning is vulnerable to \ac{AML} attacks, including in the context of wireless systems~\cite{adesina2022adversarial, liu2023exploring}. 
Depending on the attack phase, \ac{AML} can be categorized into backdoor attacks launched in the training stage~\cite{zhao2024explanation,zhao2024backdoor} and adversarial/evasion attacks launched in the inference stage~\cite{bao2021threat,liu2023robust,ma2023white,papangelo2024adversarial,li2024slpa}.

The \ac{AML} attacks can be launched during the model training stage, named backdoor attacks~\cite{zhao2024explanation,zhao2024backdoor}. Zhao~\etal propose the first backdoor attack on RFFI systems, and evaluate the algorithm on three WiFi datasets and a \ac{lora} dataset. The results demonstrate that the attack can be successfully launched in either the time domain or the time-frequency domain~\cite{zhao2024explanation}. The authors in~\cite{zhao2024backdoor} further investigate the backdoor attacks against low-earth orbit satellite fingerprinting systems. 

The AML attack can be launched during the model inference stage, named adversarial/evasion attacks~\cite{bao2021threat,liu2023robust,ma2023white,papangelo2024adversarial,li2024slpa}. For instance, Ma~\etal demonstrate that adding perturbations to the deep learning input can interfere with the identification result and can even mislead into a specific identity~\cite{ma2023white}. 

More research is required to study the \ac{AML} as well as the countermeasure. For example, the transmission of perturbation in evasion attacks will experience channel propagation, but the effect is not properly studied yet.
Adversarial training and randomized smoothing are used as countermeasures for Wi-Fi sensing~\cite{yin2025evasion}, but there is no such study for RFFI.

}

{\color{black}
\subsubsection{Limited Studies on RFF Concealment}
While the majority of the research focuses on using \ac{rff} for legitimate purposes, i.e, device authentication, it can also be used maliciously, e.g., device tracking in~\cite{givehchian2022evaluating}. Hence, it is essential to design RFF concealment approaches.

Abanto-Leon~\etal added a randomized phase to each subcarrier in a WiFi OFDM system to ensure privacy, when non-linear phase errors are used for RFFI~\cite{abanto2020stay}. They proved that when the phase is generated via a random number generator, the approach is robust against statistical attack. 
Givehchian~\etal obfuscated the CFO of \ac{BLE} devices for preventing tracking attacks~\cite{givehchian2024practical}. They implemented their CFO obfuscation method using TI CC2640 chipsets and carried out a comprehensive experimental evaluation, which demonstrated the feasibility. 

These approaches only focus on the phase errors and CFO as hardware fingerprints, and their identification is based on comparing their similarities.
However, deep learning RFFI is learning all the available hardware impairments. It is not clear how RFFI will be affected if only one hardware feature is obfuscated as other impairments remain the same.
}

\subsection{Challenges for Channel-Based Authentication}


\subsubsection{Lack of Scalability} Attacks against \ac{cb} authentication can be deployed by transmitting from different positions until the features estimated by Bob are similar to those of Alice's transmissions. An alternative is that the attacker precodes the signal before transmission to introduce the features suitable for authentication \cite{baracca2012physical}. At the moment, the search space for an attack is limited, making the attack easy. Indeed, more efforts should be focused on the factors that make the attack harder in a scalable way. Such efforts include the investigation of a) scaling laws for the attack success probability with respect to design parameters such as the number of antennas or the length of pilot signals, b) new approaches such as the \ac{cr}-\ac{pla} that introduce further randomness in the authentication process, c) new bounds on attacks based on physical constraints obtained from specific technologies (i.e., type of antennas used by Bob): a recent example is given by \cite{10437915} that proved that an effective attack on AoA-based PLA can succeed only under very stringent conditions on the attacker location and hardware capabilities.        
\subsubsection{Lack of Integration} Since the first studies, \ac{pla} has been proposed as a security technique to be integrated with other approaches for authentication. However, such an integration has not been thoroughly investigated. A full protocol for \ac{pla} that integrates cryptographic approaches, for example, to secure the identification association phase or to be deployed when \ac{pla} is under attack, is yet to be investigated. Moreover, integration of \ac{cb}-\ac{pla} with authentication based on wiretap coding is still in its early stages and deserves further investigation. Lastly, integration may also include the use of diversified features for authentication, also coming from different layers of the communication stack: this is also an area that deserves more studies. In this case, \ac{ml} techniques could be particularly beneficial to capture the relation among the features, but such solutions should be, at the same time, effective in the specific scenario of deployment and robust against adversarial attacks, which leaves many open research points.

\subsubsection{Lack of Benchmark Datasets} Also for \ac{cb}-\ac{pla} as RFFI, there are not yet well-established datasets to be used for benchmarking different approaches. The difficulty of obtaining such datasets is related to the specific technologies that can be deployed (type of antennas, operating frequencies), the different kinds of environments in which the testbed operates (indoor, outdoor, with different transmit-receive distances), and the need to obtain measurements from several positions at the same time to assess also the knowledge of the attacker and the statistical relation of his channel to the legitimate channels. In this sense, apart from more extensive data collection and the use of existing simulators that provide spatially consistent channel realization, the adoption of \ac{pla} techniques in the standard would encourage a discussion from the community on reference scenarios to be used for benchmarking, thus giving a boost to \ac{pla} adoption.

{\color{black}
\subsection{Future Directions}
\subsubsection{Generative AI Approaches}
Generative AI represents transformative AI technologies to create new content, such as \ac{gan} and \ac{llm}. Generative AI has been widely used in securing communication from the physical layer~\cite{zhao2024generative}, but its application in device fingerprinting is relatively limited.

Generative strategies/architectures include \acp{ae}, \ac{vae}, diffusion model, etc. 
They are used to design detectors in the anomaly detection context. They can also be used to generate the training dataset, e.g, \ac{vae} is used to generate satellite data~\cite{wang2024ai}. This may allow a binary classification-based detector to have an initial offline training with artificial but realistic data, later refined online. 
In the context of anomaly detection, generative models may be used to generate an artificial dataset (see Section \ref{MLartif}). Regarding diffusion models, it is used for denoising in RFFI~\cite{yin2025noise}.

Recently, \ac{llm} have proven their effectiveness in multiple fields, even in the communication context \cite{Zhou2024Large}. Still, no solution that exploits \ac{llm} has been proposed in the device fingerprinting context. Due to their generalization capabilities and if trained in a multimodal manner, thus taking as input also information concerning, for instance, the environment, \ac{llm} may be used to generate high-fidelity artificial datasets, thus leading to even more robust detectors.

On the other hand, generative models may be used by the attacker to design effective attacks, as done in the RFFI context in \cite{Merchant2019Securing}. In particular, an attacker provided with the legitimate detector (or dataset used for training it) may exploit a generative architecture to generate the attack samples that are most likely to fool the verifiers. Thus, future research directions should also include these attacks into account.

\subsubsection{Emerging Communication Technologies}
While the use of device fingerprinting for securing communication technologies such as WiFi is consolidated, for newer communication technologies,  especially in the optical domain, only a few or even no work at all considers device fingerprinting for securing communication. This is the case, for instance, in \ac{vlc}~\cite{chen2024optic} or even underwater optical communications, where, to the best of the authors' knowledge, very little research has been done.   
Thus, a research direction may involve the translation of the more consolidated solutions and algorithms into these new technologies.

\subsubsection{Interplay between RFFI and Channel-based Authentication}
RFFI and channel-based authentication represent two distinct but complementary approaches to wireless device authentication. RFFI relies on the unique hardware impairments inherent to individual devices, which are introduced during the component manufacturing process. RFFI system is implemented at the receiver side, which is well-suited for scenarios where low-cost, infrastructure-independent security solutions are required. In contrast, channel-based authentication exploits the unique properties of the wireless channel, which are influenced by the surroundings; thus, it is effective in rich scattering environments, where it is hard for the attacker to predict, replicate, and compensate the attack signal to effectively mimic the legitimate channel features. 

The combination of RFFI and channel-based authentication offers a promising solution to enhance wireless security. Hybrid authentication protocols can be designed: RFFI ensures device-level identification based on unique hardware characteristics, while channel-based authentication validates location or monitors channel characteristics within a communication session. In this case, attackers would need to simultaneously replicate both the hardware impairments and the exact channel conditions to bypass the dual-layer protection, significantly increasing the difficulty of attacks. 
}

\section{Conclusions}\label{sec:conclusion}
This article presented a comprehensive survey on physical layer-based device fingerprinting, focusing on hardware impairment-based identity authentication and channel features-based location authentication.
In particular, RFFI exploits unique hardware impairments as devices are identified. Three RFFI tasks, closed-set RFFI classification, open-set RFFI recognition, and anomaly detection, were explained. The hardware impairments of both transmitters and receivers were modelled. A deep learning-based design was described. Three RFFI research topics, channel effects elimination, noise mitigation, and receiver distortion mitigation, were reviewed as they are essential for RFFI design. Finally, the experimental methodologies for RFFI was described.
Regarding CB-based authentication, an overview of existing approaches has been provided, including both statistics-based techniques and \ac{ml} solutions. Several features used for \ac{pla} have been introduced and discussed, and implementations have been classified based on the use of simulation or experimental tools. The remaining research challenges for both topics were discussed, and future research directions were suggested to make more robust device fingerprinting approaches.


\bibliographystyle{IEEEtran}
\bibliography{IEEEabrv,refs}

\begin{thebibliography}{100}
\providecommand{\url}[1]{#1}
\csname url@samestyle\endcsname
\providecommand{\newblock}{\relax}
\providecommand{\bibinfo}[2]{#2}
\providecommand{\BIBentrySTDinterwordspacing}{\spaceskip=0pt\relax}
\providecommand{\BIBentryALTinterwordstretchfactor}{4}
\providecommand{\BIBentryALTinterwordspacing}{\spaceskip=\fontdimen2\font plus
\BIBentryALTinterwordstretchfactor\fontdimen3\font minus
  \fontdimen4\font\relax}
\providecommand{\BIBforeignlanguage}[2]{{%
\expandafter\ifx\csname l@#1\endcsname\relax
\typeout{** WARNING: IEEEtran.bst: No hyphenation pattern has been}%
\typeout{** loaded for the language `#1'. Using the pattern for}%
\typeout{** the default language instead.}%
\else
\language=\csname l@#1\endcsname
\fi
#2}}
\providecommand{\BIBdecl}{\relax}
\BIBdecl

\bibitem{iot_dev_no}
\BIBentryALTinterwordspacing
State of {IoT} 2024: Number of connected {IoT} devices growing 13\% to 18.8
  billion globally. Accessed on 7 Dec., 2024. [Online]. Available:
  \url{https://iot-analytics.com/number-connected-iot-devices}
\BIBentrySTDinterwordspacing

\bibitem{burg2017wireless}
A.~Burg, A.~Chattopadhyay, and K.-Y. Lam, ``Wireless communication and security
  issues for cyber--physical systems and the {Internet}-of-things,''
  \emph{Proc. {IEEE}}, vol. 106, no.~1, pp. 38--60, Jan. 2017.

\bibitem{trappe2015low}
W.~Trappe, R.~Howard, and R.~S. Moore, ``Low-energy security: Limits and
  opportunities in the {Internet} of things,'' \emph{{IEEE} Signal Process.
  Mag.}, vol.~13, no.~1, pp. 14--21, Jan. 2015.

\bibitem{neshenko2019demystifying}
N.~Neshenko, E.~Bou-Harb, J.~Crichigno, G.~Kaddoum, and N.~Ghani,
  ``{Demystifying IoT security: An exhaustive survey on IoT vulnerabilities and
  a first empirical look on Internet-scale IoT exploitations},'' \emph{{IEEE}
  Commun. Surveys Tuts.}, vol.~21, no.~3, pp. 2702--2733, Mar. 2019.

\bibitem{zhang2020new}
J.~Zhang, G.~Li, A.~Marshall, A.~Hu, and L.~Hanzo, ``A new frontier for {IoT}
  security emerging from three decades of key generation relying on wireless
  channels,'' \emph{{IEEE} Access}, vol.~8, pp. 138\,406--138\,446, July 2020.

\bibitem{cheng2017securing}
C.~Cheng, R.~Lu, A.~Petzoldt, and T.~Takagi, ``Securing the {Internet} of
  {Things} in a quantum world,'' \emph{{IEEE} Commun. Mag.}, vol.~55, no.~2,
  pp. 116--120, Feb. 2017.

\bibitem{zeng2010non}
K.~Zeng, K.~Govindan, and P.~Mohapatra, ``Non-cryptographic authentication and
  identification in wireless networks [security and privacy in emerging
  wireless networks],'' \emph{{IEEE} Wireless Commun.}, vol.~17, no.~5, pp.
  56--62, May 2010.

\bibitem{zhang2021radio}
J.~Zhang, R.~Woods, M.~Sandell, M.~Valkama, A.~Marshall, and J.~Cavallaro,
  ``Radio frequency fingerprint identification for narrowband systems,
  modelling and classification,'' \emph{{IEEE} Trans. Inf. Forensics Security},
  vol.~16, pp. 3974--3987, 2021.

\bibitem{baracca2012physical}
P.~Baracca, N.~Laurenti, and S.~Tomasin, ``Physical layer authentication over
  {MIMO} fading wiretap channels,'' \emph{{IEEE} Trans. Wireless Commun.},
  vol.~11, no.~7, pp. 2564--2573, May 2012.

\bibitem{xu2015device}
Q.~Xu, R.~Zheng, W.~Saad, and Z.~Han, ``Device fingerprinting in wireless
  networks: Challenges and opportunities,'' \emph{{IEEE} Commun. Surveys
  Tuts.}, vol.~18, no.~1, pp. 94--104, Jan. 2015.

\bibitem{sanchez2021survey}
P.~M.~S. S{\'a}nchez, J.~M.~J. Valero, A.~H. Celdr{\'a}n, G.~Bovet, M.~G.
  P{\'e}rez, and G.~M. P{\'e}rez, ``A survey on device behavior fingerprinting:
  Data sources, techniques, application scenarios, and datasets,'' \emph{{IEEE}
  Commun. Surveys Tuts.}, vol.~23, no.~2, pp. 1048--1077, Feb. 2021.

\bibitem{chowdhury2022survey}
R.~R. Chowdhury and P.~E. Abas, ``A survey on device fingerprinting approach
  for resource-constraint {IoT} devices: Comparative study and research
  challenges,'' \emph{Internet of Things}, vol.~20, p. 100632, Nov. 2022.

\bibitem{Kumar23device}
V.~Kumar and K.~Paul, ``Device fingerprinting for cyber-physical systems: {A}
  survey,'' \emph{ACM Comput. Surv.}, vol.~55, no. 14s, July 2023.

\bibitem{xie2020survey}
N.~Xie, Z.~Li, and H.~Tan, ``A survey of physical-layer authentication in
  wireless communications,'' \emph{{IEEE} Commun. Surveys Tuts.}, vol.~23,
  no.~1, pp. 282--310, Jan. 2020.

\bibitem{bai2020physical}
L.~Bai, L.~Zhu, J.~Liu, J.~Choi, and W.~Zhang, ``Physical layer authentication
  in wireless communication networks: {A} survey,'' \emph{J. Commun. Inf.
  Netw.}, vol.~5, no.~3, pp. 237--264, Mar. 2020.

\bibitem{Wang21survey}
W.~Wang, I.~Aguilar~Sanchez, G.~Caparra, A.~McKeown, T.~Whitworth, and E.~S.
  Lohan, ``A survey of spoofer detection techniques via radio frequency
  fingerprinting with focus on the {GNSS} pre-correlation sampled data,''
  \emph{Sensors}, vol.~21, no.~9, Sep. 2021.

\bibitem{zhang23survey}
Y.~Zhang, S.~Zhao, H.~Ji, Y.~Zhang, Y.~Shen, and X.~Jiang, ``A survey of secure
  communications for satellite {I}nternet based on cryptography and physical
  layer security,'' \emph{IET Inf. Secur.}, vol. 2023, pp. 1--15, Oct. 2023.

\bibitem{illi2024physical}
E.~Illi, M.~Qaraqe, S.~Althunibat, A.~Alhasanat, M.~Alsafasfeh, M.~de~Ree,
  G.~Mantas, J.~Rodriguez, W.~Aman, and S.~Al-Kuwari, ``Physical layer security
  for authentication, confidentiality, and malicious node detection: {A}
  paradigm shift in securing {IoT} networks,'' \emph{{IEEE} Commun. Surveys
  Tuts.}, vol.~26, no.~1, pp. 347--388, Jan. 2024.

\bibitem{hoang2024physical}
T.~M. Hoang, A.~Vahid, H.~D. Tuan, and L.~Hanzo, ``Physical layer
  authentication and security design in the machine learning era,''
  \emph{{IEEE} Commun. Surveys Tuts.}, vol.~26, no.~3, pp. 1830--1860, Feb.
  2024.

\bibitem{wang2016wireless}
W.~Wang, Z.~Sun, S.~Piao, B.~Zhu, and K.~Ren, ``Wireless physical-layer
  identification: Modeling and validation,'' \emph{{IEEE} Trans. Inf. Forensics
  Security}, vol.~11, no.~9, pp. 2091--2106, Apr. 2016.

\bibitem{robyns2017physical}
P.~Robyns, E.~Marin, W.~Lamotte, P.~Quax, D.~Singel{\'e}e, and B.~Preneel,
  ``Physical-layer fingerprinting of {LoRa} devices using supervised and
  zero-shot learning,'' in \emph{Proc. ACM Conf. Secur. Privacy Wireless Mobile
  Netw. (WiSec)}, 2017, pp. 58--63.

\bibitem{shen2021jsac}
G.~Shen, J.~Zhang, A.~Marshall, L.~Peng, and X.~Wang, ``Radio frequency
  fingerprint identification for {LoRa} using deep learning,'' \emph{{IEEE} J.
  Sel. Areas Commun.}, vol.~39, no.~8, pp. 2604--2616, June 2021.

\bibitem{shen2021towards}
G.~Shen, J.~Zhang, A.~Marshall, and J.~R. Cavallaro, ``Towards scalable and
  channel-robust radio frequency fingerprint identification for {LoRa},''
  \emph{{IEEE} Trans. Inf. Forensics Security}, vol.~17, pp. 774--787, Feb.
  2022.

\bibitem{shen2023length}
G.~Shen, J.~Zhang, A.~Marshall, M.~Valkama, and J.~R. Cavallaro, ``Toward
  length-versatile and noise-robust radio frequency fingerprint
  identification,'' \emph{{IEEE} Trans. Inf. Forensics Security}, vol.~18, pp.
  2355 -- 2367, Apr. 2023.

\bibitem{hanna2020open}
S.~Hanna, S.~Karunaratne, and D.~Cabric, ``Open set wireless transmitter
  authorization: Deep learning approaches and dataset considerations,''
  \emph{{IEEE} Trans. on Cogn. Commun. Netw.}, vol.~7, no.~1, pp. 59--72, Jan.
  2021.

\bibitem{gritsenko2019finding}
A.~Gritsenko, Z.~Wang, T.~Jian, J.~Dy, K.~Chowdhury, and S.~Ioannidis,
  ``Finding a ‘new’ needle in the haystack: Unseen radio detection in large
  populations using deep learning,'' in \emph{Proc. IEEE Int. Symp. Dynamic
  Spectr. Access Netw. (DySPAN)}, Newark, NJ, USA, 2019, pp. 1--10.

\bibitem{zhu2013blind}
Z.~Zhu, X.~Huang, M.~Caron, and H.~Leung, ``Blind self-calibration technique
  for {I/Q} imbalances and {DC}-offsets,'' \emph{{IEEE} Trans. Circuits Syst.
  {I}}, vol.~61, no.~6, pp. 1849--1859, Jun. 2013.

\bibitem{zhu2013challenges}
Z.~Zhu, H.~Leung, and X.~Huang, ``Challenges in reconfigurable radio
  transceivers and application of nonlinear signal processing for {RF}
  impairment mitigation,'' \emph{{IEEE} Circuits Syst. Mag.}, vol.~13, no.~1,
  pp. 44--65, Jan. 2013.

\bibitem{shen2023methodology}
G.~Shen, J.~Zhang, and A.~Marshall, ``Deep learning - powered radio frequency
  fingerprint identification: Methodology and case study,'' \emph{{IEEE}
  Commun. Mag.}, vol.~61, no.~9, pp. 170--176, Sep. 2023.

\bibitem{halperin2011tool}
D.~Halperin, W.~Hu, A.~Sheth, and D.~Wetherall, ``Tool release: Gathering
  802.11n traces with channel state information,'' \emph{ACM SIGCOMM Comp.
  Commun. Rev.}, vol.~41, no.~1, pp. 53--53, Jan. 2011.

\bibitem{Xie:2015:PPD:2789168.2790124}
\BIBentryALTinterwordspacing
Y.~Xie, Z.~Li, and M.~Li, ``Precise power delay profiling with commodity
  {WiFi},'' in \emph{Proc. 21st Annual International Conference on Mobile
  Computing and Networking (MobiCom)}.\hskip 1em plus 0.5em minus 0.4em\relax
  New York, NY, USA: ACM, 2015, p. 53–64. [Online]. Available:
  \url{http://doi.acm.org/10.1145/2789168.2790124}
\BIBentrySTDinterwordspacing

\bibitem{atheroscsi}
\BIBentryALTinterwordspacing
{Atheros CSI} toolkit. [Online]. Available:
  \url{https://wands.sg/research/wifi/AtherosCSI/}
\BIBentrySTDinterwordspacing

\bibitem{gringoli2019free}
F.~Gringoli, M.~Schulz, J.~Link, and M.~Hollick, ``Free your {CSI}: A channel
  state information extraction platform for modern {Wi-Fi} chipsets,'' in
  \emph{Proc. Int. Workshop Wireless Netw. Testbeds, Exp. Evaluation \&
  Characterization}, 2019, p. 21–28.

\bibitem{nexmon:project}
\BIBentryALTinterwordspacing
M.~Schulz, D.~Wegemer, and M.~Hollick. Nexmon channel state information
  extractor. [Online]. Available:
  \url{https://github.com/seemoo-lab/nexmon_csi}
\BIBentrySTDinterwordspacing

\bibitem{Hern2006:Lightweight}
S.~M. Hernandez and E.~Bulut, ``{Lightweight and Standalone {IoT} Based {WiFi}
  Sensing for Active Repositioning and Mobility},'' in \emph{Proc. 21st Int.
  Symp. a World of Wireless, Mobile and Multimedia Networks (WoWMoM)}, Cork,
  Ireland, Jun. 2020.

\bibitem{esp32csi}
\BIBentryALTinterwordspacing
{ESP32 CSI} toolkit. [Online]. Available:
  \url{https://stevenmhernandez.github.io/ESP32-CSI-Tool/}
\BIBentrySTDinterwordspacing

\bibitem{hua2018accurate}
J.~Hua, H.~Sun, Z.~Shen, Z.~Qian, and S.~Zhong, ``Accurate and efficient
  wireless device fingerprinting using channel state information,'' in
  \emph{Proc. IEEE INFOCOM}, 2018, pp. 1700--1708.

\bibitem{vo2016fingerprinting}
T.~D. Vo-Huu, T.~D. Vo-Huu, and G.~Noubir, ``Fingerprinting {Wi-Fi} devices
  using software defined radios,'' in \emph{Proc. ACM Conf. Secur. Privacy
  Wireless Mobile Netw. (WiSec)}, 2016, pp. 3--14.

\bibitem{peng2018design}
L.~Peng, A.~Hu, J.~Zhang, Y.~Jiang, J.~Yu, and Y.~Yan, ``Design of a hybrid
  {RF} fingerprint extraction and device classification scheme,'' \emph{{IEEE}
  Internet Things J.}, vol.~6, no.~1, pp. 349--360, Jan. 2018.

\bibitem{soltani2020more}
N.~Soltani, K.~Sankhe, J.~Dy, S.~Ioannidis, and K.~Chowdhury, ``More is better:
  Data augmentation for channel-resilient {RF} fingerprinting,'' \emph{{IEEE}
  Commun. Mag.}, vol.~58, no.~10, pp. 66--72, Oct. 2020.

\bibitem{matlab_fading}
\BIBentryALTinterwordspacing
Fading channels. Accessed on February 5, 2025. [Online]. Available:
  \url{https://mathworks.com/help/comm/ug/fading-channels.html}
\BIBentrySTDinterwordspacing

\bibitem{wlan_channel}
\BIBentryALTinterwordspacing
{WLAN} channel models. Accessed on February 5, 2025. [Online]. Available:
  \url{https://mathworks.com/help/wlan/gs/wlan-channel-models.html}
\BIBentrySTDinterwordspacing

\bibitem{ma2025wcnc}
J.~Ma, J.~Zhang, G.~Shen, L.~Peng, and A.~Marshall, ``Towards channel-robust
  radio frequency fingerprint identification using contrastive learning,'' in
  \emph{Proc. IEEE Wireless Commun. and Netw. Conf. (WCNC)}, 2025.

\bibitem{al2021deeplora}
A.~Al-Shawabka, P.~Pietraski, S.~B~Pattar, F.~Restuccia, and T.~Melodia,
  ``{DeepLoRa}: Fingerprinting {LoRa} devices at scale through deep learning
  and data augmentation,'' in \emph{Proc. ACM Int. Symp. Mob. Ad Hoc Netw.
  Comput. (MobiHoc)}, Shanghai, China, Jul. 2021.

\bibitem{peng2019deep}
L.~Peng, J.~Zhang, M.~Liu, and A.~Hu, ``Deep learning based {RF} fingerprint
  identification using differential constellation trace figure,'' \emph{{IEEE}
  Trans. Veh. Technol.}, vol.~69, no.~1, pp. 1091--1095, Jan. 2019.

\bibitem{merchant2018deep}
K.~Merchant, S.~Revay, G.~Stantchev, and B.~Nousain, ``Deep learning for {RF}
  device fingerprinting in cognitive communication networks,'' \emph{{IEEE} J.
  Sel. Topics Signal Process.}, vol.~12, no.~1, pp. 160--167, Jan. 2018.

\bibitem{ding2018specific}
L.~Ding, S.~Wang, F.~Wang, and W.~Zhang, ``Specific emitter identification via
  convolutional neural networks,'' \emph{{IEEE} Commun. Lett.}, vol.~22,
  no.~12, pp. 2591--2594, Dec. 2018.

\bibitem{zhang2016specific}
J.~Zhang, F.~Wang, O.~A. Dobre, and Z.~Zhong, ``Specific emitter identification
  via {Hilbert-Huang} transform in single-hop and relaying scenarios,''
  \emph{{IEEE} Trans. Inf. Forensics Security}, vol.~11, no.~6, pp. 1192--1205,
  Jun. 2016.

\bibitem{hall2005radio}
J.~Hall, M.~Barbeau, and E.~Kranakis, ``Radio frequency fingerprinting for
  intrusion detection in wireless networks,'' \emph{{IEEE} Trans. Depend. Sec.
  Comput.}, vol.~12, pp. 1--35, 2005.

\bibitem{shen2021infocom}
G.~Shen, J.~Zhang, A.~Marshall, L.~Peng, and X.~Wang, ``Radio frequency
  fingerprint identification for {LoRa} using spectrogram and {CNN},'' in
  \emph{Proc. IEEE INFOCOM}, May 2021.

\bibitem{pan2019specific}
Y.~Pan, S.~Yang, H.~Peng, T.~Li, and W.~Wang, ``Specific emitter identification
  based on deep residual networks,'' \emph{{IEEE} Access}, vol.~7, pp.
  54\,425--54\,434, Apr. 2019.

\bibitem{al2020exposing}
A.~Al-Shawabka, F.~Restuccia, S.~D’Oro, T.~Jian, B.~C. Rendon, N.~Soltani,
  J.~Dy, S.~Ioannidis, K.~Chowdhury, and T.~Melodia, ``Exposing the
  fingerprint: Dissecting the impact of the wireless channel on radio
  fingerprinting,'' in \emph{Proc. IEEE INFOCOM}, 2020, pp. 646--655.

\bibitem{wisig_dataset}
\BIBentryALTinterwordspacing
(2022) {WiSig: RF fingerprinting dataset}. [Online]. Available:
  \url{https://cores.ee.ucla.edu/downloads/datasets/wisig/}
\BIBentrySTDinterwordspacing

\bibitem{hanna2022wisig}
S.~Hanna, S.~Karunaratne, and D.~Cabric, ``{WiSig}: A large-scale {WiFi} signal
  dataset for receiver and channel agnostic {RF} fingerprinting,'' \emph{{IEEE}
  Access}, vol.~10, pp. 22\,808--22\,818, Feb. 2022.

\bibitem{kong2024deepcrf_dataset}
\BIBentryALTinterwordspacing
(2025) {DeepCRF TIFS}. [Online]. Available:
  \url{https://github.com/Oriseven/DeepCRF\_TIFS}
\BIBentrySTDinterwordspacing

\bibitem{kong2024deepcrf}
R.~Kong and H.~Chen, ``Deepcrf: Deep learning-enhanced csi-based rf
  fingerprinting for channel-resilient wifi device identification,''
  \emph{{IEEE} Trans. Inf. Forensics Security}, vol.~20, pp. 264 -- 278, 2025.

\bibitem{xie2025towards_dataset}
\BIBentryALTinterwordspacing
L.~Xie, L.~Peng, and J.~Zhang. (2024) {Wi-Fi dataset for channel-robust RFFI}.
  [Online]. Available:
  \url{https://ieee-dataport.org/documents/wi-fi-dataset-channel-robust-rffi}
\BIBentrySTDinterwordspacing

\bibitem{xie2025towards}
------, ``Towards robust {RF} fingerprint identification using spectral
  regrowth and carrier frequency offset,'' in \emph{Proc. IEEE INFOCOM}, 2025.

\bibitem{shi2023robust_dataset}
\BIBentryALTinterwordspacing
J.~Shi, L.~Peng, H.~Fu, and A.~Hu, ``{ZigBee RFF dataset},'' 2023. [Online].
  Available: \url{https://dx.doi.org/10.21227/b4qd-gv36}
\BIBentrySTDinterwordspacing

\bibitem{shi2023robust}
------, ``Robust {RF} fingerprint extraction based on cyclic shift
  characteristic,'' \emph{{IEEE} Internet Things J.}, vol.~10, no.~21, pp.
  19\,218--19\,233, Nov. 2023.

\bibitem{shen2022lora_channel}
\BIBentryALTinterwordspacing
G.~Shen, J.~Zhang, and A.~Marshall. (2022) {LoRa RFFI} dataset. [Online].
  Available: \url{https://dx.doi.org/10.21227/qqt4-kz19}
\BIBentrySTDinterwordspacing

\bibitem{shen2023length_dataset}
\BIBentryALTinterwordspacing
------. (2023) {LoRa RFFI dataset with different spreading factors}. [Online].
  Available: \url{https://dx.doi.org/10.21227/5q6q-c107}
\BIBentrySTDinterwordspacing

\bibitem{shen2024lora_rx}
\BIBentryALTinterwordspacing
------. (2024) Radio frequency fingerprint {LoRa} dataset with multiple
  receivers. [Online]. Available: \url{https://dx.doi.org/10.21227/d6vx-r538}
\BIBentrySTDinterwordspacing

\bibitem{shen2023towards}
G.~Shen, J.~Zhang, A.~Marshall, R.~Woods, J.~Cavallaro, and L.~Chen, ``Towards
  receiver-agnostic and collaborative radio frequency fingerprint
  identification,'' \emph{{IEEE} Trans. Mobile Comput.}, vol.~23, no.~7, pp.
  7618 -- 7634, Dec. 2023.

\bibitem{shen2024federated_dataset}
\BIBentryALTinterwordspacing
G.~Shen and J.~Zhang. (2024) {LoRa Federated RFFI dataset}. [Online].
  Available: \url{https://dx.doi.org/10.21227/nkdv-az07}
\BIBentrySTDinterwordspacing

\bibitem{shen2024federated}
G.~Shen, J.~Zhang, X.~Wang, and S.~Mao, ``Federated radio frequency fingerprint
  identification powered by unsupervised contrastive learning,'' \emph{{IEEE}
  Trans. Inf. Forensics Security}, vol.~19, pp. 9204 -- 9215, Sep. 2024.

\bibitem{deeplora_dataset}
\BIBentryALTinterwordspacing
{LoRa} radio data. [Online]. Available:
  \url{https://www.interdigital.com/data_sets/lora-radio-data.}
\BIBentrySTDinterwordspacing

\bibitem{oregon_dataset}
\BIBentryALTinterwordspacing
{RF} fingerprinting ({RFFP}) datasets. [Online]. Available:
  \url{https://research.engr.oregonstate.edu/hamdaoui/datasets}
\BIBentrySTDinterwordspacing

\bibitem{elmaghbub2021lora}
A.~Elmaghbub and B.~Hamdaoui, ``{LoRa} device fingerprinting in the wild:
  Disclosing {RF} data-driven fingerprint sensitivity to deployment
  variability,'' \emph{{IEEE} Access}, vol.~9, pp. 142\,893--142\,909, Oct.
  2021.

\bibitem{jagannath2023embedding_dataset}
\BIBentryALTinterwordspacing
A.~Jagannath and J.~Jagannath. (2022) {RF-fingerprint-BT-IoT: Real-world
  frequency hopping Bluetooth dataset from IoT devices for RF fingerprinting}.
  [Online]. Available: \url{https://dx.doi.org/10.21227/364j-6j73}
\BIBentrySTDinterwordspacing

\bibitem{jagannath2023embedding}
------, ``Embedding-assisted attentional deep learning for real-world {RF}
  fingerprinting of {Bluetooth},'' \emph{{IEEE} Trans. on Cogn. Commun. Netw.},
  vol.~9, no.~4, pp. 940--949, 2023.

\bibitem{ardoin2025tracking_dataset}
\BIBentryALTinterwordspacing
T.~Ardoin and M.~Kholghi. (2024) {RUFF -- Rotating UWB For Fingerprint}.
  [Online]. Available: \url{https://zenodo.org/records/11083153}
\BIBentrySTDinterwordspacing

\bibitem{ardoin2025tracking}
T.~Ardoin, N.~Pauli, B.~Gro{\ss}, M.~Kholghi, K.~Reaz, and G.~Wunder,
  ``Tracking {UWB} devices through radio frequency fingerprinting is
  possible,'' in \emph{Proc. Int. Conf. Comput., Netw. and Commun. (ICNC)},
  2025.

\bibitem{peng2024hybrid_dataset}
\BIBentryALTinterwordspacing
L.~Peng. (2024) {LTE} mobile phone {PRACH} signal. [Online]. Available:
  \url{https://ieee-dataport.org/documents/lte-mobile-phone-prach-signal}
\BIBentrySTDinterwordspacing

\bibitem{peng2024hybrid}
L.~Peng, Z.~Wu, J.~Zhang, M.~Liu, H.~Fu, and A.~Hu, ``Hybrid {RFF}
  identification for {LTE} using wavelet coefficient graph and differential
  spectrum,'' \emph{{IEEE} Trans. Veh. Technol.}, Apr. 2024.

\bibitem{oligeri2022dataset}
\BIBentryALTinterwordspacing
G.~Oligeri and S.~Sciancalepore. (2022) Physical layer data acquisition of
  {IRIDIUM} satellites broadcast messages. [Online]. Available:
  \url{https://data.mendeley.com/datasets/xcxspv8c2r/2}
\BIBentrySTDinterwordspacing

\bibitem{oligeri2022past}
G.~Oligeri, S.~Sciancalepore, S.~Raponi, and R.~Di~Pietro, ``{PAST-AI}:
  Physical-layer authentication of satellite transmitters via deep learning,''
  \emph{{IEEE} Trans. Inf. Forensics Security}, vol.~18, pp. 274--289, 2023.

\bibitem{2023watch}
\BIBentryALTinterwordspacing
J.~Smailes, S.~K{\"o}hler, S.~Birnbach, M.~Strohmeier, and I.~Martinovic.
  (2023) Dataset for "watch this space: Securing satellite communication
  through resilient transmitter fingerprinting". [Online]. Available:
  \url{https://zenodo.org/records/8220494}
\BIBentrySTDinterwordspacing

\bibitem{smailes2023watch}
------, ``Watch this space: Securing satellite communication through resilient
  transmitter fingerprinting,'' in \emph{Proc. ACM SIGSAC Conf. on Comput. and
  Commun. Secur. (CCS)}, 2023, pp. 608--621.

\bibitem{xie2023disentangled}
R.~Xie, W.~Xu, J.~Yu, A.~Hu, D.~W.~K. Ng, and A.~L. Swindlehurst,
  ``Disentangled representation learning for {RF} fingerprint extraction under
  unknown channel statistics,'' \emph{{IEEE} Trans. Commun.}, vol.~71, no.~7,
  pp. 3946--3962, Jul. 2023.

\bibitem{xing2022design}
Y.~Xing, A.~Hu, J.~Zhang, L.~Peng, and X.~Wang, ``Design of a channel robust
  radio frequency fingerprint identification scheme,'' \emph{{IEEE} Internet
  Things J.}, vol.~10, no.~8, pp. 6946--6959, Aug. 2023.

\bibitem{zheng2019fid}
T.~Zheng, Z.~Sun, and K.~Ren, ``{FID}: Function modeling-based data-independent
  and channel-robust physical-layer identification,'' in \emph{Proc. IEEE
  INFOCOM}, 2019, pp. 199--207.

\bibitem{cekic2020robust}
M.~Cekic, S.~Gopalakrishnan, and U.~Madhow, ``Wireless fingerprinting via deep
  learning: The impact of confounding factors,'' in \emph{Proc. Asilomar Conf.
  Signals, Syst., and Comput.}, 2021, pp. 677--684.

\bibitem{fu2023deep}
H.~Fu, L.~Peng, M.~Liu, and A.~Hu, ``Deep learning based {RF} fingerprint
  identification with channel effects mitigation,'' \emph{IEEE Open J. Commun.
  Soc.}, Jul. 2023.

\bibitem{piva2021tags}
M.~Piva, G.~Maselli, and F.~Restuccia, ``The tags are alright: Robust
  large-scale {RFID} clone detection through federated data-augmented radio
  fingerprinting,'' in \emph{Proc. ACM Int. Symposium Mob. Ad Hoc Netw. Comput.
  (MobiHoc)}, Shanghai, China, Jul. 2021.

\bibitem{tian2022transfer}
T.~Tian, Y.~Wang, H.~Dong, Y.~Peng, Y.~Lin, G.~Gui, and H.~Gacanin, ``Transfer
  learning-based radio frequency fingerprint identification using {ConvMixer}
  network,'' in \emph{Proc. IEEE Global Commun. Conf. (GLOBECOM)}, 2022, pp.
  4722--4727.

\bibitem{pan2024equalization}
R.~Pan, H.~Chen, H.~Chen, and W.-Q. Wang, ``Equalization assisted domain
  adaptation for radio frequency fingerprint identification,'' \emph{{IEEE}
  Wireless Commun. Lett.}, Apr. 2024.

\bibitem{wang2022radio}
W.~Wang and L.~Gan, ``Radio frequency fingerprinting improved by statistical
  noise reduction,'' \emph{{IEEE} Trans. on Cogn. Commun. Netw.}, vol.~8,
  no.~3, pp. 1444--1452, Mar. 2022.

\bibitem{xing2018radio}
Y.~Xing, A.~Hu, J.~Zhang, L.~Peng, and G.~Li, ``On radio frequency fingerprint
  identification for {DSSS} systems in low {SNR} scenarios,'' \emph{{IEEE}
  Commun. Lett.}, vol.~22, no.~11, pp. 2326--2329, Nov. 2018.

\bibitem{ohtsuji2019noise}
T.~Ohtsuji, T.~Takeuchi, T.~Soma, and M.~Kitsunezuka, ``Noise-tolerant,
  deep-learning-based radio identification with logarithmic power spectrum,''
  in \emph{Proc. IEEE Int. Conf. Commun. (ICC)}, 2019, pp. 1--6.

\bibitem{andrews2019crowdsourced}
S.~Andrews, R.~M. Gerdes, and M.~Li, ``Crowdsourced measurements for device
  fingerprinting,'' in \emph{Proc. ACM Conf. Secur. Privacy Wireless Mobile
  Netw. (WiSec)}, 2019, pp. 72--82.

\bibitem{he2020cooperative}
B.~He and F.~Wang, ``Cooperative specific emitter identification via multiple
  distorted receivers,'' \emph{{IEEE} Trans. Inf. Forensics Security}, vol.~15,
  pp. 3791--3806, Jun. 2020.

\bibitem{shen2021asilomar}
G.~Shen, J.~Zhang, A.~Marshall, M.~Valkama, and J.~Cavallaro, ``Radio frequency
  fingerprint identification for security in low-cost {IoT} devices,'' in
  \emph{Proc. Asilomar Conf. Signals, Syst., and Comput.}, 2021, pp. 309--313.

\bibitem{wu2021dsln}
W.~Wu, S.~Hu, D.~Lin, and Z.~Liu, ``{DSLN}: Securing {Internet of Things}
  through {RF} fingerprint recognition in {low-SNR} settings,'' \emph{{IEEE}
  Internet Things J.}, vol.~9, no.~5, pp. 3838--3849, Jul. 2021.

\bibitem{merchant2019toward}
K.~Merchant and B.~Nousain, ``Toward receiver-agnostic {RF} fingerprint
  verification,'' in \emph{Proc. IEEE Globecom Workshops (GC Wkshps)}, 2019,
  pp. 1--6.

\bibitem{zhao2023gan}
T.~Zhao, S.~Sarkar, E.~Krijestorac, and D.~Cabric, ``{GAN-RXA}: A practical
  scalable solution to receiver-agnostic transmitter fingerprinting,''
  \emph{{IEEE} Trans. on Cogn. Commun. Netw.}, vol.~10, no.~2, pp. 403--416,
  2023.

\bibitem{li2025receiver}
K.~Li, J.~Bao, X.~Xie, J.~Hong, and C.~Hua, ``Receiver-agnostic radio frequency
  fingerprint identification for zero-trust wireless networks,'' \emph{{IEEE}
  J. Sel. Areas Commun.}, 2025.

\bibitem{del2024fingerprint}
J.~A. Gutierrez~del Arroyo, B.~J. Borghetti, and M.~A. Temple, ``Fingerprint
  extraction through distortion reconstruction ({FEDR}): A {CNN}-based approach
  to {RF} fingerprinting,'' \emph{{IEEE} Trans. Inf. Forensics Security},
  vol.~19, pp. 9258--9269, Sep. 2024.

\bibitem{yu2019robust}
J.~Yu, A.~Hu, G.~Li, and L.~Peng, ``A robust {RF} fingerprinting approach using
  multisampling convolutional neural network,'' \emph{{IEEE} Internet Things
  J.}, vol.~6, no.~4, pp. 6786--6799, Apr. 2019.

\bibitem{bihl2016feature}
T.~J. Bihl, K.~W. Bauer, and M.~A. Temple, ``Feature selection for {RF}
  fingerprinting with multiple discriminant analysis and using {ZigBee} device
  emissions,'' \emph{{IEEE} Trans. Inf. Forensics Security}, vol.~11, no.~8,
  pp. 1862--1874, Aug. 2016.

\bibitem{givehchian2022evaluating}
H.~Givehchian, N.~Bhaskar, E.~R. Herrera, H.~R.~L. Soto, C.~Dameff,
  D.~Bharadia, and A.~Schulman, ``Evaluating physical-layer {BLE} location
  tracking attacks on mobile devices,'' in \emph{Proc. IEEE Symp. on Secur. and
  Privacy (SP)}, 2022, pp. 1690--1704.

\bibitem{givehchian2024practical}
H.~Givehchian, N.~Bhaskar, A.~Redding, H.~Zhao, A.~Schulman, and D.~Bharadia,
  ``Practical obfuscation of {BLE} physical-layer fingerprints on mobile
  devices,'' in \emph{Proc. IEEE Symp. on Secur. and Privacy (SP)}, 2024, pp.
  2867--2885.

\bibitem{yuan2025robust}
N.~Yuan, J.~Zhang, Y.~Ding, and S.~L. Cotton, ``Robust radio frequency
  fingerprint identification for {Bluetooth} low energy under low snr and
  channel variations,'' in \emph{Proc. IEEE Wireless Commun. and Netw. Conf.
  (WCNC)}, 2025.

\bibitem{zhuang2018fbsleuth}
Z.~Zhuang, X.~Ji, T.~Zhang, J.~Zhang, W.~Xu, Z.~Li, and Y.~Liu, ``{FBSleuth}:
  {Fake} base station forensics via radio frequency fingerprinting,'' in
  \emph{Proc. Asia Conf. Comp. Commun. Secur. (ASIA CCS)}.\hskip 1em plus 0.5em
  minus 0.4em\relax ACM, 2018, p. 261–272.

\bibitem{yang2023led}
X.~Yang and D.~Li, ``{LED-RFF: LTE} {DMRS} based channel robust radio frequency
  fingerprint identification scheme,'' \emph{{IEEE} Trans. Inf. Forensics
  Security}, vol.~19, pp. 1855--1869, Dec. 2023.

\bibitem{yin2024multi}
P.~Yin, L.~Peng, G.~Shen, J.~Zhang, M.~Liu, H.~Fu, A.~Hu, and X.~Wang,
  ``Multi-channel {CNN}-based open-set {RF} fingerprint identification for
  {LTE} devices,'' \emph{{IEEE} Trans. on Cogn. Commun. Netw.}, vol.~10, no.~5,
  pp. 1788--1800, Apr. 2024.

\bibitem{reus2020trust}
G.~Reus-Muns, D.~Jaisinghani, K.~Sankhe, and K.~R. Chowdhury, ``Trust in {5G}
  open {RANs} through machine learning: {RF} fingerprinting on the {POWDER}
  {PAWR} platform,'' in \emph{Proc. IEEE Global Commun. Conf.
  (GLOBECOM)}.\hskip 1em plus 0.5em minus 0.4em\relax IEEE, 2020, pp. 1--6.

\bibitem{foruhandeh2020spotr}
M.~Foruhandeh, A.~Z. Mohammed, G.~Kildow, P.~Berges, and R.~Gerdes, ``Spotr:
  {GPS} spoofing detection via device fingerprinting,'' in \emph{Proc. ACM
  Conf. Secur. Privacy Wireless Mobile Netw. (WiSec)}, 2020, pp. 242--253.

\bibitem{kong2024csi}
R.~Kong and H.~Chen, ``{CSI-RFF}: Leveraging micro-signals on {CSI} for {RF}
  fingerprinting of commodity {WiFi},'' \emph{{IEEE} Trans. Inf. Forensics
  Security}, vol.~19, pp. 5301 -- 5315, 2024.

\bibitem{yin2025noise}
G.~Yin, J.~Zhang, Y.~Ding, and S.~Cotton, ``Noise-robust radio frequency
  fingerprint identification using denoise diffusion model,'' in \emph{Proc.
  IEEE Wireless Commun. and Netw. Conf. (WCNC) Workshop}, 2025.

\bibitem{gu2024cqp}
X.~Gu, W.~Wu, Y.~Zhou, A.~Song, M.~Yang, Z.~Ling, and J.~Luo, ``{CQP-RFFI}:
  Injecting a communication-quality preserving {RF} fingerprint for {Wi-Fi}
  device identification,'' in \emph{Proc. IEEE/ACM 32nd Int. Symp. Quality of
  Service (IWQoS)}, 2024, pp. 1--10.

\bibitem{MatlabSDR}
\BIBentryALTinterwordspacing
{Supported Hardware – Software-Defined Radio}. [Online]. Available:
  \url{https://uk.mathworks.com/help/comm/supported-hardware-software-defined-radio.html}
\BIBentrySTDinterwordspacing

\bibitem{shen2021radio}
G.~Shen, J.~Zhang, A.~Marshall, L.~Peng, and X.~Wang, ``Radio frequency
  fingerprint identification for {LoRa} using deep learning,'' vol.~39, no.~8,
  pp. 2604--2616, 2021.

\bibitem{gr-ieee802-11}
\BIBentryALTinterwordspacing
{IEEE 802.11 a/g/p Transceiver}. [Online]. Available:
  \url{https://github.com/bastibl/gr-ieee802-11}
\BIBentrySTDinterwordspacing

\bibitem{li2022radionet}
H.~Li, K.~Gupta, C.~Wang, N.~Ghose, and B.~Wang, ``Radionet: Robust
  deep-learning based radio fingerprinting,'' in \emph{Proc. IEEE Conf. Commun.
  and Netw. Secur. (CNS)}, 2022, pp. 190--198.

\bibitem{gr-ieee802-15-4}
\BIBentryALTinterwordspacing
{IEEE 802.15.4 ZigBee Transceiver}. [Online]. Available:
  \url{https://github.com/bastibl/gr-ieee802-15-4}
\BIBentrySTDinterwordspacing

\bibitem{PySDR}
\BIBentryALTinterwordspacing
{PySDR: A Guide to SDR and DSP using Python}. [Online]. Available:
  \url{https://pysdr.org/index.html}
\BIBentrySTDinterwordspacing

\bibitem{PicoScenes}
\BIBentryALTinterwordspacing
{PicoScenes: Enabling Modern Wi-Fi ISAC Research!} [Online]. Available:
  \url{https://ps.zpj.io/}
\BIBentrySTDinterwordspacing

\bibitem{liu2019real}
P.~Liu, P.~Yang, W.-Z. Song, Y.~Yan, and X.-Y. Li, ``Real-time identification
  of rogue {WiFi} connections using environment-independent physical
  features,'' in \emph{Proc. IEEE INFOCOM}, 2019, pp. 190--198.

\bibitem{huang2023phyfinatt}
J.~Huang, B.~Liu, C.~Miao, X.~Zhang, J.~Liu, L.~Su, Z.~Liu, and Y.~Gu,
  ``{PhyFinAtt}: An undetectable attack framework against phy layer
  fingerprint-based {WiFi} authentication,'' \emph{{IEEE} Trans. Mobile
  Comput.}, vol.~23, no.~7, pp. 7753--7770, 2023.

\bibitem{restuccia2021deepfir}
F.~Restuccia, S.~D’Oro, A.~Al-Shawabka, B.~C. Rendon, S.~Ioannidis, and
  T.~Melodia, ``{DeepFIR}: Channel-robust physical-layer deep learning through
  adaptive waveform filtering,'' \emph{{IEEE} Trans. Wireless Commun.},
  vol.~20, no.~12, pp. 8054--8066, 2021.

\bibitem{sankhe2019oracle}
K.~Sankhe, M.~Belgiovine, F.~Zhou, S.~Riyaz, S.~Ioannidis, and K.~Chowdhury,
  ``{ORACLE}: Optimized radio classification through convolutional neural
  networks,'' in \emph{Proc. IEEE INFOCOM}, Paris, France, 2019, pp. 370--378.

\bibitem{GhazalehDataset}
\BIBentryALTinterwordspacing
G.~Kia, D.~Plets, B.~Van~Herbruggen, J.~Fontaine, L.~Verloock, E.~De~Poorter,
  and J.~Talvitie, ``{UWB} {CIR} data collected in 9 different environments in
  {Ghent}, {Belgium},'' 2023. [Online]. Available:
  \url{https://dx.doi.org/10.21227/kt06-tw72}
\BIBentrySTDinterwordspacing

\bibitem{Brighente2019Machine}
A.~Brighente, F.~Formaggio, G.~M. Di~Nunzio, and S.~Tomasin, ``Machine learning
  for in-region location verification in wireless networks,'' \emph{{IEEE} J.
  Sel. Areas Commun.}, vol.~37, no.~11, pp. 2490--2502, Nov. 2019.

\bibitem{Senigagliesi2021Comparison}
L.~Senigagliesi, M.~Baldi, and E.~Gambi, ``Comparison of statistical and
  machine learning techniques for physical layer authentication,'' \emph{{IEEE}
  Trans. Inf. Forensics Security}, vol.~16, pp. 1506--1521, Oct. 2021.

\bibitem{Ardizzon2024Learning}
\BIBentryALTinterwordspacing
F.~Ardizzon and S.~Tomasin, ``Learning the likelihood test with one-class
  classifiers for physical layer authentication,'' 2024. [Online]. Available:
  \url{https://arxiv.org/abs/2210.12494}
\BIBentrySTDinterwordspacing

\bibitem{9982485}
S.~Tomasin, H.~Zhang, A.~Chorti, and H.~V. Poor, ``Challenge-response physical
  layer authentication over partially controllable channels,'' \emph{{IEEE}
  Commun. Mag.}, vol.~60, no.~12, pp. 138--144, Dec. 2022.

\bibitem{10700780}
S.~Tomasin, T.~N. M.~M. Elwakeel, A.~V. Guglielmi, R.~Maes, N.~Noels, and
  M.~Moeneclaey, ``Analysis of challenge-response authentication with
  reconfigurable intelligent surfaces,'' \emph{{IEEE} Trans. Inf. Forensics
  Security}, vol.~19, pp. 9494--9507, 2024.

\bibitem{10615572}
L.~Crosara, A.~V. Guglielmi, N.~Laurenti, and S.~Tomasin,
  ``Divergence-minimizing attack against challenge-response authentication with
  {IRSs},'' in \emph{Proc. IEEE Int. Conf. on Commun. Workshops (ICC
  Workshops)}, 2024, pp. 1986--1991.

\bibitem{10615677}
A.~V. Guglielmi, L.~Crosara, S.~Tomasin, and N.~Laurenti, ``Physical-layer
  challenge-response authentication with {IRS} and single-antenna devices,'' in
  \emph{Proc. IEEE Int. Conf. on Commun. Workshops (ICC Workshops)}, 2024, pp.
  560--565.

\bibitem{Xiao2016PHY}
L.~Xiao, Y.~Li, G.~Han, G.~Liu, and W.~Zhuang, ``{PHY}-layer spoofing detection
  with reinforcement learning in wireless networks,'' \emph{{IEEE} Trans. Veh.
  Technol.}, vol.~65, no.~12, pp. 10\,037--10\,047, Dec. 2016.

\bibitem{Tan2023Generalized}
H.~Tan, N.~Xie, J.~Lu, and D.~Niyato, ``Generalized tag-based physical-layer
  authentication under frequency selective fading channels,'' \emph{{IEEE}
  Trans. Commun.}, vol.~71, no.~5, pp. 2876--2890, May 2023.

\bibitem{amin2024potential}
\BIBentryALTinterwordspacing
H.~Amin, W.~Aman, and S.~Al-Kuwari, ``On the potential of re-configurable
  intelligent surface ({RIS})-assisted physical layer authentication ({PLA}),''
  2024. [Online]. Available: \url{https://arxiv.org/abs/2405.00426}
\BIBentrySTDinterwordspacing

\bibitem{Xie2021Physical}
N.~Xie, J.~Chen, and L.~Huang, ``Physical-layer authentication using multiple
  channel-based features,'' \emph{{IEEE} Trans. Inf. Forensics Security},
  vol.~16, pp. 2356--2366, Jan. 2021.

\bibitem{Wang2020PhysicalKalman}
J.~Wang, Y.~Shao, Y.~Ge, and R.~Yu, ``Physical-layer authentication based on
  adaptive {Kalman} filter for {V2X} communication,'' \emph{Veh. Commun.},
  vol.~26, p. 100281, Dec. 2020.

\bibitem{10436823}
F.~Mazzo, S.~Tomasin, H.~Zhang, A.~Chorti, and H.~V. Poor, ``Physical-layer
  challenge-response authentication for drone networks,'' in \emph{Proc. IEEE
  Global Commun. Conf. (GLOBECOM)}, 2023, pp. 3282--3287.

\bibitem{piana2024challenge}
\BIBentryALTinterwordspacing
M.~Piana, F.~Ardizzon, and S.~Tomasin, ``Challenge-response to authenticate
  drone communications: A game theoretic approach,'' 2024. [Online]. Available:
  \url{https://arxiv.org/abs/2410.00785}
\BIBentrySTDinterwordspacing

\bibitem{6584940}
D.~Shan, K.~Zeng, W.~Xiang, P.~Richardson, and Y.~Dong, ``{PHY-CRAM}: Physical
  layer challenge-response authentication mechanism for wireless networks,''
  \emph{{IEEE} J. Sel. Areas Commun.}, vol.~31, no.~9, pp. 1817--1827, Sep.
  2013.

\bibitem{7501833}
X.~Wu, Z.~Yang, C.~Ling, and X.-G. Xia, ``Artificial-noise-aided physical layer
  phase challenge-response authentication for practical {OFDM} transmission,''
  \emph{{IEEE} Trans. Wireless Commun.}, vol.~15, no.~10, pp. 6611--6625, Oct.
  2016.

\bibitem{Cicek2024impact}
B.~Çiçek and H.~Alakoca, ``Impact of residual hardware impairments on
  {RIS}-aided authentication,'' in \emph{Proc. Virtual Conference on
  Communications (VCC)}, 2024, pp. 1--6.

\bibitem{Lu2020Reinforcement}
X.~Lu, L.~Xiao, T.~Xu, Y.~Zhao, Y.~Tang, and W.~Zhuang, ``Reinforcement
  learning based {PHY} authentication for {VANETs},'' \emph{{IEEE} Trans. Veh.
  Technol.}, vol.~69, no.~3, pp. 3068--3079, Mar. 2020.

\bibitem{Fang2019Learning}
H.~Fang, X.~Wang, and L.~Hanzo, ``Learning-aided physical layer authentication
  as an intelligent process,'' \emph{{IEEE} Trans. Commun.}, vol.~67, no.~3,
  pp. 2260--2273, Mar. 2019.

\bibitem{Meng2024Multidimensional}
R.~Meng, X.~Xu, G.~Li, B.~Xu, F.~Zhu, B.~Wang, and P.~Zhang, ``Multidimensional
  fingerprints-based multiattacker detection for {6G} systems,'' \emph{{IEEE}
  Internet Things J.}, vol.~11, no.~2, pp. 2665--2683, Feb. 2024.

\bibitem{Wang2019Deep}
Q.~Wang, H.~Li, D.~Zhao, Z.~Chen, S.~Ye, and J.~Cai, ``Deep neural networks for
  {CSI}-based authentication,'' \emph{{IEEE} Access}, vol.~7, pp.
  123\,026--123\,034, Aug. 2019.

\bibitem{Abdrabou2022Adaptive}
M.~Abdrabou and T.~A. Gulliver, ``Adaptive physical layer authentication using
  machine learning with antenna diversity,'' \emph{{IEEE} Trans. Commun.},
  vol.~70, no.~10, pp. 6604--6614, Oct. 2022.

\bibitem{Meng2023Physical}
R.~Meng, X.~Xu, B.~Wang, H.~Sun, S.~Xia, S.~Han, and P.~Zhang, ``Physical-layer
  authentication based on hierarchical variational autoencoder for industrial
  {Internet} of {Things},'' \emph{{IEEE} Internet Things J.}, vol.~10, no.~3,
  pp. 2528--2544, Mar. 2023.

\bibitem{Li2015Spoofing}
Y.~Li, L.~Xiao, Q.~Li, and W.~Su, ``Spoofing detection games in underwater
  sensor networks,'' in \emph{Proc. OCEANS 2015 - MTS/IEEE Washington}, 2015,
  pp. 1--5.

\bibitem{Khalid2020Node}
M.~Khalid, R.~Zhao, and X.~Wang, ``Node authentication in underwater acoustic
  sensor networks using time-reversal,'' in \emph{Proc. Global Oceans 2020:
  Singapore – U.S. Gulf Coast}, 2020, pp. 1--4.

\bibitem{Wang2017Physical}
N.~Wang, T.~Jiang, S.~Lv, and L.~Xiao, ``Physical-layer authentication based on
  extreme learning machine,'' \emph{{IEEE} Commun. Lett.}, vol.~21, no.~7, pp.
  1557--1560, Jul. 2017.

\bibitem{Xiao2019Learning}
L.~Xiao, G.~Sheng, X.~Wan, W.~Su, and P.~Cheng, ``Learning-based {PHY}-layer
  authentication for underwater sensor networks,'' \emph{{IEEE} Commun. Lett.},
  vol.~23, no.~1, pp. 60--63, Jan. 2019.

\bibitem{Zhao2023Physical}
R.~Zhao, T.~Shi, C.~Liu, X.~Shen, and O.~A. Dobre, ``Physical layer
  authentication without adversary training data in resource-constrained
  underwater acoustic networks,'' \emph{{IEEE} Sensors J.}, vol.~23, no.~22,
  pp. 28\,270--28\,281, Nov. 2023.

\bibitem{Brighente2024Physical}
A.~Brighente, S.~Xu, S.~Soderi, and M.~Conti, ``Physical layer authentication
  for distributed {RIS} ({DRIS}) enabled {VLC} systems,'' in \emph{Proc. IEEE
  Int. Conf. on Commun. (ICC)}, 2024, pp. 3340--3345.

\bibitem{5683463}
L.~Xiao, A.~Reznik, W.~Trappe, C.~Ye, Y.~Shah, L.~Greenstein, and N.~Mandayam,
  ``{PHY}-authentication protocol for spoofing detection in wireless
  networks,'' in \emph{Proc. IEEE Global Telecommun. Conf. (GLOBECOM)}, 2010,
  pp. 1--6.

\bibitem{4289438}
L.~Xiao, L.~Greenstein, N.~Mandayam, and W.~Trappe, ``Fingerprints in the
  {Ether}: Using the physical layer for wireless authentication,'' in
  \emph{Proc. IEEE Int. Conf. on Commun. (ICC)}, 2007, pp. 4646--4651.

\bibitem{7733021}
M.~Liu, A.~Mukherjee, Z.~Zhang, and X.~Liu, ``{TBAS}: Enhancing {Wi-Fi}
  authentication by actively eliciting channel state information,'' in
  \emph{Proc. Annu. IEEE Int. Conf. on Sens., Commun., and Netw. (SECON)},
  2016, pp. 1--9.

\bibitem{song2021enhancing}
Y.~Song, B.~Chen, T.~Wu, T.~Zheng, H.~Chen, and J.~Wang, ``Enhancing
  packet-level {Wi-Fi} device authentication protocol leveraging channel state
  information,'' \emph{Wireless Commun. Mob. Comput.}, vol. 2021, no.~1, p.
  2993019, Jan. 2021.

\bibitem{Liu2018Authenticating}
H.~Liu, Y.~Wang, J.~Liu, J.~Yang, Y.~Chen, and H.~V. Poor, ``Authenticating
  users through fine-grained channel information,'' \emph{{IEEE} Trans. Mobile
  Comput.}, vol.~17, no.~2, pp. 251--264, Feb. 2018.

\bibitem{Wang2022Channel}
H.-M. Wang and Q.-Y. Fu, ``Channel-prediction-based one-class mobile {IoT}
  device authentication,'' \emph{{IEEE} Internet Things J.}, vol.~9, no.~10,
  pp. 7731--7745, Oct. 2022.

\bibitem{Han2024Model}
J.~Han, Y.~Li, G.~Liu, J.~Ma, Y.~Zhou, H.~Fang, and X.~Wu, ``Model-driven
  learning for physical layer authentication in dynamic environments,''
  \emph{{IEEE} Commun. Lett.}, vol.~28, no.~3, pp. 572--576, Mar. 2024.

\bibitem{Chen2021Automated}
S.~Chen, Z.~Pang, H.~Wen, K.~Yu, T.~Zhang, and Y.~Lu, ``Automated labeling and
  learning for physical layer authentication against clone node and {Sybil}
  attacks in industrial wireless edge networks,'' \emph{{IEEE} Trans. Ind.
  Informat.}, vol.~17, no.~3, pp. 2041--2051, Mar. 2021.

\bibitem{Xie2022Weighted}
F.~Xie, Z.~Pang, H.~Wen, W.~Lei, and X.~Xu, ``Weighted voting in physical layer
  authentication for industrial wireless edge networks,'' \emph{{IEEE} Trans.
  Ind. Informat.}, vol.~18, no.~4, pp. 2796--2806, Apr. 2022.

\bibitem{pan2019threshold}
F.~Pan, Z.~Pang, H.~Wen, M.~Luvisotto, M.~Xiao, R.-F. Liao, and J.~Chen,
  ``Threshold-free physical layer authentication based on machine learning for
  industrial wireless {CPS},'' \emph{{IEEE} Trans. Ind. Informat.}, vol.~15,
  no.~12, pp. 6481--6491, Dec. 2019.

\bibitem{zhang2023cooperative}
T.~Zhang, Y.~Huo, Q.~Gao, L.~Ma, Y.~Wu, and R.~Li, ``Cooperative physical layer
  authentication with reputation-inspired collaborator selection,''
  \emph{{IEEE} Internet Things J.}, vol.~10, no.~24, Dec. 2023.

\bibitem{chen2023physical}
Y.~Chen, H.~He, S.~Liu, Y.~Zhang, Y.~Li, B.~Xing, B.~Guo, and L.~Chen,
  ``Physical layer authentication for industrial control based on convolutional
  denoising autoencoder,'' \emph{{IEEE} Internet Things J.}, vol.~11, no.~9,
  May 2023.

\bibitem{wang2022csi}
S.~Wang, K.~Huang, X.~Xu, Z.~Zhong, and Y.~Zhou, ``{CSI}-based physical layer
  authentication via deep learning,'' \emph{{IEEE} Wireless Commun. Lett.},
  vol.~11, no.~8, pp. 1748--1752, Aug. 2022.

\bibitem{liao2019security}
R.-F. Liao, H.~Wen, J.~Wu, F.~Pan, A.~Xu, H.~Song, F.~Xie, Y.~Jiang, and
  M.~Cao, ``Security enhancement for mobile edge computing through physical
  layer authentication,'' \emph{IEEE Access}, vol.~7, pp. 116\,390--116\,401,
  Aug. 2019.

\bibitem{Qiu2020Learning}
X.~Qiu, J.~Dai, and M.~Hayes, ``A learning approach for physical layer
  authentication using adaptive neural network,'' \emph{{IEEE} Access}, vol.~8,
  pp. 26\,139--26\,149, Feb. 2020.

\bibitem{Liu2014Practical}
H.~Liu, Y.~Wang, J.~Liu, J.~Yang, and Y.~Chen, ``Practical user authentication
  leveraging channel state information ({CSI}),'' in \emph{Proc. ACM Symp. on
  Inf., Comput. and Commun. Secur. (ASIA CCS)}.\hskip 1em plus 0.5em minus
  0.4em\relax ACM, 2014, p. 389–400.

\bibitem{Abdrabou2024Game}
M.~Abdrabou and T.~A. Gulliver, ``Game theoretic spoofing detection for space
  information networks using physical attributes,'' \emph{{IEEE} Trans.
  Commun.}, vol.~72, no.~7, pp. 3947--3956, Feb. 2024.

\bibitem{5934990}
K.~Zeng, K.~Govindan, D.~Wu, and P.~Mohapatra, ``Identity-based attack
  detection in mobile wireless networks,'' in \emph{Proc. IEEE INFOCOM}, 2011,
  pp. 1880--1888.

\bibitem{Kalamandeen2010Ensemble}
A.~Kalamandeen, A.~Scannell, E.~de~Lara, A.~Sheth, and A.~LaMarca, ``Ensemble:
  cooperative proximity-based authentication,'' in \emph{Proc. Int. Conf. on
  Mobile Syst., Appl., and Services (MobiSys)}, 2010, p. 331–344.

\bibitem{Wang2020NonlinearKalman}
J.~Wang, Y.~Shao, Y.~Wang, Y.~Ge, and R.~Yu, ``Physical layer authentication
  based on nonlinear {Kalman} filter for {V2X} communication,'' \emph{IEEE
  Access}, vol.~8, pp. 163\,746--163\,757, Sept. 2020.

\bibitem{Abdrabou2022Physical}
M.~Abdrabou and T.~A. Gulliver, ``Physical layer authentication for satellite
  communication systems using machine learning,'' \emph{IEEE Open J. Commun.
  Soc.}, vol.~3, pp. 2380--2389, Nov. 2022.

\bibitem{Romero2024Spoofing}
D.~Romero, T.~N. Ha, and P.~Gerstoft, ``Spoofing attack detection in the
  physical layer with robustness to user movement,'' in \emph{Proc. IEEE
  Wireless Commun. and Netw. Conf. (WCNC)}, 2024, pp. 1--6.

\bibitem{10.1145/3522783.3529522}
Y.~Jiang, L.~Jiao, L.~Zhao, and K.~Zeng, ``Beam pattern fingerprinting with
  missing features for spoofing attack detection in millimeter-wave networks,''
  in \emph{Proc. ACM Workshop on Wireless Secur. and Mach. Learn.
  (WiseML)}.\hskip 1em plus 0.5em minus 0.4em\relax ACM, 2022, p. 75–80.

\bibitem{9321212}
N.~Wang, L.~Jiao, P.~Wang, W.~Li, and K.~Zeng, ``Exploiting beam features for
  spoofing attack detection in {mmWave} {60-GHz} {IEEE} 802.11ad networks,''
  \emph{{IEEE} Trans. Wireless Commun.}, vol.~20, no.~5, pp. 3321--3335, May
  2021.

\bibitem{Du2023Physical}
R.~Du, L.~Zhen, and Y.~Liu, ``Physical layer authentication based on integrated
  semi-supervised learning in wireless networks for dynamic industrial
  scenarios,'' \emph{{IEEE} Trans. Veh. Technol.}, vol.~72, no.~5, pp.
  6154--6164, May 2023.

\bibitem{abdrabou2022authentication}
M.~Abdrabou and T.~A. Gulliver, ``Authentication for satellite communication
  systems using physical characteristics,'' \emph{IEEE Open J. Veh. Technol.},
  vol.~4, pp. 48--60, 2022.

\bibitem{chen2007detecting}
Y.~Chen, W.~Trappe, and R.~P. Martin, ``Detecting and localizing wireless
  spoofing attacks,'' in \emph{Proc. Annu. IEEE Int. Conf. on Sens., Commun.,
  and Netw. (SECON)}, 2007, pp. 193--202.

\bibitem{yang2012detection}
J.~Yang, Y.~Chen, W.~Trappe, and J.~Cheng, ``Detection and localization of
  multiple spoofing attackers in wireless networks,'' \emph{{IEEE} Trans.
  Parallel Distrib. Syst.}, vol.~24, no.~1, pp. 44--58, Jan. 2012.

\bibitem{gajbhiye2016rss}
Y.~Gajbhiye and R.~Daruwala, ``{RSS}-based spoofing detection and localization
  algorithm in {IEEE 802.11} wireless networks,'' in \emph{Proc. Int. Conf. on
  Commun. and Signal Process. (ICCSP)}.\hskip 1em plus 0.5em minus 0.4em\relax
  IEEE, 2016, pp. 1642--1645.

\bibitem{ventura2024authentication}
\BIBentryALTinterwordspacing
G.~Ventura, F.~Ardizzon, and S.~Tomasin, ``Authentication by location tracking
  in underwater acoustic networks,'' 2024. [Online]. Available:
  \url{https://arxiv.org/abs/2410.03511}
\BIBentrySTDinterwordspacing

\bibitem{8908721}
A.~Ijaz, M.~M.~U. Rahman, and O.~A. Dobre, ``On safeguarding visible light
  communication systems against attacks by active adversaries,'' \emph{IEEE
  Photonics Technol. Lett.}, vol.~32, no.~1, pp. 11--14, Jan. 2020.

\bibitem{He2024enhancing}
J.~He, M.~Niu, P.~Zhang, and C.~Qin, ``Enhancing {PHY}-layer authentication in
  {RIS}-assisted iot systems with cascaded channel features,'' \emph{IEEE
  Internet of Things Journal}, vol.~11, no.~14, pp. 24\,984--24\,997, 2024.

\bibitem{Joo2024Enhancing}
K.~Joo and W.~Choi, ``Enhancing security of {HRP} {UWB} ranging system based on
  channel characteristic analysis,'' \emph{{IEEE} Internet Things J.}, vol.~11,
  no.~24, pp. 39\,794--39\,808, 2024.

\bibitem{Diamant2019Cooperative}
R.~Diamant, P.~Casari, and S.~Tomasin, ``Cooperative authentication in
  underwater acoustic sensor networks,'' \emph{{IEEE} Trans. Wireless Commun.},
  vol.~18, no.~2, pp. 954--968, Feb. 2019.

\bibitem{Casari2022Physical}
P.~Casari, F.~Ardizzon, and S.~Tomasin, ``Physical layer authentication in
  underwater acoustic networks with mobile devices,'' in \emph{Proc. Int. Conf.
  on Underwater Netw. \& Syst. (WUWNet)}.\hskip 1em plus 0.5em minus
  0.4em\relax ACM, 2022.

\bibitem{Bragagnolo2021Authentication}
L.~Bragagnolo, F.~Ardizzon, N.~Laurenti, P.~Casari, R.~Diamant, and S.~Tomasin,
  ``Authentication of underwater acoustic transmissions via machine learning
  techniques,'' in \emph{Proc. IEEE Int. Conf. on Microw., Antennas, Commun.
  and Electron. Syst. (COMCAS)}, 2021, pp. 255--260.

\bibitem{Ardizzon2022Machine}
F.~Ardizzon, R.~Diamant, P.~Casari, and S.~Tomasin, ``Machine learning-based
  distributed authentication of {UWAN} nodes with limited shared information,''
  in \emph{Proc. Underwater Commun. and Netw. Conf. (UComms)}, 2022, pp. 1--5.

\bibitem{Ardizzon2024RNN}
F.~Ardizzon, P.~Casari, and S.~Tomasin, ``A {RNN}-based approach to physical
  layer authentication in underwater acoustic networks with mobile devices,''
  \emph{Comput. Netw.}, vol. 243, p. 110311, Apr. 2024.

\bibitem{Jedermann2021Orbit}
E.~Jedermann, M.~Strohmeier, M.~Sch\"{a}fer, J.~Schmitt, and V.~Lenders,
  ``Orbit-based authentication using {TDOA} signatures in satellite networks,''
  in \emph{Proc. ACM Conf. Secur. Privacy Wireless Mobile Netw. (WiSec)}.\hskip
  1em plus 0.5em minus 0.4em\relax New York, NY, USA: Association for Computing
  Machinery, 2021, p. 175–180.

\bibitem{Gao2019Lightweight}
S.~Gao, Y.~Ding, Y.~Lu, L.~Han, L.~Zhou, C.~Chen, X.~Yu, and X.~Huang, ``A
  lightweight fingerprint-based device authentication architecture for wireless
  industrial automation networks,'' in \emph{Proc. Int. Conf on Ind. Artif.
  Intell. (IAI)}, 2019, pp. 1--6.

\bibitem{Aman2018Impersonation}
W.~Aman, M.~M.~U. Rahman, J.~Qadir, H.~B. Pervaiz, and Q.~Ni, ``Impersonation
  detection in line-of-sight underwater acoustic sensor networks,''
  \emph{{IEEE} Access}, vol.~6, pp. 44\,459--44\,472, Aug. 2018.

\bibitem{Xiong2013SecureArray}
J.~Xiong and K.~Jamieson, ``{SecureArray}: improving wifi security with
  fine-grained physical-layer information,'' in \emph{Proc. Annu. Int. Conf. on
  Mobile Comput. \& Netw. (MobiCom)}, 2013, p. 441–452.

\bibitem{ning2020detection}
L.~Ning, B.~Li, C.~Zhao, Y.~Tao, and X.~Wang, ``Detection and localization of
  the eavesdropper in {MIMO} systems,'' \emph{{IEEE} Access}, vol.~8, pp.
  94\,984--94\,993, May 2020.

\bibitem{tao2023pilot}
Y.~Tao, X.~Wang, B.~Li, and C.~Zhao, ``Pilot spoofing attack detection and
  localization with mobile eavesdropper,'' \emph{{IEEE} Trans. Mobile Comput.},
  vol.~22, no.~3, pp. 1688--1701, Mar. 2023.

\bibitem{xiong2010secure}
J.~Xiong and K.~Jamieson, ``{SecureAngle}: improving wireless security using
  angle-of-arrival information,'' in \emph{Proc. ACM SIGCOMM Workshop on Hot
  Topics in Netw.}, 2010.

\bibitem{Bendaimi24how}
A.~Bendaimi, A.~Abdallah, A.~Celik, A.~M. Eltawil, and H.~Arslan, ``How to
  leverage double-structured sparsity of {RIS} channels to boost physical-layer
  authentication,'' \emph{IEEE Wireless Commun. Letters}, vol.~13, no.~8, pp.
  2260--2264, 2024.

\bibitem{Wang2018Efficient}
N.~Wang, L.~Jiao, P.~Wang, M.~Dabaghchian, and K.~Zeng, ``Efficient identity
  spoofing attack detection for {IoT} in {mm-Wave} and massive {MIMO} {5G}
  communication,'' in \emph{Proc. IEEE Global Commun. Conf. (GLOBECOM)}, 2018,
  pp. 1--6.

\bibitem{gao2023esanet}
N.~Gao, Q.~Huang, C.~Li, S.~Jin, and M.~Matthaiou, ``{EsaNet}: Environment
  semantics enabled physical layer authentication,'' \emph{{IEEE} Wireless
  Commun. Lett.}, vol.~13, no.~1, Jan. 2023.

\bibitem{Khalid2020Physical}
M.~Khalid, R.~Zhao, and N.~Ahmed, ``Physical layer authentication in
  line-of-sight underwater acoustic sensor networks,'' in \emph{Proc. Global
  Oceans}, 2020, pp. 1--5.

\bibitem{Yi2021Initial}
Q.-Y. Fu, Y.-H. Feng, H.-M. Wang, and P.~Liu, ``Initial satellite access
  authentication based on {Doppler} frequency shift,'' \emph{{IEEE} Wireless
  Commun. Lett.}, vol.~10, no.~3, pp. 498--502, Mar. 2021.

\bibitem{Topal2022Physical}
O.~A. Topal and G.~Karabulut~Kurt, ``Physical layer authentication for {LEO}
  satellite constellations,'' in \emph{Proc. IEEE Wireless Commun. and Netw.
  Conf. (WCNC)}, 2022, p. 1952–1957.

\bibitem{wang2013analysis}
T.~Wang and Y.~Yang, ``Analysis on perfect location spoofing attacks using
  beamforming,'' in \emph{Proc. IEEE INFOCOM}, 2013, pp. 2778--2786.

\bibitem{yilmaz2015survey}
M.~H. Yılmaz and H.~Arslan, ``A survey: Spoofing attacks in physical layer
  security,'' in \emph{Proc. IEEE Local Comput. Netw. Conf. Workshops (LCN
  Workshops)}, 2015, pp. 812--817.

\bibitem{sanders2020localizing}
C.~Sanders and Y.~Wang, ``Localizing spoofing attacks on vehicular {GPS} using
  vehicle-to-vehicle communications,'' \emph{{IEEE} Trans. Veh. Technol.},
  vol.~69, no.~12, pp. 15\,656--15\,667, 2020.

\bibitem{Neyman1933Problem}
J.~Neyman and E.~S. Pearson, ``{On the Problem of the Most Efficient Tests of
  Statistical Hypotheses},'' \emph{Phil. Trans. Roy. Soc. Lond. A}, vol. 231,
  no. 694-706, pp. 289--337, Feb. 1933.

\bibitem{Kay:1993}
S.~Kay, \emph{Fundamentals of Statistical Signal Processing: Estimation
  Theory}.\hskip 1em plus 0.5em minus 0.4em\relax Englewood Cliffs, NJ:
  Prentice-Hall, 1993.

\bibitem{Yin2021Online}
X.~Yin, X.~Fang, N.~Zhang, P.~Yang, X.~Sha, and J.~Qiu, ``Online learning aided
  adaptive multiple attribute-based physical layer authentication in dynamic
  environments,'' \emph{{IEEE} Trans. Netw. Sci. Eng.}, vol.~8, no.~2, pp.
  1106--1116, Feb. 2021.

\bibitem{wang2024transferable}
Q.~Wang, Z.~Pang, W.~Liang, J.~Zhang, and K.~Wang, ``Transferable physical
  layer authentication based on time-varying patterns toward zero training
  deployment for mobile {IIoT} devices,'' \emph{{IEEE} Trans. Ind. Informat.},
  vol.~20, no.~5, pp. 7675--7685, Feb. 2024.

\bibitem{Jaeckel2014QuaDRiGa}
S.~Jaeckel, L.~Raschkowski, K.~Börner, and L.~Thiele, ``{QuaDRiGa}: A 3-{D}
  multi-cell channel model with time evolution for enabling virtual field
  trials,'' \emph{{IEEE} Trans. Antennas Propag.}, vol.~62, no.~6, pp.
  3242--3256, Jun. 2014.

\bibitem{Kyosti2008WINNER}
P.~Kyösti, J.~Meinilä, L.~Hentila, X.~Zhao, T.~Jämsä, C.~Schneider,
  M.~Narandzi'c, M.~Milojevi'c, A.~Hong, J.~Ylitalo, V.-M. Holappa,
  M.~Alatossava, R.~Bultitude, Y.~Jong, and T.~Rautiainen, ``{IST-4-027756
  WINNER II D1.1.2 v1.2 WINNER II channel models},'' \emph{Inf. Soc. Technol},
  vol.~11, Feb. 2008.

\bibitem{bellhop}
\BIBentryALTinterwordspacing
{M.~Porter \emph{et al.}}, ``Bellhop {Gaussian} beam/finite element beam
  code,'' last accessed: Sept.~2024. [Online]. Available:
  \url{http://oalib.hlsresearch.com/Rays/index.html}
\BIBentrySTDinterwordspacing

\bibitem{Morozs2020Channel}
N.~{Morozs}, W.~{Gorma}, B.~T. {Henson}, L.~{Shen}, P.~D. {Mitchell}, and Y.~V.
  {Zakharov}, ``Channel modeling for underwater acoustic network simulation,''
  \emph{{IEEE} Access}, vol.~8, pp. 136\,151--136\,175, July 2020.

\bibitem{Qarabaqi2013Statistical}
P.~Qarabaqi and M.~Stojanovic, ``Statistical characterization and
  computationally efficient modeling of a class of underwater acoustic
  communication channels,'' \emph{{IEEE} J. Ocean. Eng.}, vol.~38, no.~4, pp.
  701--717, Apr. 2013.

\bibitem{NORAD19CelesTrak}
\BIBentryALTinterwordspacing
N.~(NORAD), ``Celestrak: {N}orad two-line element sets current data,'' 2019,
  last access: Oct. 2024. [Online]. Available:
  \url{https://celestrak.org/NORAD/elements/}
\BIBentrySTDinterwordspacing

\bibitem{ansysSTK}
\BIBentryALTinterwordspacing
Ansys, ``{Ansys System Tool Kit home Page}.'' [Online]. Available:
  \url{https://www.ansys.com/products/missions/ansys-stk}
\BIBentrySTDinterwordspacing

\bibitem{candell2016radio}
\BIBentryALTinterwordspacing
R.~Candell, ``Radio frequency measurements for selected manufacturing and
  industrial environments,'' 2016. [Online]. Available:
  \url{https://data.nist.gov/od/id/mds0139sck}
\BIBentrySTDinterwordspacing

\bibitem{wangning2025sls}
\BIBentryALTinterwordspacing
N.~Wang, ``802.11ad {SLS} {SNR} trace-based authentication,'' 2025. [Online].
  Available:
  \url{https://github.com/wangning8566/SLS-SNR-trace-based-authentication}
\BIBentrySTDinterwordspacing

\bibitem{Lohan2017Wi-Fi}
E.~S. Lohan, J.~Torres-Sospedra, H.~Leppäkoski, P.~Richter, Z.~Peng, and
  J.~Huerta, ``{Wi-Fi} crowdsourced fingerprinting dataset for indoor
  positioning,'' \emph{Data}, vol.~2, no.~4, Oct. 2017.

\bibitem{AlQahtani2023RSSI}
\BIBentryALTinterwordspacing
A.~A.~S. AlQahtani and T.~Alshayeb, ``{RSSI} measurements of beacon frames from
  {Wi-Fi} radio waves,'' 2023. [Online]. Available:
  \url{https://dx.doi.org/10.21227/2bk3-dw90}
\BIBentrySTDinterwordspacing

\bibitem{Walree2016Watermark}
P.~van Walree, R.~Otnes, and T.~Jenserud, ``Watermark: A realistic benchmark
  for underwater acoustic modems,'' in \emph{Proc. Underwater Commun. and Netw.
  Conf. (UComms)}, 2016, pp. 1--4.

\bibitem{Jianchun2019Channel}
\BIBentryALTinterwordspacing
J.~Huang and R.~Diamant, ``Channel impulse responses from {M}ar. 2019 long
  range experiment ({M}editerranean {S}ea),'' 2019. [Online]. Available:
  \url{https://dx.doi.org/10.21227/nzgr-ds72}
\BIBentrySTDinterwordspacing

\bibitem{li2020sdr}
X.~Li, J.~Liu, B.~Ding, Z.~Li, H.~Wu, and T.~Wang, ``A {SDR}-based verification
  platform for 802.11 {PHY} layer security authentication,'' \emph{World Wide
  Web}, vol.~23, pp. 1011--1034, Jan. 2020.

\bibitem{Candell2017Industrial}
R.~Candell, C.~Remley, J.~Quimby, D.~Novotny, A.~Curtin, P.~Papazian,
  G.~Koepke, J.~Diener, and M.~Hany, ``Industrial wireless systems: Radio
  propagation measurements,'' Jan. 2017.

\bibitem{quimby2017nist}
J.~Quimby, R.~Candell, C.~Remley, D.~Novotny, J.~Diener, P.~Papazian,
  A.~Curtin, and G.~Koepke, ``\BIBforeignlanguage{en}{{NIST} channel sounder
  overview and channel measurements in manufacturing facilities},'' Nov. 2017.

\bibitem{wang2016user}
W.~Wang, Z.~Sun, K.~Ren, and B.~Zhu, ``User capacity of wireless physical-layer
  identification: An information-theoretic perspective,'' in \emph{Proc. IEEE
  Int. Conf. on Commun. (ICC)}.\hskip 1em plus 0.5em minus 0.4em\relax IEEE,
  2016, pp. 1--6.

\bibitem{wang2017user}
------, ``User capacity of wireless physical-layer identification,'' \emph{IEEE
  Access}, vol.~5, pp. 3353--3368, 2017.

\bibitem{saeif2023day}
A.~Saeif, S.~Savio, and O.~Gabriele, ``The day-after-tomorrow: On the
  performance of radio fingerprinting over time,'' in \emph{Proc. Annual Comp
  Secur. Applications Conf.}, 2023, pp. 439--450.

\bibitem{elmaghbub2024no}
A.~Elmaghbub and B.~Hamdaoui, ``No blind spots: On the resiliency of device
  fingerprints to hardware warm-up through sequential transfer learning,'' in
  \emph{Proc. ACM Conf. Secur. Privacy Wireless Mobile Netw. (WiSec)}, 2024,
  pp. 134--144.

\bibitem{adesina2022adversarial}
D.~Adesina, C.-C. Hsieh, Y.~E. Sagduyu, and L.~Qian, ``Adversarial machine
  learning in wireless communications using {RF} data: A review,'' \emph{{IEEE}
  Commun. Surveys Tuts.}, vol.~25, no.~1, pp. 77--100, Jan. 2022.

\bibitem{liu2023exploring}
Z.~Liu, C.~Xu, Y.~Xie, E.~Sie, F.~Yang, K.~Karwaski, G.~Singh, Z.~L. Li,
  Y.~Zhou, D.~Vasisht \emph{et~al.}, ``Exploring practical vulnerabilities of
  machine learning-based wireless systems,'' in \emph{Proc. USENIX Symposium on
  Networked Systems Design and Implementation (NSDI 23)}, 2023, pp. 1801--1817.

\bibitem{zhao2024explanation}
T.~Zhao, X.~Wang, J.~Zhang, and S.~Mao, ``Explanation-guided backdoor attacks
  on model-agnostic {RF} fingerprinting,'' in \emph{Proc. IEEE INFOCOM}, 2024,
  pp. 221--230.

\bibitem{zhao2024backdoor}
T.~Zhao, N.~Wang, Y.~Wu, W.~Zhang, and X.~Wang, ``Backdoor attacks against
  low-earth orbit satellite fingerprinting,'' in \emph{Proc. IEEE INFOCOM
  Workshops}, 2024, pp. 01--06.

\bibitem{bao2021threat}
Z.~Bao, Y.~Lin, S.~Zhang, Z.~Li, and S.~Mao, ``Threat of adversarial attacks on
  dl-based iot device identification,'' \emph{{IEEE} Internet Things J.},
  vol.~9, no.~11, pp. 9012--9024, Sep. 2021.

\bibitem{liu2023robust}
B.~Liu, H.~Zhang, Y.~Wan, F.~Zhou, Q.~Wu, and D.~W.~K. Ng, ``Robust adversarial
  attacks on deep learning-based rf fingerprint identification,'' \emph{{IEEE}
  Wireless Commun. Lett.}, vol.~12, no.~6, pp. 1037--1041, Dec. 2023.

\bibitem{ma2023white}
J.~Ma, J.~Zhang, G.~Shen, A.~Marshall, and C.-H. Chang, ``White-box adversarial
  attacks on deep learning-based radio frequency fingerprint identification,''
  in \emph{Proc. IEEE Int. Conf. on Commun. (ICC)}, 2023, pp. 3714--3719.

\bibitem{papangelo2024adversarial}
L.~Papangelo, M.~Pistilli, S.~Sciancalepore, G.~Oligeri, G.~Piro, and
  G.~Boggia, ``Adversarial machine learning for image-based radio frequency
  fingerprinting: Attacks and defenses,'' \emph{{IEEE} Commun. Mag.}, 2024.

\bibitem{li2024slpa}
W.~Li, S.~Wang, Y.~Zhang, L.~Guo, Y.~Liu, Y.~Lin, and G.~Gui, ``Slpa:
  Single-line pixel attack on specific emitter identification using
  time-frequency spectrogram,'' \emph{{IEEE} Trans. Veh. Technol.}, 2024.

\bibitem{yin2025evasion}
G.~Yin, J.~Zhang, X.~Yi, and X.~Wang, ``Evasion attacks and countermeasures in
  deep learning-based {Wi-Fi} gesture recognition,'' \emph{{IEEE} Trans. Mobile
  Comput.}, 2025.

\bibitem{abanto2020stay}
L.~F. Abanto-Leon, A.~B{\"a}uml, G.~H. Sim, M.~Hollick, and A.~Asadi, ``Stay
  connected, leave no trace: Enhancing security and privacy in wifi via
  obfuscating radiometric fingerprints,'' \emph{Proc. ACM on Meas. and Anal. of
  Comput. Syst.}, vol.~4, no.~3, pp. 1--31, 2020.

\bibitem{10437915}
T.~M. Pham, L.~Senigagliesi, M.~Baldi, G.~P. Fettweis, and A.~Chorti, ``Machine
  learning-based robust physical layer authentication using angle of arrival
  estimation,'' in \emph{Proc. IEEE Global Commun. Conf. (GLOBECOM)}, 2023, pp.
  13--18.

\bibitem{zhao2024generative}
C.~Zhao, H.~Du, D.~Niyato, J.~Kang, Z.~Xiong, D.~I. Kim, X.~Shen, and K.~B.
  Letaief, ``Generative ai for secure physical layer communications: A
  survey,'' \emph{{IEEE} Trans. on Cogn. Commun. Netw.}, pp. 3 -- 26, 2025.

\bibitem{wang2024ai}
N.~Wang, T.~Zhao, S.~Mao, and X.~Wang, ``{AI} generated wireless data for
  enhanced satellite device fingerprinting,'' in \emph{Proc. IEEE Int. Conf. on
  Commun. Workshops}, 2024, pp. 88--93.

\bibitem{Zhou2024Large}
H.~Zhou, C.~Hu, Y.~Yuan, Y.~Cui, Y.~Jin, C.~Chen, H.~Wu, D.~Yuan, L.~Jiang,
  D.~Wu, X.~Liu, C.~Zhang, X.~Wang, and J.~Liu, ``Large language model ({LLM})
  for telecommunications: A comprehensive survey on principles, key techniques,
  and opportunities,'' \emph{{IEEE} Commun. Surveys Tuts.}, pp. 1--1, 2024.

\bibitem{Merchant2019Securing}
K.~Merchant and B.~Nousain, ``Securing {IoT} {RF} fingerprinting systems with
  generative adversarial networks,'' in \emph{Proc. of Military Commun. Conf.
  (MILCOM)}, 2019, pp. 584--589.

\bibitem{chen2024optic}
X.~Chen, Z.~Liu, X.~Zhang, Y.~Wang, D.~Shi, and X.~Liu, ``Optic fingerprint:
  Enhancing security in visible light communication networks,'' in \emph{Proc.
  IEEE INFOCOM Workshops}, 2024, pp. 1--6.

\end{thebibliography}

 \begin{IEEEbiography}[{\includegraphics[angle=0,width=1in,clip,keepaspectratio]{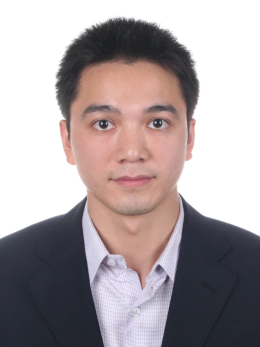}}]{Junqing Zhang}
 received a Ph.D. degree in Electronics and Electrical Engineering from Queen's University Belfast, UK in 2016. From Feb. 2016 to Jan. 2018, he was a Postdoctoral Research Fellow at Queen's University Belfast. From Feb. 2018 to Oct. 2022, he was a Tenure Track Fellow and then a Lecturer (Assistant Professor) at the University of Liverpool, UK. Since Oct. 2022, he has been a Senior Lecturer (Associate Professor) at the University of Liverpool. His research interests include the Internet of Things, wireless security, physical layer security, key generation, radio frequency fingerprint identification, and wireless sensing. Dr. Zhang is a co-recipient of the IEEE WCNC 2025 Best Workshop Paper Award. He is a Senior Area Editor of IEEE Transactions on Information Forensics and Security and an Associate Editor of IEEE Transactions on Mobile Computing.
 \end{IEEEbiography}

 \begin{IEEEbiography}[{\includegraphics[width=1in,clip,keepaspectratio]{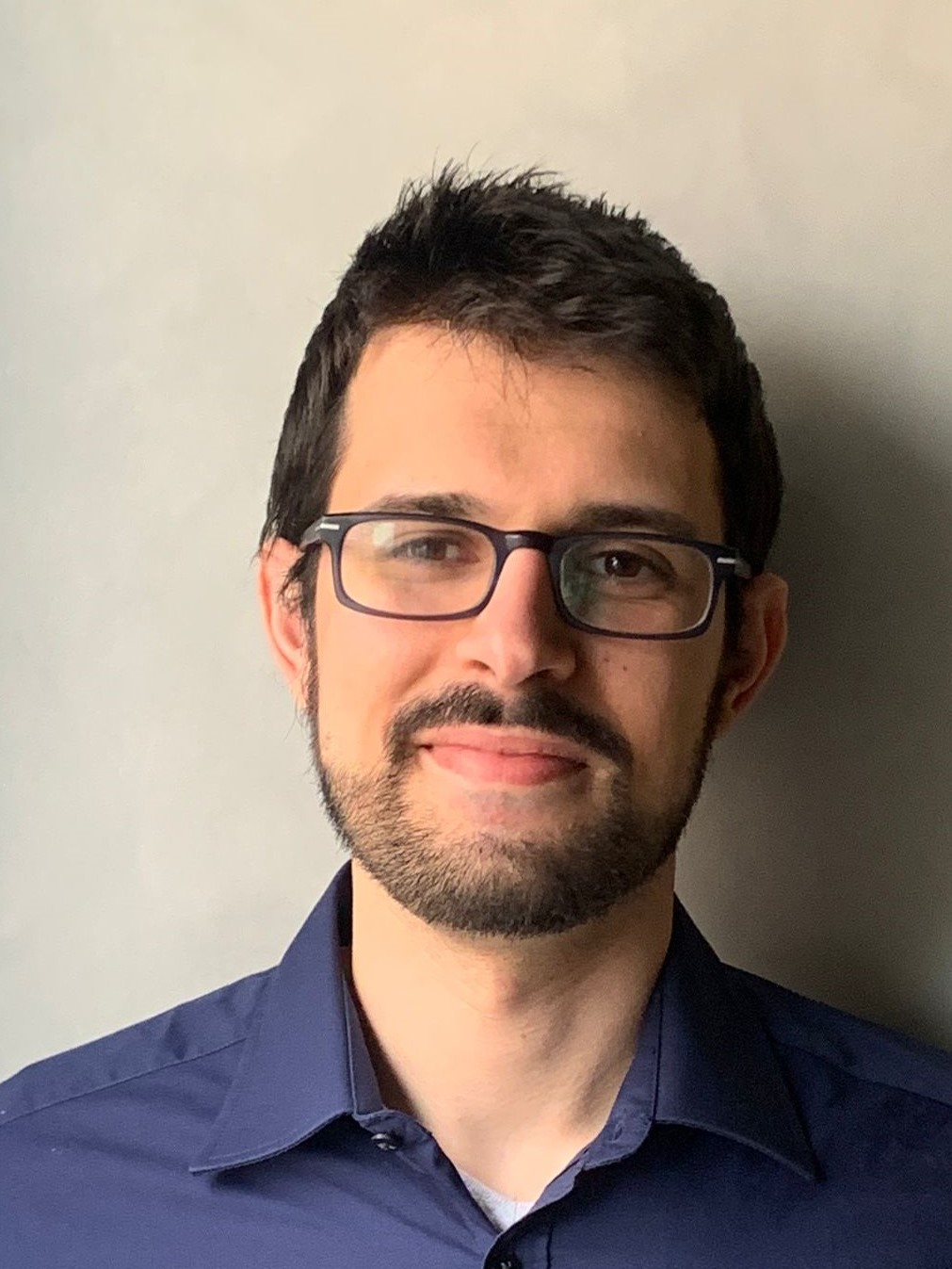}}]%
 {Francesco Ardizzon}{\space}(Member, IEEE) received the B.Sc. degree in 2016, the M.Sc. degree in 2019, and the Ph.D. degree in Information Engineering in 2023 from the University of Padova, Italy. In 2022, he was a visiting scientist at the ESA European Space Research and Technology Centre. He is currently an Assistant Professor at the University of Padova. His current research interests include authentication for global navigation satellite systems, physical layer security, and underwater acoustic communications.
 \end{IEEEbiography}

\begin{IEEEbiography}[{\includegraphics[width=1in,clip,keepaspectratio]{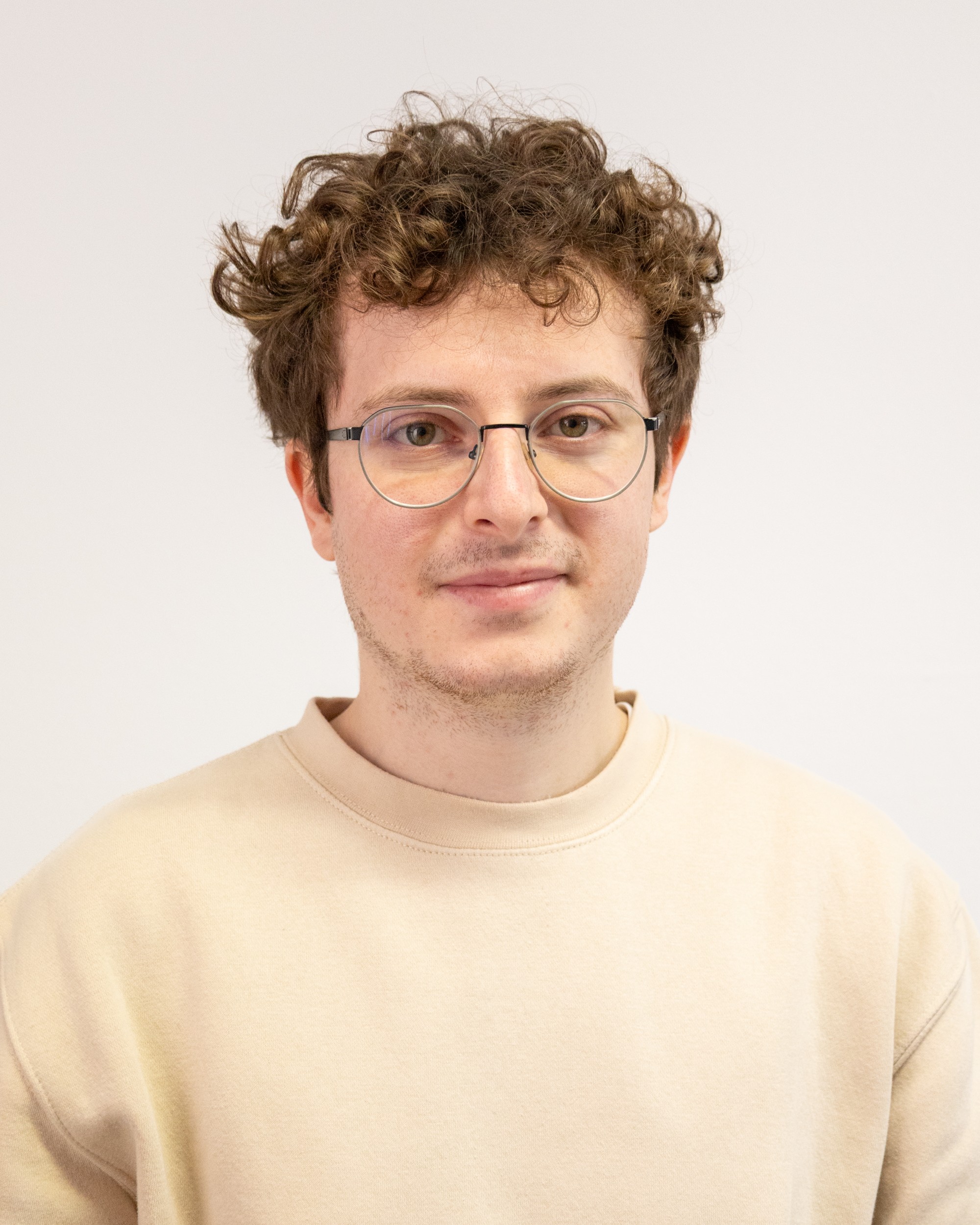}}]%
 {Mattia Piana}  (Student Member, IEEE) received a B.Sc. in Information Engineering and an M. Sc in Telecommunication Engineering from the University of Padova in 2021 and 2023, respectively. In 2023, he was at National Instruments (Dresden, Germany), where he collaborated on the development of novel techniques for mmWave antenna characterization. He is currently a PhD student at the University of Padova within the EU ROBUST-6G project, and his research interests include physical layer security and reflective intelligent surfaces. 
 \end{IEEEbiography}

 \begin{IEEEbiography}[{\includegraphics[angle=0,width=1in,clip,keepaspectratio]{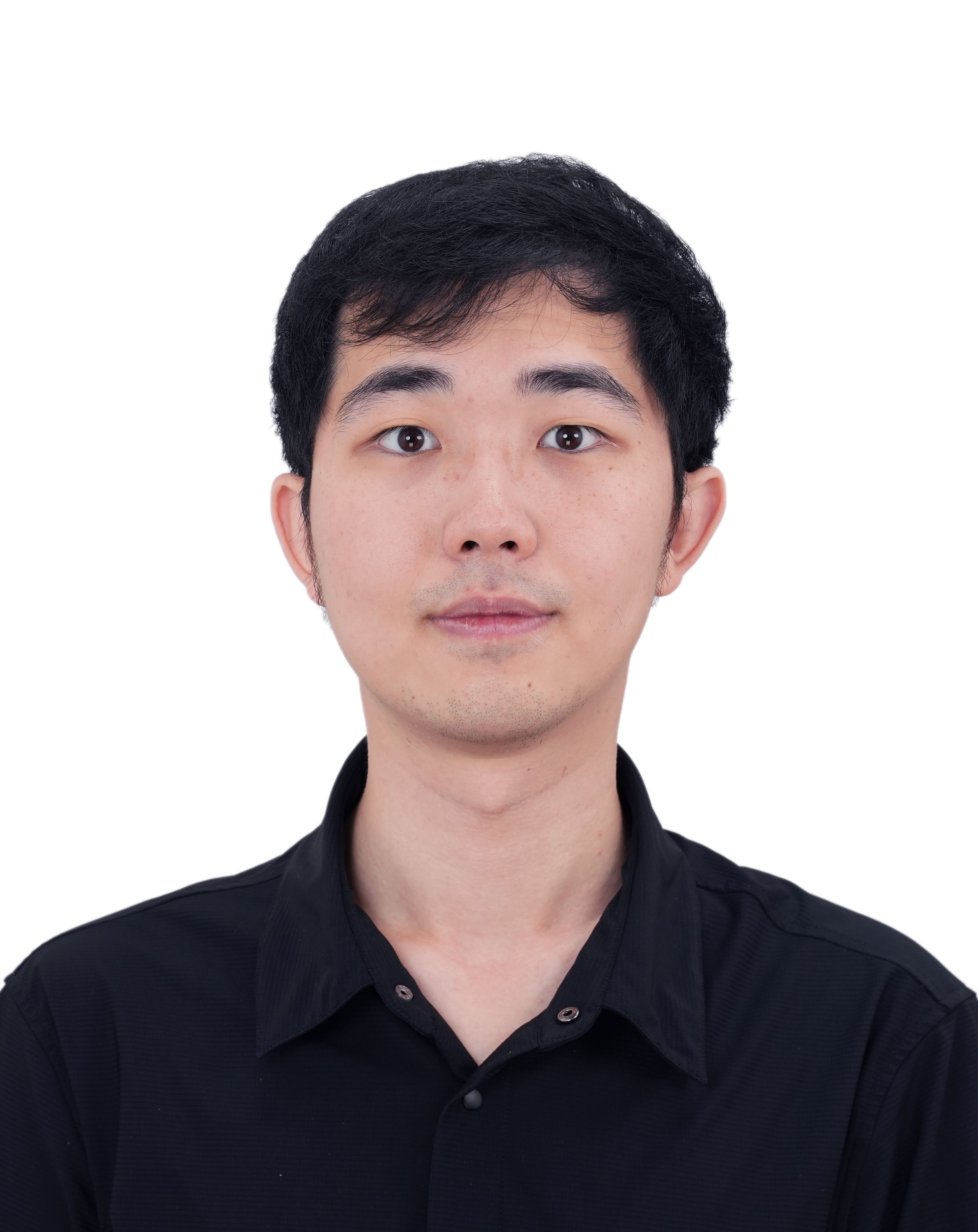}}]{Guanxiong Shen}
 received the B.Eng degree from Xidian University, Xi'an, China, in 2019, and the Ph.D degree from the University of Liverpool, UK, in 2023. He is currently an Associate professor at Southeast University, Nanjing, China. His research interests include the Internet of Things, wireless security, physical layer security, and radio frequency fingerprint identification.	
 \end{IEEEbiography}

\begin{IEEEbiography}
     [{\includegraphics[width=1in,clip,keepaspectratio]{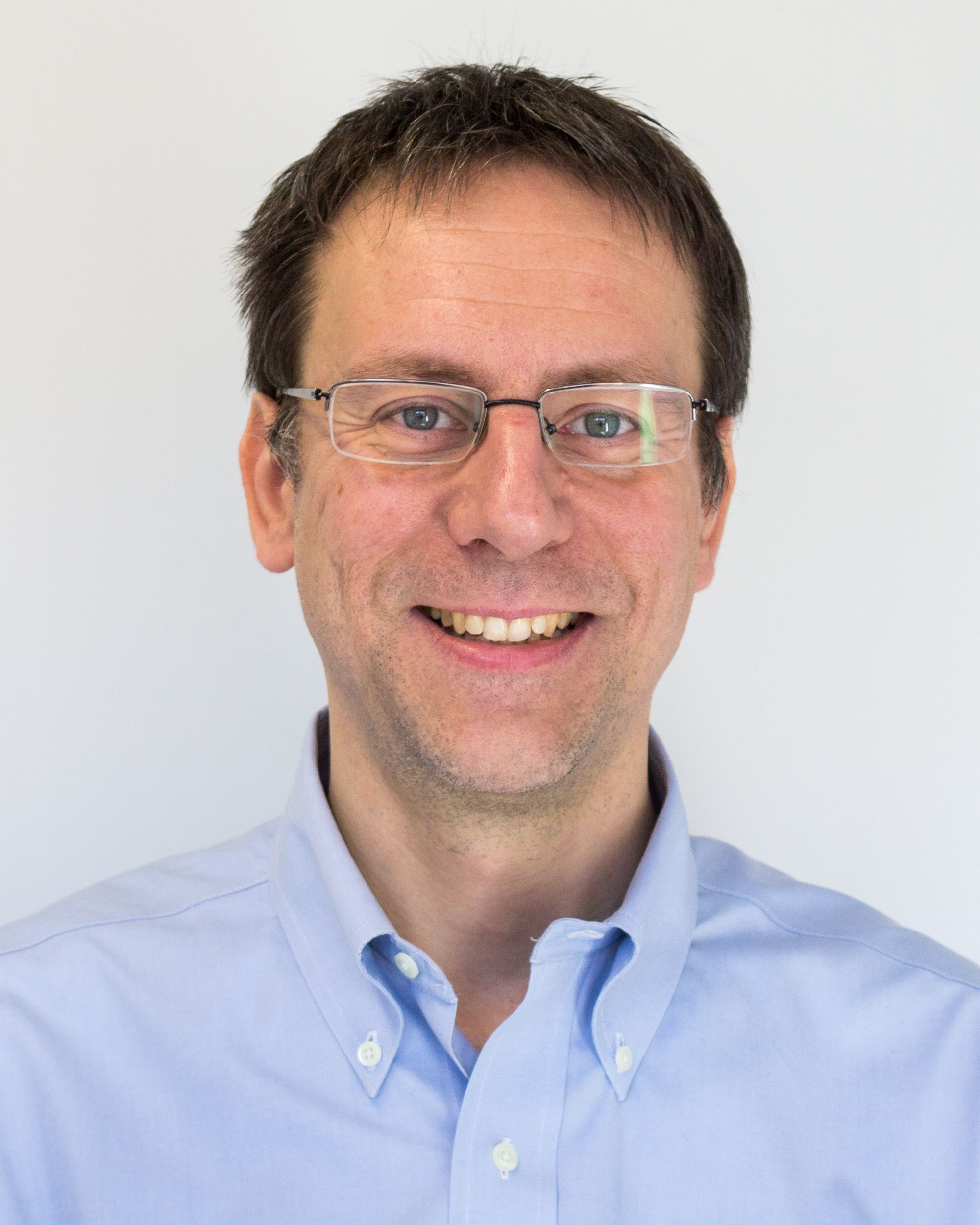}}]{Stefano Tomasin}
  received the Ph.D. degree from the University of Padova, Italy (2003), where he is now a Full Professor. During his career, he has visited IBM Research (Switzerland), Philips Research (Netherlands),  Qualcomm (California), the Polytechnic University in Brooklyn (New York), and  Huawei  (France). His current research interests include physical layer security, security of global navigation satellite systems, signal processing for wireless communications, synchronization, and scheduling of communication resources. He is a senior member of IEEE and a member of EURASIP. He is or has been an Editor of the IEEE Transactions on Vehicular Technologies, the IEEE Transactions on Signal Processing (2017-2020), the EURASIP Journal of Wireless Communications and Networking, and the IEEE Transactions on Information Forensics and Security.
 \end{IEEEbiography}

\end{document}